\documentclass[smallextended]{svjour3}
\smartqed  % flush right qed marks, e.g. at end of proof

\usepackage[ruled,vlined,lined,commentsnumbered,linesnumbered]{algorithm2e}
\usepackage{booktabs}
\usepackage{moreverb}
\usepackage{fontenc}
\usepackage{amsmath}
\usepackage{amssymb}
\usepackage{fancybox}
\usepackage{color}
\usepackage{colortbl}
\usepackage{array}
\usepackage{multirow}
\usepackage{multicol}
\usepackage{listings}
\usepackage{graphicx}
\usepackage{makecell}
\usepackage[nounderscore]{syntax}

\usepackage{balance}
\usepackage{csvsimple}
\usepackage{longtable}
\usepackage[font=small,labelfont=bf,textfont=bf]{caption}
\usepackage{url}
\usepackage{lscape}
\usepackage{rotating}
\usepackage{tikz}
\usepackage[normalem]{ulem}
\usepackage{enumitem}
\usepackage{pifont}% http://ctan.org/pkg/pifont%\usepackage{subfigure}

\usepackage{commath}
\usepackage[most]{tcolorbox}

\usepackage{threeparttable}

\usepackage{marvosym}

\usepackage{array}

\usepackage{cite}
\usepackage{subfigure}

\usetikzlibrary{intersections}
\usetikzlibrary{calc}

\usepackage{amsmath,amssymb,amsfonts}
\usepackage{algorithmic}
\usepackage{graphicx}
\usepackage{textcomp}

\usepackage{MnSymbol} % Minion Pro Symbols
\usepackage{wrapfig}

\def\BibTeX{{\rm B\kern-.05em{\sc i\kern-.025em b}\kern-.08em
    T\kern-.1667em\lower.7ex\hbox{E}\kern-.125emX}}

\setlist{noitemsep} %to leave space around whole list

\newcommand{\cmark}{\ding{51}}%
\newcommand{\xmark}{\ding{55}}%

\newcolumntype{L}[1]{>{\raggedright\let\newline\\\arraybackslash\hspace{0pt}}m{#1}}
\newcolumntype{C}[1]{>{\centering\let\newline\\\arraybackslash\hspace{0pt}}m{#1}}
\newcolumntype{R}[1]{>{\raggedleft\let\newline\\\arraybackslash\hspace{0pt}}m{#1}}

\definecolor{codegreen}{rgb}{0,0.6,0}
\definecolor{codegray}{rgb}{0.5,0.5,0.5}
\definecolor{codepurple}{rgb}{0.58,0,0.82}
\definecolor{backcolour}{rgb}{0.95,0.95,0.92}

\definecolor{yellow}{RGB}{255,255,153}
\definecolor{grey}{RGB}{224,224,224}

\newcommand{\find}[1]{
	\begin{tcolorbox}[tile,size=fbox,boxsep=2mm,boxrule=0pt,top=0pt,bottom=0pt,borderline west={1.5mm}{0pt}{blue!50!white},colback=blue!5!white]
	\em #1
\end{tcolorbox}
}

\newcommand{\quest}[1]{
	\begin{tcolorbox}[tile,size=fbox,boxsep=2mm,boxrule=0pt,top=0pt,bottom=0pt,borderline west={1.5mm}{0pt}{black!50!white},colback=black!5!white]
	\em #1
\end{tcolorbox}
}

\newboolean{showcomments}
\setboolean{showcomments}{true}
\ifthenelse{\boolean{showcomments}}
 { \newcommand{\mynote}[2]{
      \fbox{\bfseries\sffamily\scriptsize#1}
        {\small$\blacktriangleright$\textsf{\emph{#2}}$\blacktriangleleft$}}}
        { \newcommand{\mynote}[2]{}}

\definecolor{DarkOrange}{rgb}{0.8,0.3,0.0}
\definecolor{DarkCyan}{rgb}{0.0, 0.55, 0.55}

\newcommand{\toolname}{\texttt{FixMiner}\xspace}
\newcommand{\apr}{\texttt{PAR\textsubscript{FixMiner}}\xspace}
\newcommand{\ediff}{\texttt{Rich Edit Script}\xspace}

\usepackage{newfloat}

\DeclareFloatingEnvironment[
  fileext   = logr,% don't use log here!
  listname  = {List of Grammars},
  name      = Grammar,
  placement = htp
]{Grammar}

\definecolor{darkpastelgreen}{rgb}{0.01, 0.75, 0.24}

\definecolor{darkpastelred}{rgb}{0.76, 0.23, 0.13}

\definecolor{ao(english)}{rgb}{0.0, 0.5, 0.0}

\definecolor{darkolivegreen}{rgb}{0.33, 0.42, 0.18}

\lstdefinelanguage{diff}{
  morecomment=[f][\color{blue}]{@@},     % group identifier
  morecomment=[f][\color{darkpastelred}]-,         % deleted lines
  morecomment=[f][\color{ao(english)}]+,       % added lines
  morecomment=[f][\color{black}]{---}, % Diff header lines (must appear after +,-)
  morecomment=[f][\color{black}]{+++},
}

\RequirePackage[normalem]{ulem} %DIF PREAMBLE
\RequirePackage{color}\definecolor{RED}{rgb}{1,0,0}\definecolor{BLUE}{rgb}{0,0,1} %DIF PREAMBLE
\begin{document}

\title{
FixMiner: Mining Relevant Fix Patterns for Automated Program Repair
}
\titlerunning{FixMiner for Automated Program Repair}

\author{
%\alignauthor
	Anil~Koyuncu
\and
	Kui~Liu
\and
	Tegawend\'e~F.~Bissyand\'e
\and
	Dongsun~Kim
\and
	Jacques~Klein
\and
	Martin~Monperrus
\and
	Yves~Le~Traon
}%

\institute{A. Koyuncu, K. Liu, T. F. Bissyand\'e,  J. Klein, and Y. Le Traon\at
			SnT, University of Luxembourg \\
			\email{\{anil.koyuncu, kui.liu, tagewende.bissyande, jacques.klein, yves.le traon\}@uni.lu}
			\and
			D. Kim \at
			Furiosa.ai \\
			\email{darkrsw@furiosa.ai}
			\and
			M. Monperrus \at
			KTH Royal Institute of Technology \\
			\email{martin.monperrus@csc.kth.se}
}

\date{Received: date / Accepted: date}

\maketitle

\begin{abstract}
Patching is a common activity in software development.
It is generally performed on a source code base to address bugs
or add new functionalities. In this context, given the recurrence of bugs across projects, the associated similar patches can be leveraged to extract generic fix actions. While the literature includes various approaches %'works' is not allowed somehow in English when metioning different kinds of existing 'research' work.
leveraging similarity among patches to guide program repair, these approaches often do not yield fix patterns that are tractable and reusable as actionable input to APR systems.
%Code comprehension is critical in software maintenance. Towards providing tools and approaches to support maintenance tasks, researchers have investigated various research lines related to how software code can be described in an abstract form.
%So far, studies on change pattern mining, code clone detection, or semantic patch inference have mainly adopted text-, token- and tree-based representations as the basis for computing similarity among code fragments.
% Although, in general, existing techniques form clusters of  ``similar'' code, our experience in patch mining has revealed that clusters of patches formed
% by such techniques do not usually lead to actionable patterns that are leveraged by the community.

In this paper, we propose a systematic and automated approach to mining relevant and actionable fix patterns based on an iterative clustering strategy applied to atomic changes within patches. The goal of FixMiner is thus to infer separate and reusable fix patterns that can be leveraged in other patch generation systems.
Our technique, \toolname, leverages \ediff which is a specialized tree structure of the edit scripts that captures the AST-level context of the code changes. 
% \kui{It seems that the previous sentence and the following sentence are not well connected.}
\toolname uses different tree representations of \ediff{s} for each round of clustering to identify similar changes. These are abstract syntax trees, edit actions trees, and code context trees.

We have evaluated \toolname on thousands of software patches collected from open source projects. Preliminary results show that we are able to mine accurate patterns, efficiently exploiting change
information in \ediff{s}. 
% Eventually, \toolname yields patterns that are actionable and can be integrated easily to automated program repair systems. 
We further integrated the mined patterns to an automated program repair prototype, \apr, with which we are able to correctly fix 26 bugs of the Defects4J benchmark. Beyond this quantitative performance, we show that the mined fix patterns are sufficiently relevant to produce patches with a high probability of correctness: 81\% of \apr's generated plausible patches are correct.

% as well as semantic information. %Additionally, our approach yields low computation complexity, and can thus scale to large code datasets.

\end{abstract}

%
%\begin{IEEEkeywords}
%Automated program repair, fix patterns, code differencing
%\end{IEEEkeywords}

%\begin{center}
%{\em Your code could be my patch.}
%\end{center}

% !TEX root = main.tex

\section{Introduction}
\label{sec:intro}

Code change patterns have various uses in the software engineering domain. They are notably used for labeling changes~\cite{pan2009toward}, triaging developer commits~\cite{tian2012identifying} or predicting changes~\cite{ying2004predicting}.
In recent years, fix patterns have been heavily leveraged in the software maintenance community, notably for building patch generation systems, which now attract growing interest in the literature~\cite{monperrus2018automatic}.
Automated Program Repair (APR) has indeed gained incredible momentum, and various approaches~\cite{nguyen2013semfix,westley2009automatically,le2012genprog,kim2013automatic,coker2013program,ke2015repairing,mechtaev2015directfix,long2015staged,le2016enhancing,le2016history,long2016automatic,chen2017contract,le2017s3,long2017automatic,xuan2017Nopol,xiong2017Precise,jiang2018shaping,wen2018context,hua2018towards,liu2019avatar,liu2019you} have been proposed, aiming at reducing manual debugging efforts through automatically generating patches.
A common and reliable strategy in automatic program repair is to generate concrete patches based on fix patterns~\cite{kim2013automatic} (also referred to as fix templates~\cite{liu2018mining} or program transformation schemas~\cite{hua2018towards}).
Several APR systems~\cite{kim2013automatic,saha2017elixir,durieux2017dynamic,liu2018mining,hua2018towards,martinez2018ultra,liu2019avatar,liu2019you} in the literature implement this strategy by using {\bf diverse sets of fix patterns} obtained either via manual generation or automatic mining of bug fix datasets.

In PAR~\cite{kim2013automatic}, the authors mined fix patterns by inspecting 60,000  developer patches manually. Similarly, for Relifix~\cite{tan2015relifix}, a manual inspection of  73 real software regression bug fixes is performed to infer fix patterns. %\ak{do we need a new paragraph here or before with genesis?}
Manual mining is however tedious, error-prone, and cannot scale. Thus, in order to overcome the limitations of manual pattern inference, several research groups have initiated studies towards automatically inferring bug fix patterns. With Genesis~\cite{long2017automatic}, Long {\em et al.} proposed to automatically infer code transforms for patch generation.
Genesis infers 108 code transforms, from a space of 577 sampled transforms, with specific code contexts. However, this work limits the search space to previously successful patches from only three classes of defects of Java programs: null pointer, out of bounds, and class cast related defects.
% The transforms synthesizes the logic and the replacement code of the patches and the inferred transforms summarize common patterns of the patches that developers.

% Fluri et al.~\cite{fluri2008discovering} have used hierarchical clustering focusing on 41 basic change types in a semi-automated way.
% \textcolor{blue}{However, the granularity of their patterns limits to statement level, which cannot reach the code change actions with more fine-grained granularity, such as sub-expression level of statements.}\ak{not apr}

% In recent literature, the software engineering community has mostly proposed automated program repair techniques that generate patches based on templates that are built from manually identified common repair patterns (e.g., add/modify if-conditions, alter method parameters, etc.~\cite{yue2017characterization}). PAR~\cite{kim2013automatic} and Relifix~\cite{tan2015relifix} are examples of such state-of-the-art techniques which leverage manually extracted templates to either help build human-readable patches or to fix common errors.
% % }

% \textcolor{blue}{Manually summarizing fix patterns is, however, a tedious and unrealistic endeavour that will take a lot of manual efforts and might yield incorrect patterns. Therefore, automatically discovering bug fix patterns is proposed to address this issue and explore the challenges in achieving the diversity and reliability of fix patterns.}
% \textcolor{red}{Automatically discovering bug fix patterns is, however, a challenging endeavour.}
% \textcolor{blue}{

% \ak{where is QAFix?}
Liu and Zhong~\cite{liu2018mining} proposed SOFix to explore fix patterns for Java programs from Q\&A posts in Stack Overflow, which mines patterns based on GumTree~\cite{falleri2014Fine} edit scripts, and builds different categories based on repair pattern isomorphism. SOFix then mines a repair pattern from each category. However, the authors note that most of the categories are redundant or even irrelevant, mainly due to two major issues: (1) a considerable portion of code samples are designed for purposes other than repairing bugs; (2) since the underlying GumTree tool relies on structural positions to extract modifications, these ``modifications do not present the desirable semantic mappings''. They relied on heuristics for manually filtering categories (e.g., categories that contain several modifications), and then after SOFIX mines repair patterns they have to manually select useful ones (e.g., merging some repair patterns due to their similar semantics).

% Their mined fix patterns contain the ambiguous context, such as the fix pattern ``IfChecker: Insert If under $\ast$'', where $\ast$ could be any code entities and ``IfChecker'' could be any kinds of {\tt If-Statement}, which will enlarge the search space of patch candidates and further increase the quantity of nonsensical patches and possibility of generating more plausible but incorrect patches~\cite{wen2018context}\ak{what is this?}

Liu et al.~\cite{liu2018mining2} and Rolim et al.~\cite{rolim2018learning} proposed to mine fix patterns from static analysis violations from FindBugs and PMD respectively. Both approaches, leverage a similar methodology in the inference process. Rolim et al.~\cite{rolim2018learning} rely on the distance among edit scripts: edit scripts with low distances among them are grouped together according to a defined similarity threshold. Liu et al.~\cite{liu2018mining2}, on the other hand, leverage deep learning to learn features of edit scripts, to find clusters of similar edit scripts. Eventually, both works  do not consider code context in their edit scripts and manually derive the fix patterns from the clusters of similar edit scripts of patches.
%\ak{revisar part is too weak}
%\ak{avatar does not seems to be manual?}

%extract actionable repair patterns in these patches so that the repair patterns can be applied in different places to repair the same type of bug
%On the other hand, the main goal of our abstract space design is to exclude the unlikely patches by analyzing a small set of patches. As a result, our abstract space def- inition is more coarse-grained and is not actionable
%simfix
%
%
%. HDRepair prioritizes patches based
%on their similarities with previous fix patterns that are mined from
%software histories [19]. Prophet [25] also compares the generated
%patch with existing human patches to investigate its probability
%of being correct but leverages different features as HDRepair [19].
%ACS leverages the information mined from other projects and Java
%documentations to prioritize the variables and predicates used in
%the synthesized conditions [53]. S3 prioritizes the generated patches
%by comparing their similarities with the original buggy program in
%terms a set of features [55]. Different from them, CapGen directly
%leverages the context information extracted from the buggy code
%elements and the fixing ingredients to prioritize the generated
%patches. To the best of our knowledge, incorporating the context
%information of fixing ingredients is new to program repair.
%

In another vein, CapGen~\cite{wen2018context} and SimFix~\cite{jiang2018shaping} propose to use frequency of code change actions. The former uses it to drive patch selection, while the latter uses it in computing donor code similarity for patch prioritization. In both cases, however, the notion of patterns is not an actionable artefact, but rather a supplementary information that guides their patch generation system. Although we concurrently\footnote{The initial version of this paper was written concurrently to SimFix and CapGen.} share with SimFix and CapGen the idea of adding more contextual information for patch generation, our objective is to infer actionable fix patterns that are tractable and reusable as input to other APR systems.

Table~\ref{tab:compLiterature} presents an overview of different automated mining strategies implemented in literature to obtain diverse sets of fix patterns. Some of the strategies are directly presented as part of APR systems, while others are independent approaches. We characterize the different strategies by considering the diff representation format, the use of contextual information, the tractability of patterns (i.e., what extent they are separate and reusable components in patch generation systems), and the scope of mining (i.e., whether the scope is limited to specific code changes).
Overall, although the literature approaches can come handy for discovering diverse sets of fix patterns, the reality is that the intractability of the fix patterns and the generalizability of the mining strategies remain a challenge for deriving relevant patterns for program repair.%\ak{check previous sentence}

%with statistic code change actions that are however mainly used to select and prioritize donor code from the buggy program for patch generation.\ak{ our position is not clear against these 2 paper. I even think, what we do is very similar to SimFix. if you say statistic, we also use statistics, we select top 50 action patterns. Could you please help positioning our work here?}

\begin{table}[!h]

\centering

\caption{Comparison of fix pattern mining techniques in the literature.}%
\label{tab:compLiterature}
%\subfloat[]
%{\parbox{0.4\linewidth}{%
\resizebox{\linewidth}{!}{%
\begin{threeparttable}
		\begin{tabular}{l|cccccc>{\columncolor[gray]{0.8}}c}
		\toprule
			  \makecell{} & Genesis~\cite{long2017automatic} & SOFix~\cite{liu2018mining} & Liu et al.~\cite{liu2018mining2} & Rolim et al.~\cite{rolim2018learning} & CapGen~\cite{wen2018context} & SimFix~\cite{jiang2018shaping} & FixMiner \\
			  \hline\hline
       % \noalign{\smallskip}\hline\noalign{\smallskip}

        \makecell[l]{Diff \\notation} & Transform & Edit Script & Edit Script& Edit Script & Edit Script & Edit Script & Edit Script\\
        \hline
        Scope & \makecell{Three defect\\ classes}& Any bug type &\makecell{Static analysis\\ violations} &\makecell{Static analysis\\ violations}& Any bug type& \makecell{Insert and update \\changes only} & Any bug type \\
        \hline
		\makecell[l]{Context \\information} & \xmark& \xmark & \xmark & \xmark & \cmark & \cmark & \cmark\\
		\hline
        \makecell[l]{Tractability of \\Patterns*} &Medium & High& High& High & Low  & Low & High\\
		%\noalign{\smallskip}
		\bottomrule

		\end{tabular}
		\begin{tablenotes}
    \item[*]
{{\bf High: } Patterns are part of output and reusable as input to APR systems \newline
	 {\bf Medium: } Patterns are not readily usable \newline
	 {\bf Low:} Patterns are not separate or available as output.
	 
	 }

  \end{tablenotes}
  \end{threeparttable}
		}
\end{table}

\noindent
{\bf This paper.} We propose to investigate the feasibility of mining relevant fix patterns that can be easily integrated into an automated pattern-based program repair system. To that end, we propose an iterative and three-fold clustering strategy, \toolname,
to discover relevant fix patterns automatically from atomic changes within real-world developer fixes. \toolname is a pattern mining approach to produce fix patterns for program repair systems. We present in this paper the concept of \ediff which is a specialized tree data structure of the edit scripts that captures the AST-level context of code changes. To infer patterns, \toolname leverages identical trees, which are computed based on the following information encoded in \ediff{s} for each round of the iteration: abstract syntax tree, edit actions tree, and code context tree.

\noindent
{\bf Contribution.} We propose the \toolname pattern mining tool as a separate and reusable component that can be leveraged in other patch generation systems.
% Eventually,  contrary to concurrently proposed and related work,

% leveraging a specific information encoded in AST trees in terms of context, operations, and the programming tokens.

% We make no assumptions on the simplicity or semantics of fixes.

\noindent
{\bf Paper content.} Our contributions are:
\begin{itemize}[leftmargin=*]
\item We present the architecture of a pattern inference system, \toolname, which builds on a three-fold clustering strategy where we iteratively discover similar changes based on different tree representations encoding contexts, change operations and code tokens.
	\item We assess the capability of \toolname to discover patterns by mining fix patterns among 11\,416 patches addressing user-reported bugs in 43 open source projects. We further relate the discovered patterns to those that can be found in a dataset used by the program repair community \cite{just2014defects4j}. We assess the compatibility of \toolname patterns with patterns in the literature.

	\item Finally, we investigate the relevance of the mined fix patterns by embedding them as part of an Automated Program Repair system. Our experimental results on the Defects4J benchmark show that our mined patterns are effective for fixing 26 bugs. We find that the \toolname  patterns are relevant as they lead to generating plausible patches that are mostly correct.
\end{itemize}

% The remainder of this paper is structured as follows. In Section~\ref{sec:bg}, we provide the intuition behind our approach
%  as well as background information on Abstract syntax trees and code differencing. Section~\ref{sec:method} details the process of \toolname while experiments are presented in Section~\ref{sec:exp}. We discuss insights and threats to validity in Section~\ref{sec:discussion} and related work in Section~\ref{sec:relatedwork} before concluding in Section~\ref{sec:conclusion}.

\section{Motivation}

Mining, enumerating and understanding code changes have been a key challenge of software maintenance in recent years. Ten years ago, Pan et al. have contributed with a manually-compiled catalog of 27 code change patterns related to bug fixing~\cite{pan2009toward}. Such ``bug fix patterns'' however are generic patterns (e.g., IF-RMV: removal of an If Predicate) which represent the type of changes that are often fixing bugs. More recently, thanks to the availability of new AST differencing tools, researchers have proposed to automatically mine change patterns~\cite{martinez2013automatically,osman2014mining,oumarou2015identifying,lin2016empirical}. Such patterns have been mostly leveraged for analysing and towards understanding characteristics of bug fixes. In practice, however, the inferred patterns may turn out to be irrelevant and intractable.%\ak{how do we support this claim?}
%In practice, they were not \textcolor{blue}{However, those fix patterns are in a high-level form and have not been used in automated program repair.}\ak{why?}\kui{Is it clear now?}

% Eventually, mining fix patterns yields clusters of recurrent sequences of edit actions (edit scripts). Unfortunately, to date, such ``patterns'' have been of little use to improve automated repair studies. Following their fix pattern mining exercise, Xuan et al. have developed NOPOL~\cite{xuan2017Nopol} focused on repairing buggy conditional statements (i.e., applying changes to IF-then-Else statements). However, even in this case, insights from their study were not leveraged in the fixing process.\ak{now we have genesis, capgen, ...}

We argue however that mining fix patterns can help for guiding mutation operations for patch generation. In this case, there is a need to mine truly recurrent change patterns to which repair semantics can be attached, and to provide accurate, fine-grained patterns that can be actionable in practice, i.e., separate and reusable as inputs to other processes.
% \ak{do we need to reflect the reviewers concers about actionable also here, we did in abstract.}
    % \item {\em Automating the generation of repair templates for template-driven automated program repair}. In this case, there is a need to provide patterns that go beyond high-level syntactic operations, but also include contextual information (e.g., qualified token names), which can be specific to projects, to ensure the derivation of actionable
 Our intuition is that relevant patterns cannot be mined globally since bug fixes in the wild are subject to noisy details due to tangled changes~\cite{kim2013Impact}. There is thus a need to break patches into atomic units (contiguous code lines forming a hunk) and reason about the recurrences of the code changes among them.
 To mine changes, we propose to rely on the edit script format, which provides a fine-grained representation of code changes, where different layers of information are included:
 
\begin{itemize}
  \item the context, i.e., AST node type of the code element being changed (e.g., a modifier in declaration statements, should not be generalized to other types of statements);
  \item the change operation (e.g., a ``remove then add'' sequence should not be confused with ``add then remove'' as it may have a distinct meaning in a hierarchical model such as the AST);
  \item and code tokens (e.g., changing calls to ``{\em Log.warn}'' should not be confused to any other API method).
\end{itemize}

Our idea is to iteratively find patterns within the contexts, and patterns of change operations for each context, and patterns of recurrently affected literals in these operations.

% bug fix patterns to be relevant and informative must consider several layers of source code information:
% \textcolor{blue}{1) the code change actions that can guide how to mutate buggy code entities (e.g., update or insert); 2) AST nodes related to change actions; and 3) relationships among AST nodes reflecting the contextual information of bug fixes. Their instances are referred to AST-based actionable edit scripts in this work, which represent the actual code change details of bug fixes with their AST contexts.}
% \textcolor{red}{\sout{
% the shape/type of code being modified (e.g., a modifier change pattern in declaration statements, should not be generalized to other types of statements); the actual change patterns (i.e., a ``remove then add'' sequence should not be confused with ``add then remove'' as it may have distinct meaning in a hierarchical model such as the AST); and the context of the change (i.e., token information that can help specialize the patterns into templates).
% }}

\label{sec:bg}
We now provide background information for understanding the execution as well as the information processed by \toolname.
\subsection{Abstract Syntax Tree}
\label{bg:ast}
Code representation is an essential step in the analysis and verification of programs. Abstract syntax trees (ASTs), which are generally produced for program analysis and transformations, are data structures that provide an efficient form of representing program structures to reason about syntax and even semantics.  An AST indeed represents all of the syntactical elements of the programming language and focuses on the rules rather than elements like braces or semicolons that terminate statements in some popular languages like Java or C. The AST is a hierarchical representation where the elements of each programming statement are broken down recursively into their parts. Each node in the tree thus denotes a construct occurring in the programming language.

\begin{figure}[!h]
	\includegraphics[width=\linewidth]{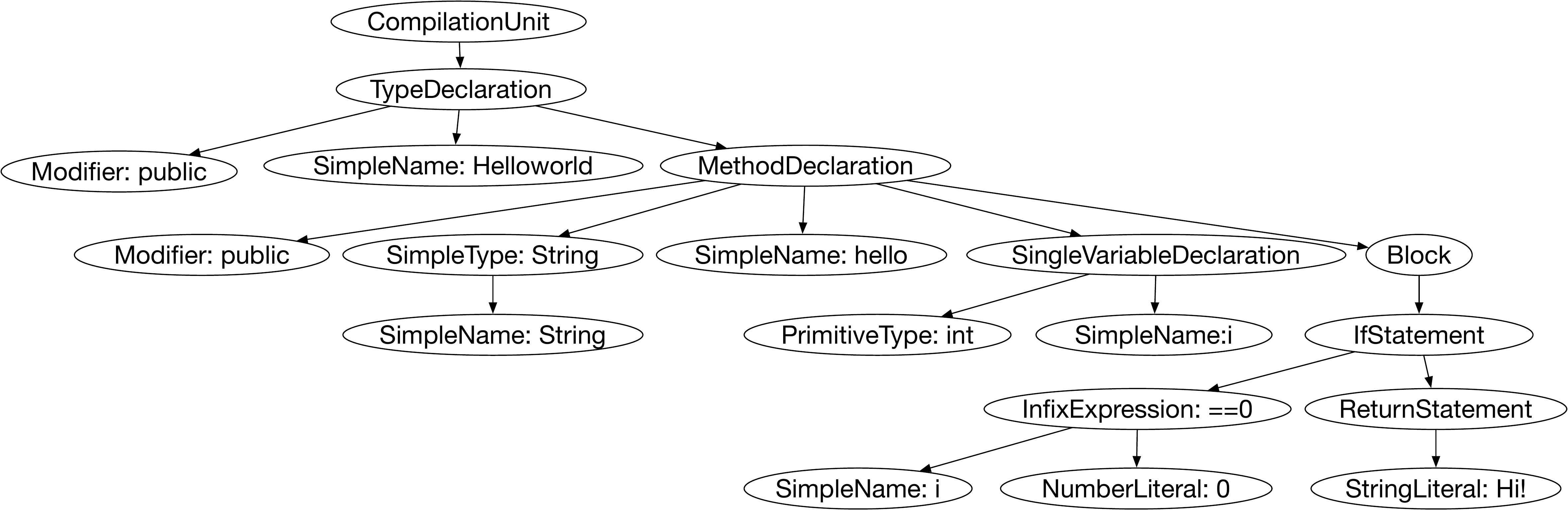}
	\caption{AST representation of the {\tt Helloworld} class.}
	\label{fig:ast}
\end{figure}

% Labels of nodes correspond to the name of their production rule in the grammar. Labels indeed indicate the AST node types while values of the nodes correspond to the raw tokens in the code.
Formally, let $t$ be an AST and $N$ be a set of AST nodes in $t$.
An AST $t$ has a root that is a node referred to as $root(t) \in N$.
Each node $n \in N$ (and $n \neq root(t)$) has a parent denoted as  $parent(n) = p \in N$. Note that there is no parent node of $root(t)$.
Furthermore, each node $n$ has a set of child nodes
(denoted as $children(n) \subset N$).
A label $l$ (i.e., AST node type) is assigned to each node from a given alphabet $L$
($label(n) = l \in L$). Finally, each node has a string value $v$ ($token(n) = v$ where $n \in N$ and $v$ is an arbitrary string) representing the corresponding raw code token.
Consider the AST representation in Figure~\ref{fig:ast} of the Java code in Figure~\ref{ast}.
We note that the illustrated AST has nodes with labels
matching structural elements of the Java language (e.g., {\tt MethodDeclaration}, {\tt IfStatement} or
{\tt StringLiteral}) and can be associated with values representing the raw tokens in the code
(e.g., A node labelled {\tt StringLiteral} from our AST is associated to value ``Hi!'').

\begin{figure}[!h]
    \centering
    \vspace{2mm}
\begin{lstlisting}[language=java,linewidth={\linewidth},frame=tb,basicstyle=\footnotesize\ttfamily]
public class Helloworld {
    public String hello(int i) {
        if (i == 0) return "Hi!";
    }
}
\end{lstlisting}
    \caption{Example Java class.}
    \label{ast}
\end{figure}

%\begin{center}
%\lstinputlisting[boxpos=c, caption=Example Java class, label=ast]{figs/Test.java}
%\end{center}

%\vspace{-8mm}
\subsection{Code Differencing}
\label{sec:codediff}

Differencing two versions of a program is the key pre-processing step of all studies on software evolution.  The evolved parts must be captured in a way that makes it easy for developers to understand or analyze the changes. Developers generally deal well with text-based differencing tools, such as the GNU Diff represents changes as addition and removal of source code lines as shown in Figure~\ref{gnudiff}.
The main issue with this text-based differencing is that it does not provide a fine-grained representation of the change (i.e., {\tt StringLiteral Replacement}) and thus it is poorly suited for systematically analysing the changes. %: e.g., {\tt -} and {\tt +} lines would be the same if the change was on other code elements such as the boolean expression, or even if a spacing was added at the end of the line.

\begin{figure}[!h]
    \centering
    \vspace{2mm}
\begin{lstlisting}[language=diff,linewidth={\linewidth},frame=tb,basicstyle=\footnotesize\ttfamily]
--- Helloworld_v1.java	2018-04-24 10:40:19.000000000 +0200
+++ Helloworld_v2.java	2018-04-24 11:43:24.000000000 +0200
@@ -1,5 +1,5 @@
 public class Helloworld {
     public String hello(int i) {
-        if (i == 0) return "Hi!";
+        if (i == 0) return "Morning!";
     }
 }
\end{lstlisting}
    \caption{GNU diff format.}
    \label{gnudiff}
\end{figure}
%\lstinputlisting[language=diff,caption=GNU diff format, label=gnudiff]{figs/patch}

To address the challenges of code differencing, recent algorithms have been proposed based on tree structures (such as the AST). ChangeDistiller and GumTree are examples of such algorithms which produce {\em edit scripts} that detail the operations to be performed on the nodes of a given AST (as formalized in Section~\ref{bg:ast}) to yield another AST corresponding to the new version of the code. In particular, in this work, we build on GumTree's core algorithms for preparing an edit script. An edit script is a sequence of edit actions describing the following code change actions:
\begin{itemize}
	\item {\tt UPD} where an $upd(n,v)$ action transforms the AST by replacing the old value of an AST node $n$ with the new value $v$.
	\item {\tt INS} where an $ins(n, n_p, i, l, v)$ action inserts a new node $n$ with $v$ as value and $l$ as label. If the parent $n_p$ is specified, $n$ is inserted as the $i^{th}$ child of $n_p$, otherwise $n$ is the root node.
	\item {\tt DEL} where a $del(n)$ action removes the leaf node $n$ from the tree.
	\item {\tt MOV} where a $mov(n, n_p ,i)$ action moves the subtree having node $n$ as root to make it the $i^{th}$ child of a parent node $n_p$.
\end{itemize}

An edit action, embeds information about the node (i.e., the relevant node in the whole AST tree of the parsed program), the operator (i.e., {\tt UPD}, {\tt INS}, {\tt DEL}, and {\tt MOV}) which describes the action performed, and the raw tokens involved in the change.

\subsection{Tangled Code Changes}
\label{sec:tangling}

Solving a single problem per patch is often considered as a best practice to facilitate maintenance tasks. However, often patches in real-world projects address multiple problems in a patch~\cite{tao2015partitioning,koyuncu2017impact}.
Developers often commit bug fixing code changes together with changes unrelated to fix such as functionality enhancements, feature requests, refactorings, or documentation. Such patches are called tangled patches~\cite{kim2013Impact} or mixed-purpose fixing commits~\cite{nguyen2013filtering}. Nguyen et al. found that 11\% to 39\% of all the fixing commits used for mining archives were tangled~\cite{nguyen2013filtering}.

Consider the example patch from GWT illustrated in Figure~\ref{fig:tangled}. The patch is intended to fix the issue\footnote{\url{https://github.com/gwtproject/gwt/issues/676}} that reported a failure in some web browsers when the page is served with a certain mime type (i.e., application/xhtml+xml). The developer fixes the issue by showing a warning when such mime type is encountered. However, in addition to this change, a typo has been addressed in the commit. Since the typo is not related to the fix, the fixing commit is tangled.
There is thus a need to separately consider single code hunks within a commit to allow the pattern inference to focus on finding recurrent atomic changes that are relevant to bug fixing operations.

% "Creating a relatively large (greater then 60 rows) FlexTable causes IE \& FireFox to abort script". By analyzing the commit log of the developer, "Fixes issue \#676
% .xhtml files do not load properly in the hosted mode browser. This is due to
% the fact that GWT does not support browsers running in full xhtml+xml mode
% and the current version of the GWTShellServlet returns application/xhtml+xml
% as the content type for files that end in .xhtml.  This change adds a
% TreeLogger warning when mime types of application/xhtml+xml are encountered
% and recommends an appropriate file renaming. Also, this fixes a a little typo."

\begin{figure}[!h]
    \centering
    \vspace{2mm}
\begin{lstlisting}[language=diff,linewidth={\linewidth},frame=tb,basicstyle=\footnotesize\ttfamily]
--- a/dev/core/src/com/google/gwt/dev/shell/GWTShellServlet.java
+++ b/dev/core/src/com/google/gwt/dev/shell/GWTShellServlet.java
@@ -72,6 +72,8 @@

+   private static final String XHTML_MIME_TYPE = "application/xhtml+xml";
   private final Map loadedModulesByName = new HashMap();
   private final Map loadedServletsByClassName = new HashMap();
@@ -110,7 +112,7 @@
       writer.println("<html><body><basefont face='arial'>");
-      writer.println("To launch an an application, specify a URL of the form <code> 
		(@\textcolor{darkpastelred}{/<i>module</i>/<i>file.html</i></code>");}@)
+      writer.println("To launch an application, specify a URL of the form <code>/<i> 
		(@\textcolor{ao(english)}{module</i>/<i>file.html</i></code>");}@)
       writer.println("</body></html>");
     }
@@ -407,6 +409,8 @@
     }
+    maybeIssueXhtmlWarning(logger, mimeType, partialPath);

@@ -755,6 +759,25 @@
 
+  private void maybeIssueXhtmlWarning(TreeLogger logger, String mimeType,
+      String path) {
+    if (!XHTML_MIME_TYPE.equals(mimeType)) {
+      return;
+    }
+
+    String msg = "File was returned with content-type of \"" + mimeType
+        + "\". GWT requires browser features that are not available to "
+        + "documents with this content-type.";
+
+    int ix = path.lastIndexOf('.');
+    if (ix >= 0 && ix < path.length()) {
+      String base = path.substring(0, ix);
+      msg += " Consider renaming \"" + path + "\" to \"" + base + ".html\".";
+    }
+
+    logger.log(TreeLogger.WARN, msg, null);
+  }
\end{lstlisting}
    \caption{Tangled commit.}
    \label{fig:tangled}
\end{figure}

% However, developers do not necessarily follow the best practice of creating only atomic commits [9]. For example, a small bug may be quickly fixed while working on another feature and placed in the same commit. Floss refactoring is another problem: refactoring in order to implement a new feature. These situations make for larger commits in which many unrelated changes are tangled together. Such commits are called tangled commits.

% When interacting with version control systems, de- velopers often commit unrelated or loosely related code changes in a single transaction. When analyzing the version history, such tangled changes will make all changes to all modules appear related, possibly compromising the resulting analyses through noise and bias.

% What is a tangled change? Assume a developer is assigned multiple tasks A, B, and C, all with a separate purpose: A is a bug fix, B is a feature request, and C is a refactoring or code cleanup. Once all tasks are completed, the developer commits her changes to the source code management system (SCM), such that her changes be visible to other developers and integrated into the product. However, when committing changes, developers frequently group separate changes into a single commit, resulting in a tangled change.

% commits are often tangled, meaning that a commit may not solely consist of changes of a single purpose (i.e., only changes that fix dependency- related build breakage).

\section{Approach}
\label{sec:method}

\begin{figure}[!h]
	\includegraphics[width=\linewidth]{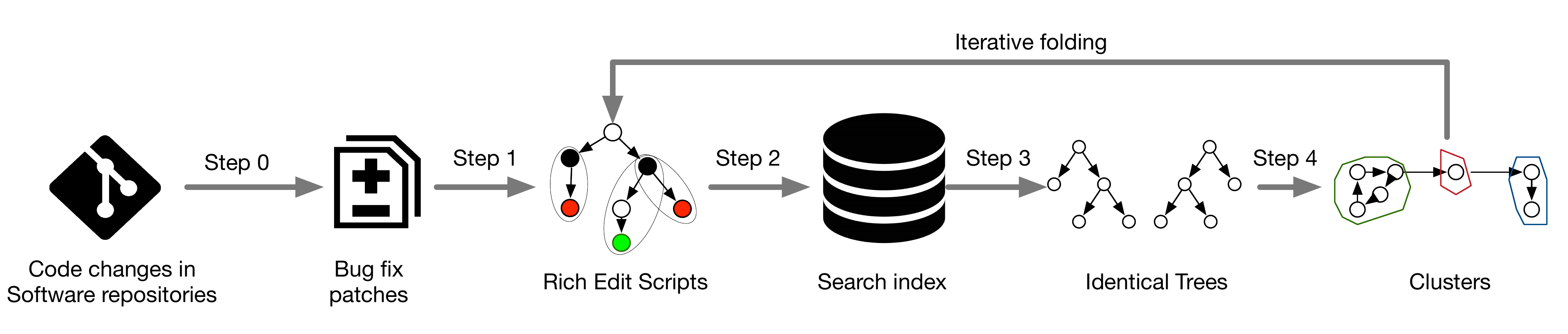}
	\caption{The \toolname Approach. {\normalfont At each iteration, the search index is refined, and the computation of tree similarity is specialized in specific AST information details.}}
	\label{fig:approach}
	\vspace{0.2cm}
\end{figure}

\toolname aims to discover relevant fix patterns from the atomic changes within bug fixing patches in software repositories. To that end, we mine code changes that are similar in terms of context, operations, and the programming tokens that are involved. Figure~\ref{fig:approach} illustrates an overview of the \toolname approach.
% between the buggy and fixed versions of the programs (cf. Section~\ref{step1}). \kui{Do we need to highlight the ``ENHANCED''? If so, reviews argue how we could know the ``ENHANCED'' GumTree works well.}

\subsection{Overview}

In Step 0, as an initial step, we collect the relevant bug-fixing patches (cf. Definition~\ref{def:patch}) from project change tracking systems. Then, in Step 1, we compute a \ediff representation (cf. Section~\ref{step1}) to describe a code change in terms of the context, operations performed and tokens involved. Accordingly, we consider three specialized tree representations of the \ediff (cf. Definition~\ref{def:re}) carrying information about either the impacted AST node types, or the repair actions performed, or the program tokens affected. \toolname works in an iterative manner considering a single specialized tree representation in each pattern mining iteration, to discover similar changes: first, changes affecting the same code context (i.e., on identical abstract syntax trees) are identified; then among those identified changes, changes using the same actions (i.e., identical sequence of operations) are regrouped; and finally within each group, changes affecting the same tokens set are mined. Therefore, in \toolname, we perform a three-fold strategy, carrying out the following steps in a pattern mining iteration:
%\ak{what does fold driver in the iteration?, in step 3 i wrote for a given fold, but eventually all the steps are for a given fold.}
    \begin{itemize}
        \item Step 2: We build a search index (cf. Definition~\ref{def:si}) to identify the \ediff{s} that must be compared.
        \item Step 3: We detect identical trees (cf. Definition~\ref{def:identical}) by computing the distance between two representations of \ediff{s}.
        \item Step 4: We regroup identical trees into clusters (cf. Definition~\ref{def:clusters}).
        % \item Eventually, the trees in a yielded cluster are used as input of the next iteration.
    \end{itemize}
The initial pattern mining iteration uses \ediff{s} computed in Step 1 as its input, where the following rounds use clusters of identical trees yielded in Step 4 as their input.
% Each pattern mining iteration round forms clusters of identical trees : abstract syntax tree, edit actions tree, and code context tree.

In the following sections, we present the details of Steps~1-4, considering that a dataset of bug fix patches is available.

\subsection{Step 0 - Patch Collection}
\label{sec:buglink}

\quest{
\begin{definition}
    \label{def:patch}
({\bf Patch}) A program patch is a transformation of a program into another program, usually to fix a defect. Let $\mathbb{P}$ being a set of programs, a patch is represented by a pair ($p, p'$), where $p, p' \in \mathbb{P}$ are
programs before and after applying the patch, respectively.
Concretely, a patch implements changes in code block(s) within source code file(s).
\end{definition}
}%

To identify bug fix patches in software repositories projects, we build on the bug linking strategies implemented in the Jira issue tracking software. We use a similar approach to the ones proposed by Fischer et al.~\cite{fischer2003populating} and Thomas et al.~\cite{thomas2013impact} in order to link commits to relevant bug reports.
Concretely, we crawl the bug reports for a given project and assess the links with a two-step search strategy: (i) we check project commit logs to identify bug report IDs and associate the corresponding bug reports to commits; then (ii) we check for bug reports that are indeed considered as such (i.e., tagged as ``BUG'') and are further marked as resolved (i.e., with tags ``RESOLVED'' or ``FIXED''), and completed (i.e., with status ``CLOSED'').

We further curate the patch set by considering bug reports that are fixed by a single commit. This provides more guarantees that the selected commits are indeed fixing the bugs in a single shot (i.e., the bug does not require supplementary patches~\cite{park2012empirical}). Eventually, we consider only changes that are made on the source code files: changes on configuration, documentation, or test files are excluded.

\subsection{Step 1 -- \ediff Computation}
\label{step1}

\quest{
\begin{definition}
\label{def:re}
(\ediff) A \ediff $r \in RE$ represents a patch as a specialized tree of changes. This tree describes which operations are made on a given AST, associated with the code block before patch application, to transform it
into another AST, associated with the code block after patch application: i.e., $r: \mathbb{P} \rightarrow \mathbb{P}$. Each node in the tree is an AST node affected by the patch.
Every node in \ediff has three different types
of information: {\bf Shape}, {\bf Action}, and {\bf Token}.
% ``Shape'' corresponds to AST node types (e.g., Statement, Expression, etc). ``Action'' refers to an element of \{UPD, INS, DEL, MOV\} (cf. Section~\ref{sec:codediff}). For each node, the ``Token'' contains a pair of text token sets (before and after applying the patch).
% The root of a \ediff must be an element of \{TypeDeclaration, FieldDeclaration, MethodDeclaration, SwitchCase, CatchClause, ConstructorInvocation, SuperConstructorInvocation, Statement\}.
\end{definition}
}%

A bug-fix patch collected in open source change tracking systems is represented in the GNU diff format based on addition and removal of source code lines as shown in Figure~\ref{patch1}. This representation is not suitable for fine-grained analysis of changes.

 \begin{figure}[!h]
    \centering
    \vspace{2mm}
\begin{lstlisting}[language=diff,linewidth={\linewidth},frame=tb,basicstyle=\footnotesize\ttfamily]
// modules. We need to move this code up to a common module.
-          int indexOfDot = namespace.indexOf(`.');
+          int indexOfDot = namespace.lastIndexOf(`.');
           if (indexOfDot == -1) {
\end{lstlisting}
    \caption{Patch of fixing bug Closure-93 in Defects4J dataset.}
    \label{patch1}
\end{figure}

To accurately reflect the change that has been performed, several algorithms have been proposed based on tree structures (such as the AST)~\cite{bille2005survey,pawlik2011rted,chawathe1996change,hashimoto2008diff,duley2012vdiff,fluri2007change,falleri2014Fine}. ChangeDistiller~\cite{fluri2007change} and GumTree~\cite{falleri2014Fine} are state-of-the-art examples of such algorithms which produce edit scripts that detail the operations to be performed on the nodes of a given AST in order to yield another AST corresponding to the new version of the code. In particular, in this work, we selected the GumTree AST differencing tool which has seen recently a momentum in the literature for computing edit scripts.
GumTree is claimed to build in a fast, scalable and accurate way the sequence of AST edit actions (a.k.a edit scripts) between the two associated AST representations (the buggy and fixed versions) of a given patch.

\begin{figure}[!h]
    \centering
    \vspace{2mm}
\begin{lstlisting}[language=java,linewidth={\linewidth},frame=tb,basicstyle=\footnotesize\ttfamily]
UPD SimpleName ``indexOf'' to ``lastIndexOf''
\end{lstlisting}
    \caption{GumTree edit script corresponding to Closure-93 bug fix patch represented in Figure~\ref{patch1}.}
    \label{patch_gumtree}
\end{figure}

Consider the example edit script computed by GumTree for the patch of {\rm Closure-93} bug from Defects4J illustrated in Figure~\ref{patch_gumtree}. The intended behaviour of the patch is to fix the wrong variable declaration of {\em indexOfDot} due to a wrong method reference ({\em lastIndexOf} instead of {\em  indexOf}) of java.lang.String object. GumTree edit script summarizes the change as an update operation on an AST node simple name (i.e., an identifier other than a keyword) that is modifying the identifier label (from {\em indexOf} to {\em lastIndexOf}).

Although GumTree edit script is accurate in describing the bug fix operation at fine-grained level, much of the contextual information describing the intended behaviour of the patch is missing. The information regarding method invocation, the method name ({\em java.lang.String}), the variable declaration fragment which assigns the value of the method invocation to {\em indexOfDot}, as well as the type information ({\em int} for {\em indexOfDot} - cf. Figure~\ref{patch1}) that is implied in the variable declaration statement are all missing in the GumTree edit script. Since such contextual information is lost, the yielded edit script fails to convey the full syntactic and semantic meaning of the code change.

To address this limitation, we propose to enrich GumTree-yielded edit scripts by retaining more contextual information. To that end, we construct a specialized tree structure of the edit scripts which captures the AST-level context of the code change. We refer to this specialized tree structure as \ediff. A \ediff is computed as follows:

Given a patch, we start by computing the set of edit actions (edit script) using GumTree, where the set contains an edit action for each contiguous group of code lines (hunks) that are changed by a patch. In order to capture the context of the change, we re-organize edit actions under new AST (minimal) subtrees building an AST hierarchy. 
For each edit action in an edit script, we extract a minimal subtree from the original AST tree which has the GumTree edit action as its leaf node, and one of the following predefined node types as its root node: TypeDeclaration, FieldDeclaration, MethodDeclaration, SwitchCase, CatchClause, ConstructorInvocation, SuperConstructorInvocation or any Statement node. The objective is to limit the scope of context to the encompassing statement, instead of going backwards until the compilation unit (cf. Figure~\ref{fig:ast}). We limit the scope of parent traversal mainly for two reasons: first, the pattern mining must focus on the program context that is relevant to the change; second, program repair approaches, which \toolname is built for, generally target statement-level fault localization and patch generation.

\begin{figure}[!h]
    \centering
    \includegraphics[width=0.7\linewidth]{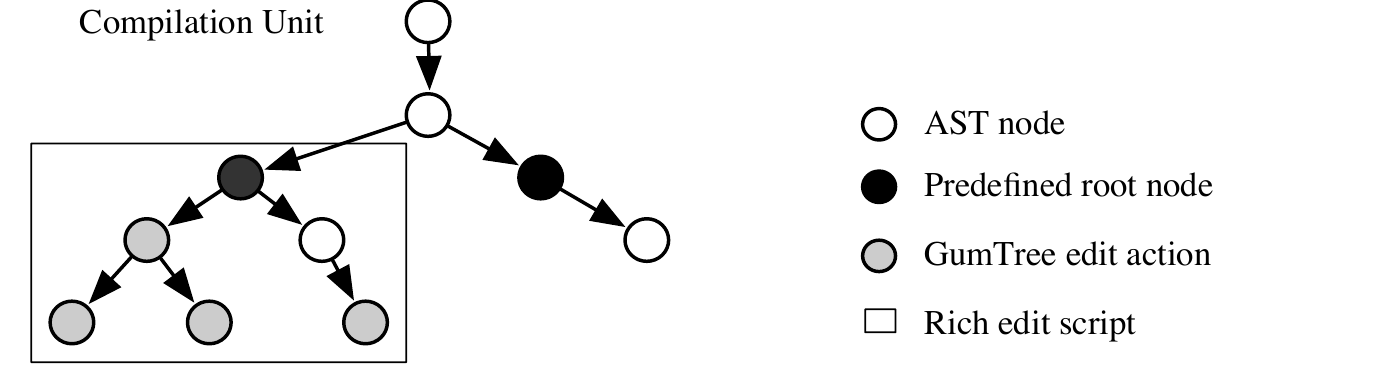}
    \caption{Illustration of subtree extraction.}
    \label{fig:richEdit}
    \vspace{0.2cm}
\end{figure}

Consider the AST differencing tree presented in Figure~\ref{fig:richEdit}. From this diff tree, GumTree yields the leaf nodes (gray) of edit actions as the final edit script. To build the \ediff, we follow these steps:

\begin{itemize}
	\item[i)] For each GumTree-produced edit action, we remap it to the relevant node in the program AST;
	\item[ii)] Then, starting from the \textit{GumTree edit action} nodes, we traverse the AST tree of the parsed program from bottom to top until we reach a node of \textit{predefined root node} type.
	\item[iii)] For every \textit{predefined root node} that is reached, we extract the AST subtree between the discovered \textit{predefined root node} down to the leaf nodes mapped to the \textit{GumTree edit actions}.
	\item[iv)] Finally, we create an ordered\footnote{The order of AST subtrees follows the order of hunks of the GNU diff format.} sequence of these extracted AST subtrees and store it as \ediff.
\end{itemize}

\begin{figure}[!h]
    \includegraphics[width=\linewidth]{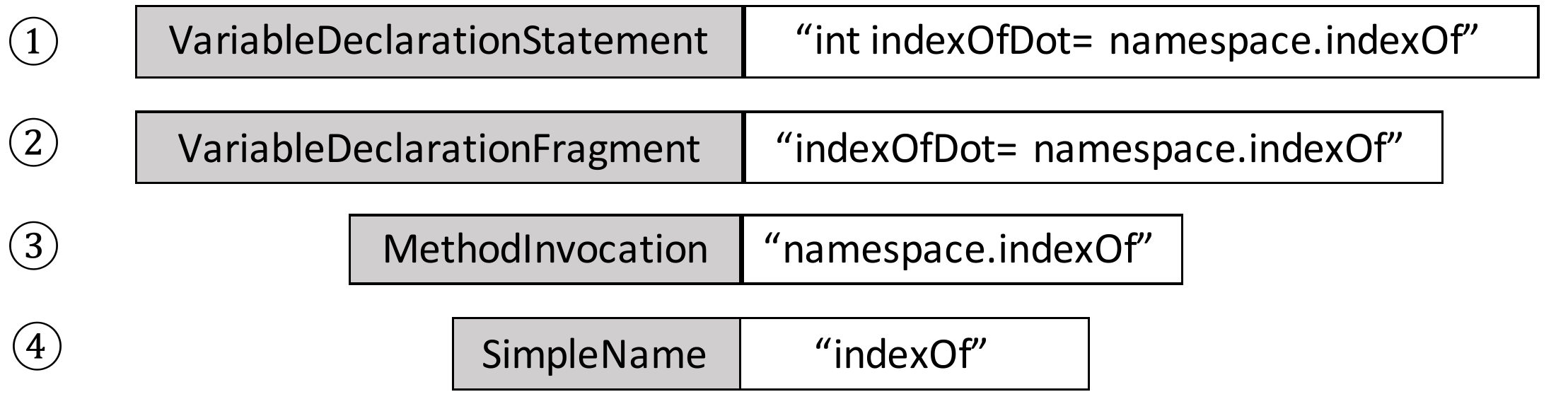}
    \caption{Excerpt AST of buggy code (Closure-93).}
    \label{fig:ediff}
\end{figure}

Concretely, with respect to our running example, consider the case of {\rm Closure-93} illustrated in Figure~\ref{patch1}. The construction of the \ediff starts by generating the GumTree edit script (cf. Figure~\ref{patch_gumtree}) of the patch. The patch consists of a single hunk, thus we expect to extract a single AST subtree, which is illustrated by Figure~\ref{fig:ediff}. To extract this AST subtree, first, we identify the node of the edit action ``SimpleName'' at position 4 in the AST Tree of program. Then, starting from this node, we traverse backward the AST tree until we reach the node ``VariableDeclarationStatement'' at position 1. We extract the AST subtree, by creating a new tree, setting ``VariableDeclarationStatement'' as root node of the new tree, and adding the intermediate nodes at positions 2,3 until we reach the corresponding node of the edit action ``{\tt UPD SimpleName}'' at position 4. We create a sequence, and add the extracted AST subtree to the sequence.

\ediff{s} are tree data structures. They are used to represent changes. In order to provide tractable and reusable patterns as input to other APR systems, we define the following string notation (cf. Grammar~\ref{gra:rich}) based on syntactic rules governing the formation of correct \ediff.

\begin{Grammar}[!h]
    \centering
    \vspace{2mm}
    \begin{grammar}
<richEditScript> ::= <node>+ 

<node> ::= `-\,-\,-'* <move> 
\alt `-\,-\,-'* <delete>
\alt `-\,-\,-'* <insert>
\alt `-\,-\,-'* <update> 

<move> ::= `MOV ' <astNodeType> `@@' <tokens> `@TO@' <astNodeType> `@@' <tokens> `@AT@'

<delete> ::= `DEL ' <astNodeType> `@@' <tokens> `@AT@'

<insert> ::= `INS ' <astNodeType> `@@' <tokens> `@TO@' <astNodeType> `@@' <tokens> `@AT@'

<update> ::= `UPD ' <astNodeType> `@@' <tokens> `@TO@' <tokens> `@AT@'

\end{grammar}
    \caption{Notation of \ediff}\label{gra:rich}
    \vspace{0.2cm}
\end{Grammar}

Figure~\ref{patch_Enhanced} illustrates the computed \ediff. The first line indicates the root node (no dashed line). `UPD ' indicates the action type of the node, VariableDeclarationStatement corresponds to ast node type of the node, tokens between `@@' and `@TO@' contains the corresponding code tokens before the change, where as tokens between `@TO@' and `@AT' corresponding to new code tokens with the change. The three dashed  (-\,-\,-) lines indicate a child node. Immediate children nodes contain three dashes while their children add another three dashes (-\,-\,-\,-\,-\,-) preserving the parent-child relation.

\begin{figure}[!h]
    \centering
    \vspace{2mm}
        % \begin{adjustbox}{width={\linewidth}}
    \lstset{prebreak=\raisebox{0ex}[0ex][0ex]{\ensuremath{\hookleftarrow}}}
    \lstset{breaklines=true, breakatwhitespace=true}
\begin{lstlisting}[language=java,linewidth={\linewidth},frame=tb,basicstyle=\footnotesize\ttfamily]
UPD VariableDeclarationStatement@@int indexOfDot = namespace.indexOf('.'); @TO@ int indexOfDot = namespace.lastIndexOf('.'); @AT@
---UPD VariableDeclarationFragment@@indexOfDot = namespace.indexOf('.') @TO@ indexOfDot = namespace.lastIndexOf('.') @AT@
-----UPD MethodInvocation@@namespace.indexOf('.') @TO@ namespace.lastIndexOf('.') @AT@
--------UPD SimpleName@@MethodName:indexOf:['.'] @TO@ MethodName:lastIndexOf:['.'] @AT@  
\end{lstlisting}
    % \end{adjustbox}
    \caption{\ediff for Closure-93 patch in Defects4J. $\protect\hookleftarrow$ represents the carriage return character which is necessary for presentation reasons. }
    \label{patch_Enhanced}
    \vspace{0.2cm}
\end{figure}

An edit action node carries the following three types of information: the AST node type (Shape), the repair action (Action), the raw tokens (Token) in the patch. For each of these three information types, we create separate tree representations from the \ediff, named as ShapeTree, ActionTree and TokenTree, each carrying respectively the type of information indicated by its name. Figures~\ref{shape},~\ref{action}, and~\ref{token} show ShapeTree, ActionTree, and TokenTree, respectively, generated for Closure-93.

\begin{figure}[!h]
    \centering
    \vspace{2mm}
\begin{lstlisting}[language=diff,linewidth={\linewidth},frame=tb,basicstyle=\footnotesize\ttfamily]
VariableDeclarationStatement
---VariableDeclarationFragment
------MethodInvocation
---------SimpleName
\end{lstlisting}
    \caption{ShapeTree of Closure-93.}
    \label{shape}
    \vspace{0.2cm}
\end{figure}

\begin{figure}[!h]
    \centering
    \vspace{2mm}
\begin{lstlisting}[language=diff,linewidth={\linewidth},frame=tb,basicstyle=\footnotesize\ttfamily]
UPD root
---UPD child1
------UPD child1_1
---------UPD child1_1_1
\end{lstlisting}
    \caption{ActionTree of Closure-93.}
    \label{action}
    \vspace{0.2cm}
\end{figure}

\begin{figure}[!h]
    \centering
    \vspace{2mm}
\begin{lstlisting}[language=java,linewidth={\linewidth},frame=tb,basicstyle=\footnotesize\ttfamily]
@@int indexOfDot = namespace.indexOf('.'); @TO@ int indexOfDot = namespace.la...
---@@indexOfDot = namespace.indexOf('.') @TO@ indexOfDot = namespace.lastInde...
------@@namespace.indexOf('.') @TO@ namespace.lastIndexOf('.') 
---------@@MethodName:indexOf:['.'] @TO@ MethodName:lastIndexOf:['.'] 
\end{lstlisting}
    \caption{TokenTree of Closure-93.}
    \label{token}
    \vspace{0.2cm}
\end{figure}

\subsection{Step 2 -- Search Index Construction}
\label{step2}

\quest{
\begin{definition}
\label{def:si}
(Search Index) To reduce the effort of matching similar patches, a search index (SI)
is used to confine the comparison space. Each fold (\{Shape, Action, Token\}) defines a search index: $SI_{Shape}$, $SI_{Action}$, and $SI_{Token}$, respectively. Each is defined as $SI_{\ast}: Q_{\ast} \rightarrow 2^{RE}$, where $Q$ is a query set specific to each fold and $\ast \in \{Shape, Action, Token\}$.
\end{definition}
}

Given that \ediff{s} are computed for each hunk in a patch, they are spread inside and across different patches. A direct pairwise comparison of these \ediff{s} would lead to a combinatorial explosion of the comparison space. In order to reduce this comparison space and enable a fast identification of \ediff{s} to compare, we build search indices. A search index is a set of comparison sub-spaces created by grouping the \ediff{s} with criteria that depend on the information embedded the used tree representation (Shape, Action, Token) for the different iterations.

The search indices are built as follows:
%\vspace{-0.5cm}

%\paragraph{``Shape'' search index.}
\noindent
{\bf ``Shape'' search index.} The construction process takes the ShapeTree representations of the \ediff{s} produced by Step 1 as input, and groups them based on their tree structure in terms of AST node types.
Concretely, \ediff{s} having the same root node (e.g., IfStatement, MethodDeclaration, ReturnStatement) and same depth are grouped together.
For each group, we create a comparison space by enumerating the pairwise combinations of the group members.
Eventually, the ``Shape'' search index is built by storing an identifier per group, denoted as root node/depth (e.g., IfStatement/2, IfStatement/3, MethodDeclaration/4), and a pointer to its comparison space (i.e., the pairwise combinations of its members).

%The ``Shape'' search index is used in Step 3 of first pattern mining iteration, by retrieving the identifiers from the search index, which contains pairs of {\tt ShapeTrees} to compare.

\noindent
\textbf{``Action'' search index.} The construction process follows the same principle as for ``Shape'' search index, except that the regrouping is based on the clustering output of ShapeTrees. Thus, the input is formed by ActionTree representations of the \ediff{s} and the group identifier for each comparison space is generated as node/depth/ShapeTreeClusterId  (e.g., IfStatement/2/1,MethodDeclaration/2/2) where ShapeTreeClusterId represents the id of the cluster yielded by the clustering (Steps~3-4) based on the ShapeTree information. Concretely, this means that the ``Action'' search index is built on groups of trees having the same shape.

\noindent
\textbf{``Token'' search index.} The construction process follows the same principle as for ``Action'' search index, using this time the clustering output of ActionTrees. Thus, the input is formed by TokenTree representations of the \ediff{s} and the group identifier for each comparison space is generated as node/depth/ShapeTreeClusterId/ActionTreeClusterId  (e.g., IfStatement/2/1/3,MethodDeclaration/2/2/1) where ActionTreeClusterId represents the id of the cluster yielded by the clustering (Steps~3-4) based on the ActionTree information.

%The enhanced AST Diff form calculation of all the patches is computationally expensive. A patch can have several rooted trees, where each rooted tree could have separate repair actions. In our study we considered only the UPDATE repair actions, which reduces significantly the among of the patches to be considered for the enhanced AST Diff form calculation, thus resulting lower computation time. Additionally, the computed enhanced AST Diffs stored in memory cache as rooted trees in order to avoid recalculation in the further steps of the approach.

\subsection{Step 3 -- Tree Comparison}
\label{step3}

\quest{
\begin{definition}
\label{def:identical}
(Pair of identical trees) Let $a = (r_{i}, r_{j}) \in R_{\mbox{\scriptsize identical}}$ be a pair of \ediff specialized tree representations if
$d(r_{i}, r_{j})=0$, where $r_{i}, r_{j} \in RE$ and $d$ is a distance function. $R_{\mbox{\scriptsize identical}}$ is a subset of $RE\times RE$.
\end{definition}
}

% As mentioned earlier, each iteration of \toolname is focusing solely on a single type of information available in the \ediff. Instead of using heuristics, as in typical hierarchical clustering scenarios where granularity is constructed by the extent of similarity among sets of change operations, we leverage the variety of information types to build levels of abstractions with regards to code shape, the actions and the specificity of tokens.\ak{I dont like the previous sentence.}
The goal of tree comparison is to find identical tree representations of \ediff{s} for a given fold. There are several straightforward approaches for checking whether two \ediff{s} are identical. For example, syntactical equality could be used. However, we aim at making \toolname a flexible and extensible framework where future research may tune threshold values for defining similar trees. Thus, we propose a generic approach for comparing \ediff{s}, taking into account the diversity of information to compare for each specialized tree representation. To that end, we compute tree edit distances for the three representations of \ediff{s} separately. The tree edit distance is defined as the sequence of edit actions that transform one tree into another. When the edit distance is zero (i.e., no operation is necessary to transform one tree to another) the trees are considered as identical. In Algorithm~\ref{isoAlgo} we define the steps to compare \ediff{s}.

\begin{algorithm}[!h]
\scriptsize
    \SetKwInOut{Input}{input}
    \SetKwInOut{Output}{output}
    \SetKw{Return}{return}

    \Input{$SI$: Search Index where $SI$ $\in$ \{$SI_{Shape}$,$SI_{Action}$,$SI_{Token}$\}}
    \Input{$fold$ $\in$ \{$Shape$,$Action$,$Token$\}}
    \Input{$threshold$: Set to 0 to retrieve identical trees.  }
    % \Input{$threshold$: Text similarity threshold }
    % \BlankLine
    \Output{$R_{\mbox{identical}}$: A set of pairs tagged to be identical}
    \BlankLine
    \SetKwProg{Fn}{Function}{}{end}
    \SetKwFunction{main}{main}
    \BlankLine
    \DontPrintSemicolon

  \Fn{\main($SI$,$fold$)}{
      $R_{\mbox{identical}}\leftarrow\emptyset$\;
      $I\leftarrow SI$.getIdentifiers() \tcc*{$I$: list of identifiers in the index}
      \ForEach{$i$ $\in$ $I$ }{
        $R$ $\leftarrow$ $SI$.getPairs($i$) \tcc*{$R$: list of tree pairs to compare for identifier $i$}
        \ForEach{$a$ $\in$ $R$ }{
          \If{compareTree($a$,$fold$)} {
            $R_{\mbox{identical}}$.add($a$)   \tcc*{add if $a$ is a pair of identical trees}
            }
      }
     }
     \Return $R_{\mbox{identical}}$\;
   }

   \Fn{compareTree($a$,$fold$)}{
     ($sTree1$,$sTree2$) $\leftarrow$ specializedTree($a$,$fold$)\;
     \uIf{$Fold$ != $Token$}{
       $editActions$ $\leftarrow$ GumTree($sTree1$, $sTree2$)\;
       $editDistance$ $\leftarrow$ size($editActions$) \;
     }
     \Else{
        $tokens1$,$tokens2$  $\leftarrow$ getTokens($sTree1$,$sTree2$) \;
        $editDistance$ $\leftarrow$ $d_{w}$($tokens1$,$tokens2$)  \tcc*{$d_{w}$: Jaro-Winkler distance}
     }
     \uIf{ $editDistance$ \textless= $threshold$} {
        \Return true\;
     }
     \Else{
        \Return false\;
     }

    }

    \Fn{specializedTree($a$,$fold$)}{
        ($eTree1$, $eTree2$) $\leftarrow$ getRichEditScripts($a$) \tcc*{restore Rich Edit Scripts of a given pair from the cache}

       \uIf{$fold$ == $Shape$}{
          $sTree1$,$sTree2$ $\leftarrow$ getASTNodeTrees($eTree1$, $eTree2$)\;
        }
       \uElseIf{$fold$ == $Action$}{
          $sTree1$,$sTree2$ $\leftarrow$ getActionTrees($eTree1$, $eTree2$)\;
        }
       \Else{
          $sTree1$,$sTree2$ $\leftarrow$ getTokenTrees($eTree1$, $eTree2$) \tcc*{$fold$ == $Token$}
        }
        \Return($sTree1$,$sTree2$)\;

    }

    \caption{\ediff Comparison.}\label{isoAlgo}
\end{algorithm}
%\ak{should i make the editDistance == 0 parametric??}

%
%considering the rooted trees of a given pair to be compared from the cache (which is already holding all the rooted trees).

%Following our intuition, we iteratively refine the clusters of changes into more and more fine-grained patterns.
% , referred as the $fold$ in the algorithm.

The algorithm starts by retrieving the identifiers from the search index $SI$ corresponding to the $fold$. An identifier is a pointer to a comparison sub-space that contains
pairwise combinations of tree representation of \ediff{s} to compare (cf. Section~\ref{step2}). Concretely, we restore the \ediff{s} of a given pair from the cache, and their corresponding specialized tree representation according to the $fold$: At the first iteration, we consider only trees denoted ShapeTrees, whereas in the second iteration we focus on ActionTrees, and TokenTrees for the third iteration. We compute the edit distance between the restored trees in two distinct ways.
%\ak{should i hide these details, as we said that in the response letter?}
\begin{itemize}[leftmargin=*]
	\item In the first two iterations (i.e, Shape and Action) we leverage again the edit script algorithm of GumTree~\cite[Section~3]{gumtree}. We compute the edit distance by simply invoking GumTree on restored trees as input, given that \ediff{s} are indeed AST subtrees that are compatible with GumTree.  Concretely, GumTree takes the two AST trees as input, and generates a sequence of edit actions (a.k.a edit script) that transform one tree into another, where the size of the edit script represents the edit distance between the two trees.
	\item For the third iteration (i.e., Token), since the relevant information in the tree is text, we use a text distance algorithm (Jaro-Winkler~\cite{jaro1989advances,winkler1990string}) to compute the edit distance between two tokens extracted from the trees. We use the implementation of  Jaro-Winkler edit distance from Apache Commons Text library\footnote{\url{https://commons.apache.org/proper/commons-text/}}, which computes the Jaro-Winkler edit distance of two strings ${d_{w}}$ as defined in Equation \ref{eq:jaro}. The equation consists of two components; Jaro's original algorithm ($j_{sim}$) and Winkler's extension($w_{sim}$). The Jaro similarity is the weighted sum of percentage of matched characters $c$ from each file and transposed characters $t$. Winkler increased this measure for matching initial characters, by using a prefix scale $p$ that is is set to 0.1 by default, which gives more favorable ratings to strings that match from the beginning for a set prefix length $l$. The algorithm produces a similarity score (${w_{sim}}$) between 0.0 to 1.0, where 0.0 is the least likely and 1.0 is a positive match. Finally, this similarity score is transformed to distance (${d_{w}}$).
\begin{equation}
%	\footnotesize
\begin{array}{l}
{d_{w}}(s_1,s_2)=1 - {w_{sim}}(s_1,s_2) \\
{w_{sim}}(s_1,s_2) = {j_{sim}}(s_1,s_2) + l*p(1-{j_{sim}}(s_1,s_2)) \\

{j_{sim}}(s_1,s_2) =
\begin{cases}
    {0} &  \text{if } c = 0; \\
    \frac{1}{3}(\frac{c}{\abs{s_1}}+\frac{c}{\abs{s_2}}+\frac{c-t}{c}) & otherwise.
 \end{cases} \\

\scriptsize \mbox{$l$: The number of agreed characters at the
beginning of two strings}. \\
\scriptsize \parbox{0.7\linewidth}{$p$: is a constant scaling factor for how much the score is adjusted upwards for having common prefixes, which is set to 0.1 in Winkler's work~\cite{winkler1990string}.}\\

\end{array}
\label{eq:jaro}
\end{equation}

\end{itemize}
As the last step of comparison, we check the edit distance of the tree pair and tag the pairs having the distance zero as identical pairs, since the distance zero implies that no operation is necessary to transform one tree to another, or for the third fold ($Token$) the tokens in the tree are the same.
Eventually, we store and save the set of identical tree pairs produced in each iteration, which would be used in Step 4.
% The similarity score is 0.0 to 1.0, where 0.0 is the least likely and 1.0 is a positive match.
% Jaro-Winkler distance $d_{w}$ is defined as  $ d_{w}=1-sim_{w}} $

 % For our purposes, anything below a 0.8 is not considered as similar. We remove the tokens that are similar from the sequences of the token change actions, in order to produce action sets as in the different folds, containing the sequences of the token change actions that are different from them.
	% Finally, we check the size of actions set, and tag the pairs having the size zero, since the size zero implies that no operation is necessary to transform one tree to another, thus the trees are similar.

%
%Then, for the first and second iterations, the relevant trees are matched in order to establish the matrix mappings between the tree nodes to infer a tree edit script (which lists the necessary operations to transform one tree to another).

%
%Finally, in order to tag the similar trees, we check the edit distance.
\subsection{Step 4 -- Pattern Inference}
\label{step4}

\quest{
\begin{definition}
	\label{def:clusters}
(Pattern) Let $g$ be a graph in which nodes are elements of $RE$ and edges are defined by $R_{\mbox{\scriptsize identical}}$. \\
$g$ consists of a set of connected subgraphs $SG$ (i.e., clusters of specialized tree representations of \ediff{s}) where $sg_{i}$ and $sg_{j}$ are disjoint $\forall sg_{i}, sg_{j} \in SG$. A pattern is defined by $sg_{i} \in SG$ if $sg_{i}$ has at least two nodes (i.e., there are \underline{recurrent trees}).
%We define a fix pattern as a recurrent edit script over the AST representations of code changes.
\end{definition}
}

%
%\dongsun{seriously, we need a running example here.}
%In this step, we take the list of pairs that are tagged as identical in Step 3, and form clusters of \ediff{s} that have the same specialized tree representations according to the $fold$.
%The identical tree pairs are stored as follows:
%MethodDeclaration/2/2/1/0\_1,MethodDeclaration/2/2/1/0\_2,MethodDeclaration/2/2/2/0\_1.In order to form clusters of trees from the list of tree pairs that are tagged as identical, it is necessary to distinguish the trees that are identical.
%
%
Finally, to infer patterns, we resort to clustering of the specialized tree representations of \ediff{s}. First, we start by retrieving the set of identical tree pairs produced in Step 3 for each iteration. Following Algorithm~\ref{clusterAlgo}, we extract the corresponding specialized tree representations according to the fold (i.e., ShapeTrees, ActionTrees, TokenTrees) since the trees are identical only in a given fold. In order to find groups of trees that are identical among themselves (i.e., clusters) we leverage graphs. Concretely, we implement a clustering process based on the theory of connected components (i.e., subgraph) identification in a graph~\cite{skiena1997stony}. We create an undirected graph from the list of tree pairs, where the nodes of the graph are the trees and the edges represent trees that are associated (i.e., identical tree pairs). From this graph, we identify clusters as the subgraphs, where each subgraph contains a group of trees that are identical among themselves and disjoint from others.

% includes \ediff{s} that are connected (i.e., an edge in the graph implies that the tree representations associated to the relevant graph nodes are identical), which is qualified as a cluster.

% we collect the indices of the tagged pairs and use these indices as the vertices of a graph. A graph is then formed by connected vertices (edges) that represent pairs of indices that are associated. 

 % where the subgraph are considered as the clusters, which are connecting identical \ediff{s}.

\begin{algorithm}
\scriptsize
    \SetKwInOut{Input}{input}
    \SetKwInOut{Output}{output}
    \SetKw{Return}{return}

    \Input{$R_{\mbox{identical}}$: A list of identical \ediff pairs}
    \Input{$fold$ $\in$ \{$Shape$,$Action$,$Token$\}}
    %\BlankLine
    \Output{$C$: A list of clusters}
    %\BlankLine
    \SetKwProg{Fn}{Function}{}{end}
    \SetKwFunction{main}{main}
    %BlankLine

    \DontPrintSemicolon

	\Fn{main($R_{\mbox{identical}}$,$fold$)}{
        $C\leftarrow\emptyset$\;
        $TP$ $\leftarrow$ getTreePairs($R_{\mbox{identical}}$,$fold$)\;
        $E$ $\leftarrow$ transformPairsToEdges($TP$) \tcc*{E: edges created from tree pairs $TP$}
        %\tcp{ [(1, 2), (2, 3), (3, 4)]  list of edges}
        $g$ $\leftarrow$ createGraph($E$)\;
        $SG$ $\leftarrow$ g.connectedComponents() \tcc*{$SG$: list of subgraphs found in graph $g$}
        \ForEach{$sg$ in $SG$}{
            $c\leftarrow$ s.nodes() \tcc*{c: cluster formed from the nodes of subgraph $sg$}
            $C$.add($c$)\;
        }
        \Return $C$\;
    }

    \Fn{getTreePairs($R_{\mbox{identical}}$,$fold$)}{
        $P\leftarrow\emptyset$ \tcc*{P: list of tree pairs}
        \ForEach{$a$ in $R_{\mbox{identical}}$}{
            ($eTree1$, $eTree2$) $\leftarrow$ getRichEditScripts($a$) \tcc*{restore Rich Edit Scripts of a given pair from the cache}

           \uIf{$fold$ == $Shape$}{
              $sTree1$,$sTree2$ $\leftarrow$ getASTNodeTrees($eTree1$, $eTree2$)\;
            }
           \uElseIf{$fold$ == $Action$}{
              $sTree1$,$sTree2$ $\leftarrow$ getActionTrees($eTree1$, $eTree2$)\;
            }
           \Else{
              $sTree1$,$sTree2$ $\leftarrow$ getTokenTrees($eTree1$, $eTree2$) \tcc*{$fold$ == $Token$}
            }
            $P$.add($sTree1$,$sTree2$)\;
        }
        \Return $P$\;

    }

    \caption{Clustering based on subgraph identification.}\label{clusterAlgo}
\end{algorithm}

A cluster contains a list of \ediff{s} sharing a common specialized tree representations according to the $fold$. Finally, a cluster is qualified as a pattern, when it has at least two members.
 % which are spread in several patches, while the members that are spread within a single patch (as the common specialized tree representation is repeat only inside the patch as hunks) are considered as a single member \ak{need to explain better}.
 The patterns for each $fold$ are defined as follows:

% The step 3 for each fold of the iteration produces a list of pairs that are tagged as identical, which is the subset of the search indices indicating the indices of the hunks that are similar. In each pattern mining iteration, we form clusters of fix patterns targets a specific specialized representation, listed as follows:

\paragraph{Shape patterns.}\label{shapedBasedFixPattern} The first iteration attempts to find patterns in the ShapeTrees associated to developer patches. We refer to them as Shape patterns, since they represent the shape of the changed code in a structure of the tree in terms of node types. Thus, they are not fix patterns per se, but rather the context in which the changes are recurrent.

%\tb{Give examples of what that could be}
% \begin{figure}[!h]
%     \centering
%     \vspace{2mm}
%     \lstinputlisting[language=diff,linewidth={\linewidth},frame=tb,basicstyle=\footnotesize\ttfamily]{386ab6_level1}
%     \caption{Shape-based fix pattern for  Figure~\ref{patch1}.}
%     \label{shape}
% \end{figure}
%\begin{center}
%	\lstinputlisting[caption=Shape-based fix pattern for  Listing~\ref{patch1}, label=shape]{386ab6_level1}
%\end{center}

\paragraph{Action patterns.} The second iteration considers samples associated to each shape pattern and attempts to identify reoccurring repair actions from their ActionTrees. This step produces patterns that are relevant to program repair as they refer to recurrent code change actions. Such patterns can indeed be matched to dissection studies performed in the literature~\cite{defects4J-dissection}. We will refer to Action patterns as the sought fix patterns. Nevertheless, it is noteworthy that, in contrast with literature fix patterns, which can be generically applied to any matching code context, our Action patterns are specifically mapped to a code shape (i.e., a shape pattern) and is thus applicable to specific code contexts. This constrains the mutations to relevant code contexts, thus yielding more likely precise fix operations.
%\ak{i fell that the action patterns are not very clear.}
% \ak{maybe we can discuss about holes here,fix templates contain "holes", that can be filled with arbitrary program variables {by selecting the donor code from the local file}}
%\tb{Give examples of pattern here}

% \begin{figure}[!h]
%     \centering
%     \vspace{2mm}
%     \lstinputlisting[language=diff,linewidth={\linewidth},frame=tb,basicstyle=\footnotesize\ttfamily]{386ab6_level2}
%     \caption{Action-driven fix pattern for Figure~\ref{patch1}.}
%     \label{action}
% \end{figure}
%\lstinputlisting[caption=Action-driven fix pattern for Listing~\ref{patch1}, label=action]{386ab6_level2}

\paragraph{Token patterns.} The third iteration finally considers samples associated to each action pattern and attempts to identify more specific patterns with respect to the tokens available. Such token-specific patterns, which include specific tokens, are not suitable for implementation into pattern-based automated program repair systems from the literature. We discuss however their use in the context of deriving collateral evolutions (cf. Section~\ref{rq2:cocci}).

\section{Experimental Evaluation}
\label{sec:expSetup}

We now provide details on the experiments that we carry out for \toolname.
Notably, we discuss the dataset, and present the implementation details.
Then, we overview the statistics on the mining steps, and eventually enumerate the research questions for the assessment of \toolname.

\subsection{Dataset}

We collect code changes from 44 large and popular open-source projects from Apache-Commons, JBoss, Spring and Wildfly communities with the following selection criteria: we focused on projects (1) written in Java, (2) with publicly available bug reports, (3) having at least 20 source code files in at least one of its version; finally, to reduce selection bias, (4) we choose projects from a wide range of categories - middleware, databases, data warehouses, utilities, infrastructure.
This is a process similar to Bench4bl~\cite{lee2018bench4bl}.
 Table~\ref{tab:dataset}
% \dongsun{we can probably transpose this table since currently it is too small.}
details the number of bug fixing patches that we considered in each project. Eventually, our dataset includes 11\,416 patches.

\begin{table}[!h]
\centering

\caption{Dataset.}%
\label{tab:dataset}
%\subfloat[]

\resizebox{0.9\linewidth}{!}{%

		\begin{tabular}{l||l|r|l|r}
\toprule
  Community & Project & \makecell{\# Patches} & Project & \makecell{\# Patches} \\
   \hline\noalign{\smallskip}

\multirow{5}{*}{Apache} & {camel}                  &1366 &    {commons codec}    & 11 \\
 & {commons collections} &     56 &    {commons compress} & 73 \\
 & {commons configuration} &   89 &    {commons crypto}   & 9  \\
 & {commons csv} &             18 &    {common io}        & 58 \\
 & {hbase} &                 2169 &    {hive}             &2641 \\
      \hline\noalign{\smallskip}
JBoss &  {entesb} &                   15&    {jbmeta}           & 14 \\
      \hline\noalign{\smallskip}
\multirow{13}{*}{Spring} & {amqp} &                    89 &   {android}           &5 \\
& {batch} &                  224 &  {batchadm}           &11 \\
& {datacmns} &               151 &  {datagraph}          &19 \\
& {datajpa} &                112 &  {datamongo}          &190 \\
& {dataredis} &               65 &  {datarest}           &91 \\
& {ldap}  &                26   & {mobile}              &11  \\
& {roo} &               414     & {sec}                 &304     \\
& {secoauth}  &              66 & {sgf}                 &   35     \\
& {shdp}  &              35     & {shl}                 &   11     \\
& {social}  &            14     & {socialfb}            &   12     \\
& {socialli}  &             2  & {socialtw}            &   9     \\
& {spr} &             1098      & {swf}                 &   84     \\
& {sws} &               101     &                       & \\
      \hline\noalign{\smallskip}
\multirow{3}{*}{Wildfly} & {ely} &   217                 & {swarm}               &   131     \\
 & {wfarq} &   8                 & {wfcore}              &   547     \\
 & {wfly}  &   802               & {wfmp}                &   13     \\
  \hline\noalign{\smallskip}
    \hline\noalign{\smallskip}

\multicolumn{4}{l}{Total}  &{\bf 11416}\\
\bottomrule
    \end{tabular}
		}

\end{table}

\subsection{Implementation Choices}
We recall that we have made the following parameter choices in the \toolname workflow:

\begin{itemize}
	\item The ``Shape'' search index considers only \ediff{s} having a depth greater than 1 (i.e., the AST sub-tree should include at least one parent and one child).%\ak{we we need to say to capture changes with surrounding context information??}
	\item Comparison of \ediff{s} is designed to retrieve identical trees (i.e., tree edit distance is 0).
	%\item We consider that Action patterns are the relevant fix patterns for program repair.
	% are spread among several patches. The members that are spread within a single patch (as the common specialized tree representation is repeat only inside the patch as hunks) are considered as a single member.
\end{itemize}

\subsection{Statistics}
\toolname is a pattern mining approach to produce fix patterns for program repair systems.
Its evaluation (cf. Section~\ref{sec:assessment}) will focus on evaluating the relevance of the yielded patterns. Nevertheless, we provide statistics on the mining process to provide a basis of discussion on the implications of \toolname's design choices.

\paragraph{Search Indices.}
\toolname mines fix patterns through comparison of hunks (i.e., contiguous groups of code lines). 11\,416 patches in our database are eventually associated to 41\,823 hunks. A direct pairwise comparison of these hunks would lead to 874\,560\,753 tree comparison computations.
The combinatorial explosion of the comparison space is overcome by building search indices as previously described in Section~\ref{step2}. Table~\ref{tab:statistics} shows the details on the search indices built for each fold in the \toolname iterations. From the 874+ million tree pairs to be compared (i.e., $C_{41823}^2$), the construction of the Shape index (implements criteria on the tree structure  to focus on comparable trees) lead to 670 relevant comparison sub-spaces yielding a total of only 12+ million tree comparison pairs. This represents a reduction of 98\% of the comparison space. Similarly, the Action index and the Token index reduce the associated comparison spaces by 88\% and 72\% respectively.

% Overall the space space is reduced respectively 98\%, 88\%, 72\% using the search indices Shape, Action, Token.

\begin{table}[!h]

\centering

\caption{Comparison space reduction.}% 
\label{tab:statistics}
%\subfloat[]
%{\parbox{0.4\linewidth}{%
\resizebox{\linewidth}{!}{%
		\begin{tabular}{lrrr}
		\toprule
			  \makecell{Search Index} & \makecell{\# of hunks (in fold) }& \makecell{\# Comparison sub-spaces} & \makecell{\# Tree comparison pairs}  \\
			  \hline
       % \noalign{\smallskip}\hline\noalign{\smallskip}
        Shape  & 41\;823 &  670 & 12\;601\;712 \\%& 1\;370\;406 \\
        Action & 25\;290 & 2\;457 & 1\;427\;504 \\%& 628\;531\\
        Token  & 6759 &  411 & 401\;980 \\%& 18\;471\\    

		%\noalign{\smallskip}
		\bottomrule
		\end{tabular}

		}
\end{table}

\paragraph{Clusters.} We infer patterns by considering recurrence of trees: the clustering process groups together only tree pairs that are identical among themselves. Table~\ref{tab:clusterStatistics} overviews the statistics of clusters yielded for the different iterations: Shape patterns (which represent code contexts) are the most diverse. Action patterns (which represent fix patterns that are suitable as inputs for program repair systems) are substantially less numerous. Finally, Token patterns (which may be codebase-specific) are significantly fewer. We recall that we consider all possible clusters as long as it includes at least 2 elements. A practitioner may however decide to select only large clusters (i.e., based on a threshold).

 \begin{table}[!h]

\centering

\caption{Statistics on clusters.}% 
\label{tab:clusterStatistics}
%\subfloat[]
%{\parbox{0.4\linewidth}{%
%\resizebox{\linewidth}{!}{%
		\begin{tabular}{lrrr}
		\toprule
			  \makecell{Pattern} & \makecell{\# Trees (clustering input)} &\makecell{\# Corresponding change hunks} &\makecell{\# Clusters} \\
			  \hline
       % \noalign{\smallskip}\hline\noalign{\smallskip}
        Shape  & 1\;370\;406&  25\;290 & 2947 \\ 
        Action & 628\;531&  6\;759 & 428  \\
        Token &18\;471 &  1\;562  & 326 \\

		%\noalign{\smallskip}
		\bottomrule

		\end{tabular}

%		}
\end{table}

%, and the pruning of trees that are spread among several patches.

Because \toolname considers code hunks as the unit for building \ediff{s}, a given pattern may represent a repeating context (i.e., Shape pattern) or change  (i.e., Action or Token pattern) that  is only {\em part} of the patch (i.e., this patch includes other change patterns) or that is the {\em full} patch (i.e., the whole patch is made of this change pattern). Table~\ref{tab:patternStatistics} provides statistics on partial and full patterns. The numbers represent the disjoint sets of patterns that can be identified as always full or as always partial. Patterns that may be {\em full} for a given patch but {\em partial} for another patch are not considered.%\ak{is this clear?}
Overall, the statistics indicate that, from our dataset of over 40 thousand code hunks, only a few (e.g., respectively 278 and 7\;120 hunks) are associated with patterns that are always {\em full} or always {\em partial} respectively. In the remaining cases, the pattern is associated to a code hunk that may form alone the patch or may be tangled with other code. This suggests that \toolname is able to cope with tangled changes during pattern mining.
%\ak{maybe not very clear this part. what does it implies if the pattern is full, it means that the change is not tangled??}

\begin{table}[!h]

\centering

\caption{Statistics on Full vs Partial patterns.}% 
\label{tab:patternStatistics}
%\subfloat[]
%{\parbox{0.4\linewidth}{%
%\resizebox{\linewidth}{!}{%
		\begin{tabular}{lrrr|rrr}
		\toprule
				& \multicolumn{3}{c}{Partial patterns} & \multicolumn{3}{c}{Full patterns}  \\\cline{2-7}
			   & \makecell{\# Patterns} &\makecell{\# Patch} &\makecell{\# Hunk} & \makecell{\# Patterns} &\makecell{\# Patch} &\makecell{\# Hunk} \\
			  \hline
       % \noalign{\smallskip}\hline\noalign{\smallskip}
        Shape  & 1358&  3140 & 7120 & 120 &  223 & 278\\ 
        Action & 124&  554 & 1153 & 10 &  20 & 25 \\
        Token &75 &  148  & 277 & 14 &  22  & 32\\    
        \bottomrule
        % \hline
        % Shape-based  & 120 &  223 & 278 \\ 
        % Action-driven & 10 &  20 & 25  \\
        % Token-specific & 14 &  22  & 32 \\   

		% %\noalign{\smallskip}
		% \hline
  %       Shape-based  & 47&  56 & 127 \\ 
  %       Action-driven & 6&  14 & 22  \\
  %       Token-specific &3 &  5  & 6 \\    
  %       \hline
  %       Shape-based  & 9 &  17 & 21 \\ 
  %       Action-driven & 3 &  9 & 9  \\
  %       Token-specific & 3 &  5  & 6 \\   

		\end{tabular}

%		}
\end{table}

 Similarly, we investigate how the patterns are spread among patches. Indeed, a pattern may be found because a given patch has made the same change in several code hunks. We refer to such pattern as {\em vertical}. In contrast, a pattern may be found because the same code change is spread across several patches. We refer to such pattern as {\em horizontal}.
 Table~\ref{tab:spreadStatistics} shows that vertical and horizontal patterns occur in similar proportions for Shape and Action patterns. However, Token patterns are significantly more vertical than horizontal (65 vs 224). This is in line with studies of collateral evolutions in Linux, which highlight large patches making repetitive changes in several locations at once~\cite{padioleau2008documenting} (i.e., collateral evolutions are applied through vertical patches).
%\ak{is the conclusion clear?}
 \begin{table}[!h]

\centering

\caption{Statistics on Pattern Spread.}%
\label{tab:spreadStatistics}
%\subfloat[]
%{\parbox{0.4\linewidth}{%
%\resizebox{\linewidth}{!}{%
\begin{threeparttable}
		\begin{tabular}{lrrr|rrr}
		\toprule
				& \multicolumn{3}{c}{Vertical } & \multicolumn{3}{c}{Horizontal}  \\\cline{2-7}
			   & \makecell{\# Patterns} &\makecell{\# Patch} &\makecell{\# Hunk} & \makecell{\# Patterns} &\makecell{\# Patch} &\makecell{\# Hunk} \\
			  \hline
       % \noalign{\smallskip}\hline\noalign{\smallskip}
        Shape  & 881&  881 & 2432 & 1194 & 3808  & 3808\\
        Action & 148& 148  & 488 & 132 & 574 & 574 \\
        Token &224 &  224   &709  & 65 & 170  & 170 \\
        \bottomrule
        % \hline
        % Shape-based  & 120 &  223 & 278 \\
        % Action-driven & 10 &  20 & 25  \\
        % Token-specific & 14 &  22  & 32 \\

		% %\noalign{\smallskip}
		% \hline
  %       Shape-based  & 47&  56 & 127 \\
  %       Action-driven & 6&  14 & 22  \\
  %       Token-specific &3 &  5  & 6 \\
  %       \hline
  %       Shape-based  & 9 &  17 & 21 \\
  %       Action-driven & 3 &  9 & 9  \\
  %       Token-specific & 3 &  5  & 6 \\

		\end{tabular}

%		}
\begin{tablenotes}

    \item[*]
 A pattern can simultaneously be vertical (when it is associated to several changes in hunks of the same patch) and horizontal (when it appears as well within other patches).
 \end{tablenotes}
\end{threeparttable}
\end{table}

\subsection{Research Questions}
\label{sec:exp}
% This section details the experiments for assessing \toolname
%. After enumerating the research questions, we present the experimental setup and describe our findings.
%\subsection{Research Questions}
% Data collection.
%We assess the \toolname approach

The assessment experiments are performed with the objective of investigating the usefulness of the patterns mined by \toolname. To that end, we focus on the following research questions (RQs):
\begin{enumerate}[leftmargin=9.5mm]
	\item[RQ-1] Is automated patch clustering of \toolname consistent with human manual dissection?
%	\item[RQ-2] \textcolor{red}{\sout{Do clusters}} \textcolor{blue}{Are patterns} generated by \toolname \textcolor{red}{\sout{match}}\textcolor{blue}{consistent with} community-provided fix pattern dissections?
	\item[RQ-2] Are patterns inferred by \toolname compatible with known fix patterns?
	\item[RQ-3] Are the mined patterns effective for automated program repair?
\end{enumerate}

\section{Results}
\label{sec:assessment}
\subsection{RQ1: Comparison of \toolname Clustering against Manual Dissection}
\noindent

\noindent
{\bf Objective.} We propose to assess relevance of the clusters yielded by \toolname in terms of whether they represent patterns which practitioners would view as recurrent changes that are indeed relevant to the patch behaviour. In previous section, the statistics showed that several changes are recurrent and are mapped to \toolname's clusters. In this RQ, we validate whether they are relevant to the practitioner's viewpoint. For example, if \toolname was not leveraging AST information, removal of blank lines would have been seen as a recurrent change (hence a pattern); however, a practitioner would not consider it as relevant.

\noindent
{\bf Protocol.} We consider an oracle dataset of patches with change patterns that are labelled by humans. Then we associate each of these patches to the relevant clusters mined by \toolname on our combined study datasets. This way, we ensure that the clustering does not overfit to the oracle dataset labelled by humans. Eventually, we check whether each set of patches (from the oracle dataset) that are associated to a given \toolname cluster, consists of patches having the same labels (from the oracle).

\noindent
{\bf Oracle.} For our experiments, we leverage the manual dissection of Defects4J~\cite{just2014defects4j} provided by Sobreira et. al~\cite{defects4J-dissection}.
%is a popular dataset which includes a large and manually reviewed set of real-world Java bugs.
%It has been proposed to enable reproducible studies in the software testing community, but is becoming recently a de-facto benchmark for repair approaches targeting Java programs. All 395 real bugs from our snapshot of the Defects4J dataset are provided with the associated fixes collected from the change history of the associated projects, namely JFreeChart, Google Closure Compiler, Commons Lang, Commons Math, Mockito and Joda-Time.
%Sobreira et. all~\cite{defects4J-dissection} later presented a study of the anatomy of DefectsJ4 patches,
This oracle dataset associates the developer patches of 395 bugs in the Defects4J datasets with 26 repair pattern labels (one of which is being ``Not classified'').
%for enumerating nine repair patterns at a first level. Some of the higher level patterns are further split into several sub-patterns, such as {\em Expression Fix} is further regrouped into two sub repair patterns.

\noindent
{\bf Results.} Table~\ref{tab:dataset4j} provides statistics that describe the proportion\footnote{In this experiment, we excluded 34 patches from Defects4J dataset which affect more than 1 file.} of \toolname's patterns that can be associated to change patterns in the Defects4J patches.

\begin{table}[!h]
\centering

\caption{Proportion of shared patterns between our study dataset and Defects4J.}% 
\label{tab:dataset4j}
%\subfloat[]
%{\parbox{0.4\linewidth}{%
\resizebox{\linewidth}{!}{%
		\begin{tabular}{l|cr|cr}
%			\hline\noalign{\smallskip}
\toprule
%\cline{2-5}
			 & \multicolumn{2}{c|}{Study dataset} & \multicolumn{2}{c}{Defects4J}  \\\cline{2-5}
%			&  \multicolumn{2}{FixMiner } & \multicolumn{2}{ Defects4J } & \multicolumn{2}{BugsDotJar } \\
						&  {\# corresponding hunks}  & \makecell{\# Patterns} & \makecell{\# corresponding hunks} &  \makecell{\# Patterns} \\ \cline{2-5}
       % \noalign{\smallskip}\hline\noalign{\smallskip}
        Shape  &  25272 & 2947 & 479 &214 \\
        Action  & 6755 & 428 & 103 & 37 \\
        Token  & 1562  & 326 & 23 & 13 \\

		%\noalign{\smallskip}
		\bottomrule

		\end{tabular}

		}

% 		tokens 1562
% actions 6755
% shapes 25272
\end{table}

\paragraph{Diversity.} We check the number of patterns that can be found in our study dataset and Defects4J. In absolute numbers, Defects4J patches include a limited set of change patterns (i.e., $\sim7\%=\frac{214}{2947}$) in comparison to what can be found in our study dataset.

\paragraph{Consistency.} We check for consistency of \toolname's pattern mining by assessing whether all Defects4J patches associated to a \toolname cluster are indeed sharing a common dissection pattern label. We have found that the clustering to be consistent for $\sim78\%=\frac{166}{214}$, $\sim73\%=\frac{27}{37}$ and $\sim92\%=\frac{12}{13}$ of Shape, Action and Token clusters respectively.
% \ak{where did you get this numbers? not from table 7}

%\tb{Todo}
%Shapes $\sim78\%=\frac{166}{214}$  Actions $\sim73\%=\frac{27}{37}$  Tokens $\sim92\%=\frac{12}{13}$

 \find{{\bf RQ1-Consistency} $\blacktriangleright$ \toolname can produce patterns that are matching patches that are labeled similarly by humans. The patterns are thus largely consistent with manual dissection. }

\paragraph{Granularity.} The human dissection provides repair pattern labels for a given patch. Nonetheless, the label is not specifically associated to any of the various changes in the patch. \toolname however yields patterns for code hunks. Thus, while \toolname links a given hunk to a single pattern, the dissection data associates several patterns to a given patch. We investigate the granularity level with respect to human-provided patterns. Concretely, several patterns of \toolname can actually be associated (based on the corresponding Defects4J patches) to a single human dissection pattern. Consider the example cases in Table~\ref{tab:exampleDissection}. Both patches consist of nested InfixExpression under the IfStatement. The first \toolname pattern indicates that the change operation (i.e., update operator) should be performed on the children InfixExpression. On the other hand, the second pattern implies a change operation in the parent InfixExpression. Thus, eventually, \toolname  patterns are finer-grained and associates the example patches to two distinct patterns each pointing the precise node to update, while manual dissection considers them under the same coarse-grained repair pattern. %\ak{the last sentence is somehow not very clear}
% \mm{the first pattern description is super specific (the if condition has to have two operators). Why?}\ak{I updated the text(not caption) explaining thew why, is it clear now?}
 % !TEX root = main.tex
\begin{table}[!h]
\centering
		\scriptsize
\caption{Granularity example to \toolname mined patterns.}
\label{tab:exampleDissection}

\resizebox{\linewidth}{!}{%
		\begin{tabular}{l|l|l}

			\toprule
			& Pattern & Example patch from \toolname dataset  \\
        \noalign{\smallskip}\hline\noalign{\smallskip}

\multirow{2}{*}{\toolname} & \multirow{2}{*}{
\begin{lstlisting}[language=diff]
UPD IfStatement
---UPD InfixExpression
------UPD InfixExpression
---------UPD Operator
\end{lstlisting}} & \multirow{4}{*}{
\begin{lstlisting}[language=diff]
@@ -83,7 +83,7 @@ public BoundedInputStrea ...
     @Override
     public int read() throws IOException {
-        if (max >= 0 && pos == max) {
+        if (max >= 0 && pos >= max) {
             return -1;
\end{lstlisting}} \\ 
% \makecell[l]{Logic expression\\ modification} & & \\
& & \\
& & \\
& & \\
& & \\ \noalign{\smallskip}\cline{1-2}
\noalign{\smallskip}
						Dissection~\cite{defects4J-dissection} & \makecell[l]{Logic expression modification \\ Single Line} &\\ 
						% Single Line & &\\ 
						%  & &\\ 
						% \makecell[l]{Conditional\\ expression\\ modification} & &\\ 	
% \multirow{2}{*}{Actions} &\makecell[l]{Conditional\\ expression\\ modification} & & \\
% \multirow{1}{*}{Ids} &Closure 62\\ Closure 63\\ Closure 73& & \\

\noalign{\smallskip}\hline\noalign{\smallskip}
\multirow{2}{*}{\toolname} & \multirow{3}{*}{
\begin{lstlisting}[language=diff]
UPD IfStatement
---UPD InfixExpression
------UPD Operator
\end{lstlisting}} & \multirow{3}{*}{
\begin{lstlisting}[language=diff]
@@ -145,7 +145,7 @@ private void moveFile(Path s ...
   private Path createTargetPath(Path targetPath ...
     Path deletePath = null;
     Path mkDirPath = targetPath.getParent();
-    if (mkDirPath != null & !fs.exists(mkDirPath)) {
+    if (mkDirPath != null && !fs.exists(mkDirPath)) {
       Path actualPath = mkDirPath;
\end{lstlisting}}\\ 
& & \\
& & \\
& & \\

& & \\ \noalign{\smallskip}\cline{1-2}
\noalign{\smallskip}
Dissection~\cite{defects4J-dissection} & \makecell[l]{Logic expression modification \\ Single Line} &\\ 

% \makecell[l]{Conditional\\ expression\\ modification} & &\\ 

% \multirow{2}{*}{Actions} &\makecell[l]{Conditional expression modification} & & \\
% \multirow{1}{*}{Ids} &Chart 1 - Math 82 -Math 85 -Time 19& & \\
\bottomrule		
		
		\end{tabular}

}
\end{table}

We have investigated the differences between \toolname patterns and dissection labels and found several granularity mismatches similar to the previous example: 
{\tt condBlockRetAdd} ({\em condition block addition with return statement}) from manual dissection is associated to 14 fine-grained Shape patterns of \toolname: this suggests that the repair-potential of this pattern could be further refined depending on the code context. Similarly, {\tt expLogicMod} ({\em logic expression modification}), is associated to 2 separate Action patterns (see Table~\ref{tab:exampleDissection}) of \toolname: this suggests that the application of this repair pattern can be further specialized to reduce the repair search space and the false positives.

Overall, we found in total 37, 3 and 1 dissection repair patterns are further refined into several \toolname's Shape, Action and Token patterns respectively.

 \find{{\bf RQ1-Granularity} $\blacktriangleright$ We observe that manually-dissected patterns are more coarse-grained compared to \toolname's patterns.}
%  the threshold that we have set as a constraint on the recurrence of changes to derive a pattern. Overall a maximum of 15 change instances of Defects4J are associated with our derived patterns.
% This is due to the fact that most of the Defects4J changes implement repair actions which are not considered in this version of \toolname. Indeed, we have focused on mining patterns from changes that UPDATE only existing statements. However, the large number of Defects4J patches are about INSERTING new conditional blocks (79 patches), new return statement (77 patches), etc.

% We have performed a similar study with Bugs.jar~\cite{sahaMSR18}, a more comprehensive dataset of 1082 bugs and associated fixes. We find much more cases of UPDATE change patterns overlapping with our clusters as detailed in Table~\ref{tab:dataset4j}. Figure~\ref{fig:distOfDefect4JandBugs} details the distribution differences of patches for Bugs.jar and Defects4J in all \toolname update change patterns.

\paragraph{Assessment of \toolname's patterns with respect to associated bug reports.} Beyond assessing the consistency of \toolname's patterns based on human-built oracle dataset of labels, we further propose to investigate the relevance of the patterns in terms of the semantics that can be associated to the intention of the changes.
To that end, we consider bug reports associated to patches as a proxy to characterize the intention of the code changes. We expect bug reports sharing textual similarity to be addressed by patches that are syntactically similar. This hypothesis drives the entire research direction on Information retrieval-based bug localization~\cite{lee2018bench4bl}.

\begin{figure}[!htb]

\centering
\includegraphics[width=1\textwidth]{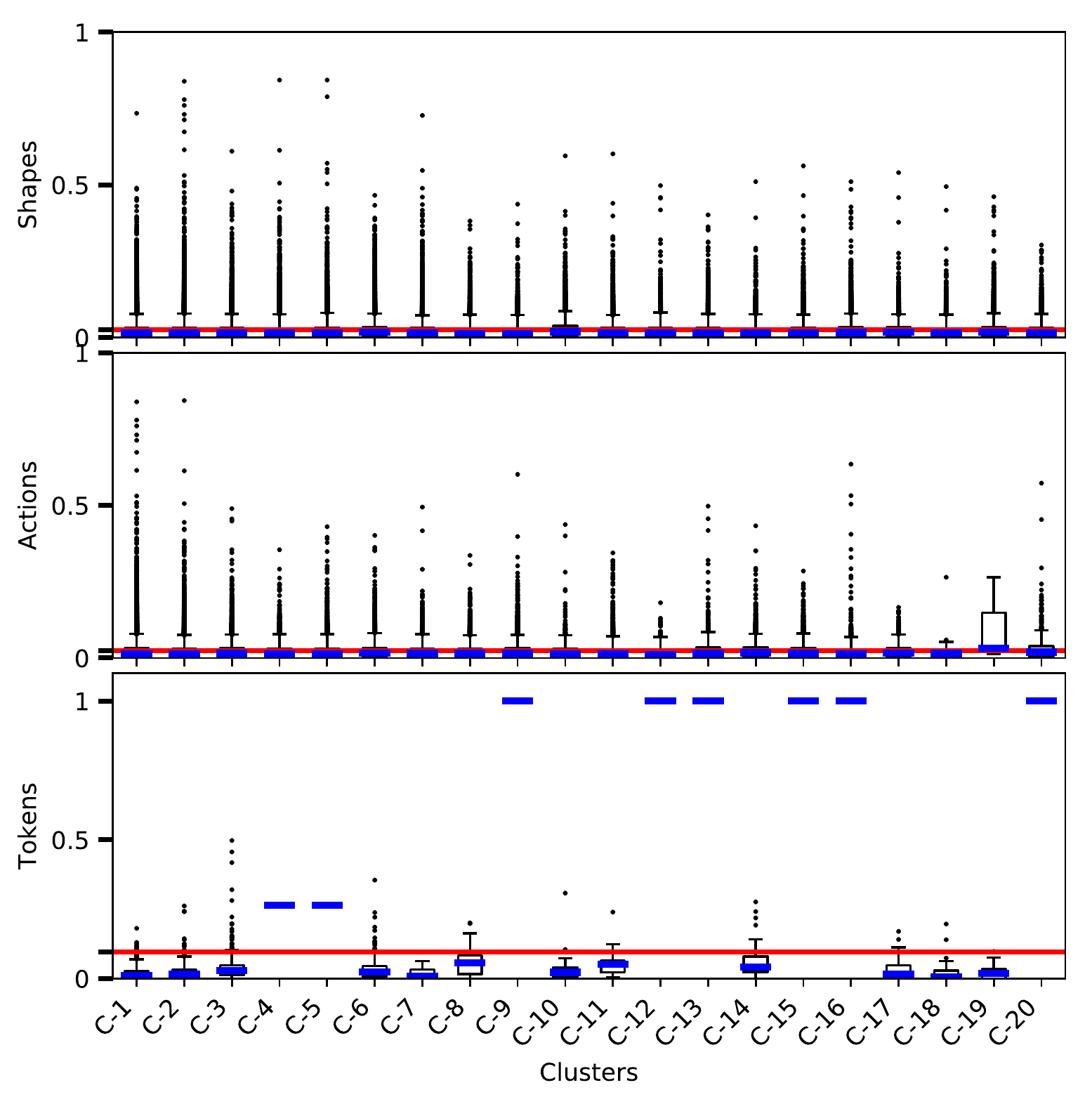}
\caption{ Distribution of pairwise bug report similarity. {\normalfont \footnotesize Note: A \textcolor{red}{red line} represents an average similarity for all bug reports in fold, and \textcolor{blue}{blue line} represents average similarity bug reports within a cluster.}}
\label{fig:bugSimiCluster2}
\end{figure}

Figure~\ref{fig:bugSimiCluster2} provides the distribution of pairwise bug report (textual) similarity values for the bug reports corresponding to patches associated to each cluster. For clear presentation, we focus on the top-20 clusters (in terms of size). We use TF-IDF to represent each bug report as a vector, and leverage Cosine similarity to compute similarity scores among vectors. The represented boxplots display all pairwise bug report similarity values, including outliers. Although for Shape and Action patterns the similarities are near 0 for all clusters, we note that there are fewer outliers for Action patterns. This suggests a relative increase in the similarity among bug reports. As expected, similarity among bug reports is the highest with Token patterns.

\subsection{RQ2: Compatibility between \toolname's patterns and APR literature patterns}

\noindent
{\bf Objective.} Given that \toolname aims to automatically produce fix patterns that can be used by automated program systems, we propose to assess whether the yielded patterns are compatible with patterns in the literature.

\noindent
{\bf Protocol.} We consider the set of patterns used by literature APR systems and compare them against \toolname's patterns. Concretely, we systematically try to map \toolname's patterns with patterns in the literature.  To that end, we rely on the comprehensive taxonomy of fix patterns proposed by Liu et al.~\cite{liu2019tbar}: if a given \toolname pattern can be mapped to a type of change in the taxonomy, then this pattern is marked as {\em compatible} with patterns in the literature. 

Recall that, as described earlier, fix patterns used by APR tools abstract changes at the form of \toolname's Action patterns~(Section~\ref{sec:method} - Step~4). In the absence of common language for specifying patterns, the comparison is performed manually. For the comparison, we do not conduct exact mapping between literature patterns and the ones yielded by \toolname as fix patterns yielded by \toolname have
more context information. We rather consider whether the context information yielded by \toolname
patterns  matches with the context of literature patterns. We discuss the related threats to validity in Section~\ref{sec:threats}. Given that the assessment is manual and thus time-consuming, we limit the comparisons to the top 50 patterns (i.e., Action patterns) yielded by FixMiner.%\ak{how do we define compatible ??}

\noindent
{\bf Oracle.} We build on the patterns enumerated by Liu et al.~\cite{liu2019tbar} who systematically reviewed fix patterns used by Java APR systems in the literature. They summarised 35 fix patterns in GNU format, which we refer to for comparing against \toolname patterns.

\noindent
{\bf Results.} Overall, among the 35 fix patterns used by the total of 11 studied APR systems, 16 patterns are also included in the fix patterns (i.e., Action patterns) yielded by \toolname when mining our study dataset. We recall that these patterns are often manually inferred and specified by researchers for their APR tools. Table~\ref{tab:literaturePatterns} illustrates examples of \toolname's fix patterns associated to some of the patterns used in literature. We note that fix patterns identified by \toolname are specific (e.g., for {\tt FP4: Insert Missed Statement}, the corresponding \toolname's fix pattern specifies which type of statement must be inserted). 

%2.1, 2.2, 2.5, 3, 4.1, 7.1, 9.1, 10.1, 10.2, 10.3, 10.4, 11.1, 12, 13.1, 14, 15.1

% !TEX root = main.tex
\begin{table*}[!h]
\centering
		\scriptsize

\caption{Example \toolname fix-patterns associated to APR literature patterns.}% 
\label{tab:literaturePatterns}
%\subfloat[]
%{\parbox{0.4\linewidth}{%
%\resizebox{\linewidth}{!}{%
\begin{threeparttable}
		\begin{tabular}{l|l}
%			\hline\noalign{\smallskip}
%
%			\noalign{\smallskip}\hline\noalign{\smallskip}
			\toprule
			Patterns enumerated by Liu et al.~\cite{liu2019tbar} & Example fix pattern from \toolname (*) \\
        \noalign{\smallskip}\hline\noalign{\smallskip}

FP2. Insert Null Pointer Checker  &\makecell[l]{INS IfStatement\\--- INS InfixExpression\\ ------ INS SimpleName
\\------ INS Operator\\------ INS NullLiteral\\--- INS ReturnStatement\\------ INS NullLiteral} \\ %cluster 3,0,0
% FP3. Insert Range Checker &\lstinputlisting[language=diff]{646e03} \\ %cluster 3,0,0
        \noalign{\smallskip}\hline\noalign{\smallskip}
FP4. Insert Missed Statement  &\makecell[l]{INS ExpressionStatement\\---INS MethodInvocation\\------INS SimpleName} \\ %cluster 3,0,0
%  INS ExpressionStatement TO MethodDeclaration
% --- INS MethodInvocation TO ExpressionStatement
        \noalign{\smallskip}\hline\noalign{\smallskip}
FP7. Mutate Data Type & \makecell[l]{UPD CatchClause\\--- UPD SingleVariableDeclaration\\------ UPD SimpleType}\\
% \makecell{	 * UPD VariableDeclarationStatement
% 	 * ---UPD PrimitiveType}
        \noalign{\smallskip}\hline\noalign{\smallskip}
FP9. Mutate Literal Expression & \makecell[l]{UPD FieldDeclaration\\--- UPD VariableDeclarationFragment\\------ UPD StringLiteral}\\
        \noalign{\smallskip}\hline\noalign{\smallskip}

% FP10. Mutate Method Invocation Expression   &\makecell[l]{UPD ExpressionStatement\\--- UPD MethodInvocation\\------ UPD SimpleName} \\ %cluster 3,0,0
% \hline

% FP10. Mutate Method Invocation Expression   &\makecell[l]{UPD ExpressionStatement\\--- UPD MethodInvocation\\------ UPD SimpleName\\--------- UPD SimpleName} \\ %cluster 3,0,0
% \hline
FP10. Mutate Method Invocation Expression   &\makecell[l]{UPD ExpressionStatement\\--- UPD MethodInvocation\\------ UPD SimpleName\\--------- INS SimpleName} \\ %cluster 3,0,0

        \noalign{\smallskip}\hline\noalign{\smallskip}
FP11. Mutate Operators  &\makecell[l]{	UPD IfStatement\\--- UPD InfixExpression\\------ UPD Operator} \\ %cluster 3,0,0
        \noalign{\smallskip}\hline\noalign{\smallskip}
FP12. Mutate Return Statement  &\makecell[l]{UPD ReturnStatement\\--- UPD MethodInvocation\\------ UPD SimpleName} \\ %cluster 3,0,0
% FP13. Mutate Variable  &\lstinputlisting[language=diff]{646e03} \\ %cluster 3,0,0
% FP14. Move Statement  &\lstinputlisting[language=diff]{646e03} \\ %cluster 3,0,0
% FP15. Remove Buggy Statement  &\lstinputlisting[language=diff]{646e03} \\ %cluster 3,0,0
% \cline{2-3}

%		\noalign{\smallskip}\hline
\bottomrule

		\end{tabular}

		\begin{tablenotes}
    \item[*]
{\bf Complete list of 16 Fix Patterns from literature that match \toolname's patterns:} FP2. Insert Null Pointer Checker (i.e., 2.1, 2.2 and 2.5), FP3. Insert Range Checker, FP4. Insert Missed Statement (i.e., 4.1),
FP7. Mutate Data Type (i.e., 7.1), FP9. Mutate Literal Expression (i.e., 9.1),
FP10. Mutate Method Invocation Expression (i.e., 10.1, 10.2, 10.3, and 10.4),
FP11. Mutate Operators (i.e., 11.1), FP12. Mutate Return Statement, FP13. Mutate Variable (i.e., 13.1), FP14. Move Statement and FP15. Remove Buggy Statement (i.e., 15.1).

  \end{tablenotes}
\end{threeparttable}

%		}
\end{table*}

% FP2. Insert Null Pointer Checker (i.e., 2.1, 2.2 and 2.5), FP3. Insert Range Checker, FP4. Insert Missed Statement (i.e., 4.1), FP7. Mutate Data Type (i.e., 7.1), FP9. Mutate Literal Expression (i.e., 9.1), FP10. Mutate Method Invocation Expression (i.e., 10.1, 10.2, 10.3, and 10.4), FP11. Mutate Operators (i.e., 11.1), FP12. Mutate Return Statement, FP13. Mutate Variable (i.e., 13.1), FP14. Move Statement and FP15. Remove Buggy Statement (i.e., 15.1).

Table~\ref{tab:fpConsistency} illustrates the proportion of \toolname's patterns that are compatible with patterns in the literature. In this comparison, we select the top-50 fix patterns yielded by \toolname and verify their presence within the fix patterns used in the APR systems.

\begin{table}[!ht]
	\scriptsize
	\centering
	\setlength\tabcolsep{2pt}
	\caption{Compatibility of Patterns: \toolname vs Literature Patterns.}
	\label{tab:fpConsistency}
	\resizebox{1\linewidth}{!}{
	\begin{threeparttable}
			\begin{tabular}{c|c|c|c|c|c|c|c|c|c|c|c}
			\toprule
			% \makecell[l]{{\bf APR}\\ {\bf tool}}
			  {\bf PAR}& {\bf HDRepair} & {\bf ssFix} & {\bf ELIXIR}& {\bf S3} & {\bf NPEfix}  & {\bf SketchFix} & {\bf SOFix} & {\bf Genesis} & {\bf CapGen} & {\bf SimFix} & {\bf AVATAR} \\
			\hline
			 7/16 & 7/12 & 6/34 & 8/11 & 3/4 & 1/9 & 5/6 & 9/12 & 1/108 & 12/30 & 8/16 & 6/13 \\
			\bottomrule
		\end{tabular}
		{We provide x/y numbers: x is the number of fix patterns in the corresponding APR tool that are mined by \toolname; y is the number of fix patterns used by the corresponding APR tool.}
    \end{threeparttable}
    }
\end{table}
We observed that
\begin{itemize}
	\item 7 patterns are compatible with fix patterns that are mined manually from bug fix patches (i.e., fix patterns in PAR~\cite{kim2013automatic}).
	\item between 1 and 8  patterns are compatible with researcher-predefined fix patterns used in ssFix~\cite{xin2017leveraging}, ELIXIR~\cite{saha2017elixir}, S3~\cite{le2017s3}, NEPfix~\cite{durieux2017dynamic}, and SketchFix~\cite{hua2018towards}, respectively.
	\item 7 patterns are compatible with fix pattern mined from history bug fixes by HDRepair~\cite{le2016history}, 9 patterns are compatible with fix patterns mined from StackOverflow by SOFix~\cite{liu2018mining}, and 1 fix pattern is compatible with 1 fix pattern mined by Genesis~\cite{long2017automatic} that focuses on mining fix patterns for three kinds of bugs.
	\item 12 and 8 patterns are compatible with the patterns used by CapGen~\cite{wen2018context} and SimFix~\cite{jiang2018shaping}, respectively, which extract patterns in a statistic way similar to the empirical studies of bug fixes~\cite{martinez2015mining, liu2018closer}.
	\item 6 patterns are compatible with the fix patterns used in AVATAR~\cite{liu2019avatar}, which are presented in a study for inferring fix patterns from FindBugs\cite{hovemeyer2004finding} static analysis violations~\cite{liu2018mining2}.
\end{itemize}

%These observation results indicate that \toolname can effectively extract useful fix patterns.\ak{should we remove prev sentence?}

\find{{\bf RQ2}$\blacktriangleright$ \toolname effectively yields Action patterns that are compatible for 16 over 35 patterns used in the literature of pattern-based program repair.}

%\subsubsection*{Are the patterns consistent with some bugs?}
%\kui{TODO: What is the answer for this question?}
%\find{{\bf RQ2}$\blacktriangleright$ Any conclusion?}

\paragraph{Manual (but Systematic) Assessment of Token patterns. }
\label{rq2:cocci}
Action and Token patterns are the two types of patterns that relate to code changes. In the assessment scenario above, we only considered Action patterns since they are the most appropriate for comparison with the literature patterns. We now focus on Token patterns to assess whether our hypothesis on their usefulness for deriving collateral evolutions holds (cf. Section~\ref{sec:method} - Step~4). To that end, we consider the various Token clusters yielded by \toolname and manually verify whether the recurrent change (i.e., the pattern) is relevant (i.e., a human can explain whether the intentions of the changes are the same). Eventually, if the pattern is validated, it should be presented as a generic/semantic patch~\cite{padioleau2008documenting,andersen2010generic} written in SmPL\footnote{Semantic Patch Language}.

%Thus, given a cluster and its change instances, we try to manually derive a SmPL\footnote{Semantic Patch Language} patch which would be a single generic representation of all the changes in the cluster. Due to space limitation we only include in Figure~\ref{fig:smpl} a couple of examples of SmPL patches corresponding to Action-driven patterns. Overall, we managed to specify the SmPL patches for 18 (i.e., 92\%) clusters selected from top50 Action-driven clusters. %\ak{it seems that I can label 32 out them manually, but for 34 of them I can write SmPL.}
In Table~\ref{tab:level1CLusters}, we list some of the patches that we found to be relevant. Among the top 50 Token patterns investigated, 12 patterns correspond to a modifier change, 4 patterns target changes in logging methods, and 1 pattern is about fixing the infix operator (e.g., {\em {\tt >}  $\rightarrow$ {\tt >=} } ). The remaining cases mainly focus on changes that complete the implementation of code {\em finally} block logic (e.g., missing call to {\em closeAll} for opened files), changes in Exception handling, updates to wrong parameters passed to method invocations, as well as wrong method invocations. As mentioned earlier, these patterns are spread mostly vertically (i.e. change is recurrent in several code hunks of a given patch) and the semantic behaviour of these patterns are specific to project nature.

Overall, our manual investigations on the top 50 Token patterns confirm that many of the recurrent changes associated to specific tokens are indeed relevant. We even found several cases where collateral evolution changes are regrouped to form a pattern as exhibited by the corresponding  pattern example presented in Figure~\ref{fig:smpl}. In this example, we illustrate the pattern using the SmPL specification language, which was designed for specifying collateral evolutions. This finding suggests that \toolname can be leveraged to systematically mined collateral evolutions in the form of Token patterns which could be automatically rewritten as semantic patches in SmPL format. This endeavour is however out of the scope of this paper, and will be investigated in future work.
 \begin{table*}[!h]
\centering
		\scriptsize

\caption{Example changes associated to \toolname mined patterns.}% 
\label{tab:level1CLusters}
%\subfloat[]
%{\parbox{0.4\linewidth}{%
\resizebox{\linewidth}{!}{%
		\begin{tabular}{l|l}
%			\hline\noalign{\smallskip}
%
%			\noalign{\smallskip}\hline\noalign{\smallskip}
			\toprule
			Semantic Behaviour of Pattern & Example change in developer patch  \\
        \noalign{\smallskip}\hline\noalign{\smallskip}
	
%method declaration statement method name change & Update method name & \lstinputlisting{1cff32} & \\ %cluster 0,0,1 
%method invocation: string literal parameter change & Update String literal name & \lstinputlisting{8dc752} & \\%cluster 2,0,0
%method invocation: string literal parameter change & Update String literal name & \lstinputlisting{8267de} & \\%cluster 2,0,1
Missing field modifier  &
\begin{lstlisting}[language=diff]
-    private boolean closed = true;
+    private volatile boolean closed = true;
\end{lstlisting} \\ %cluster 3,0,0
% \cline{2-3}
% & \lstinputlisting[language=diff]{ec4adb} \\ %cluster 3,0,0
% \cline{2-3}
% & \lstinputlisting[language=diff]{5058a8b} \\ %cluster 3,2,1
% \hline

% Variable type change & \lstinputlisting[language=diff]{ca1ca9} \\ %cluster 32,0,0 defects4j

%\cline{3-3}
%&  & \lstinputlisting{ca1ca9} & \\ %cluster 32,0,0
%>>>>>>> 316e8cdac20c83a40585348683171d4085f870a5
% \hline
%local variable declaration statement:methodInvocation change && Update method reference & \\
%if condition infix operator change & &
%Method call parameter value modification: %cluster 1

% Wrong Variable Reference & \makecell{Method call parameter value change \\ Replaceme variable by another variable}& \lstinputlisting[language=diff]{29c3b7} &\\
% \hline
% Wrong Method Reference & \lstinputlisting[language=diff]{386ab6}  \\
\hline
Wrong Log level &
\begin{lstlisting}[language=diff]
       } catch (Exception e) {
-        LOG.fatal("Could not append. Requesting close of wal", e);
+        LOG.warn("Could not append. Requesting close of wal", e);
         requestLogRoll();
\end{lstlisting} \\
\bottomrule		
		
		\end{tabular}

		}
\end{table*}

%Figure~\ref{fig:smpl} illustrates SmPL patches corresponding to the semantically relevant patterns.

%\ak{could you please check? }
%
%
\begin{figure}[!h]
\centering
% \tcbox[boxrule=0.5pt,arc=.3em,boxsep=-1mm,top=0pt,bottom=3pt]{\lstinputlisting{646e03_cocci}}
	\tcbox[boxrule=0.5pt,arc=.3em,boxsep=-1mm,top=0pt,bottom=3pt,width=0.8\linewidth]{
\begin{lstlisting}
// [caption=Wrong Log level]
@@
Logger log;
@@
...
- log.fatal(...);
+ log.warn(...);

\end{lstlisting}}
	\caption{Example SmPL patch corresponding to generic representation of the pattern associated to \toolname pattern.}
	\label{fig:smpl}
\end{figure}

\subsection{RQ3: Evaluation of Fix Patterns' Relevance for APR}
\label{sec:RQ3}
\noindent
{\bf Objective.} We propose to assess whether fix patterns yielded by \toolname are effective for automated program repair.

\begin{figure}[!ht]
    \centering
    \includegraphics[width=\linewidth]{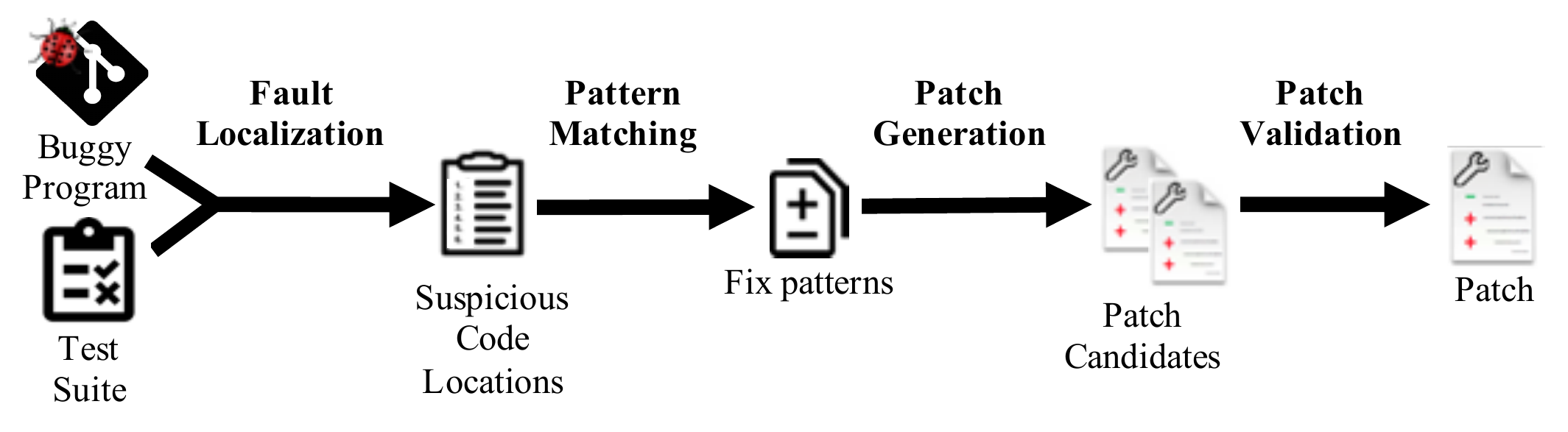}
    \vspace{-1mm}
    \caption{The overall workflow of \apr program repair pipeline.}
    \label{fig:fixMinerAPR}
\end{figure}

\noindent
{\bf Protocol.} We implement a prototype APR system that uses the fix patterns mined by \toolname to generate patches for bugs by following the principles of the PAR~\cite{kim2013automatic}, which is referred to as \apr in the remainder of this paper.
% state-of-the-art approach, since we have the experience of its implementation which is similar to the implementation of \toolname fix patterns. 
In contrast with PAR where the templates were engineered by a manual investigation of example bug fixes, in \apr, the templates for repair are engineered based on fix patterns mined by \toolname. Figure~\ref{fig:fixMinerAPR} overviews the workflow of \apr.

% automatically inferred fix patterns such as the fix pattern in Figure~\ref{fig:fpEg} of updating  method names in a return statement. 

\underline{\em Fault Localization.} \apr uses spectrum-based fault localization. We use the GZoltar\footnote{We used GZoltar version 0.1.1}~\cite{campos2012GZoltar} dynamic testing framework and leverage Ochiai~\cite{abreu2007accuracy} ranking metric to predict buggy statements based on execution coverage information of passing and failing test cases. This setting is widely used in the repair community~\cite{martinez2016astor, xiong2017Precise, xin2017leveraging, wen2018context,liu2018lsrepair}, allowing for comparable assessment of \apr against the state-of-the-art. 

\underline{\em Pattern Matching and Patch Generation.} Once the spectrum-based fault localization (or ir-based fault localization~\cite{koyuncu2019d,wen2016locus}) process yields a list of suspicious code locations,
\apr attempts to select fix patterns for each statement in the list. The selection of fix patterns is conducted by matching the context information of suspicious code locations and fix patterns mined by \toolname. Concretely, first, we parse the suspicious statement and traverse each node of its AST from its first child node to its last leaf node and form an AST subtree to represent its context. 
Then, we try to match the context (i.e., shape) of the AST subtree (from a suspicious statement) to  the fix patterns' shapes.

If a matching fix pattern is found, we proceed with the generation of a patch candidate. Some fix patterns require donor code (i.e., source code extracted from the buggy program) to generate patch candidates with fix patterns. These are also often referred to as part of fix ingredients. Recall that, to integrate with repair tools, we leverage \toolname Action patterns, which do not contain any code token information: they have “holes”. Thus we search the donor code locally from the file which contains the suspicious statement. We select relevant donor code among the ones that are applicable to the fix pattern and the suspicious statement (i.e., data type(s) of variable(s), expression types, etc. that are matching to the context) to reduce the search space of donor code and further limit the generation of nonsensical patch candidates. For example, the fix pattern in Figure~\ref{fig:fpEg} can only be matched to a suspicious return statement that has a method invocation expression: thus, the suspicious return statement will be patched by replacing its method name with another one (i.e., donor code). The donor code is searched by identifying all method names from the suspicious file having the same return type and parameters with the suspicious statement. Finally, a patch candidate is generated by mutating suspicious statements with identified donor code following the actions indicated in the matched fix pattern. We generate as many patches as the number of identified pieces of donor code. Patches are generated consecutively in the order of matching within the AST.

Note: We remind the reader that in this study, we do not perform a specific patch prioritization strategy. We search donor code from the AST tree of the local file that contains the suspicious statement by traversing each node of the AST of the local file from its first child node to its last leaf node in a breadth-first strategy (i.e., left-to-right and top-to-bottom). In case of multiple donor code options for a given fix pattern, the candidate patches are generated (each with a specific donor code) following the positions of donor codes in the AST tree (of the local file which contains the suspicious statement). 
\begin{figure}[!h]
    \centering
    \vspace{2mm}
\begin{lstlisting}[language=java,linewidth={\linewidth},frame=tb,basicstyle=\footnotesize\ttfamily]
UPD ReturnStatement
---UPD MethodInvocation
------UPD Simple@MethodName
\end{lstlisting}
    \caption{Example of fix patterns yielded by \toolname.}
    \label{fig:fpEg}
\end{figure}

% The fix patterns yielded by \toolname are presented in terms of hierarchical AST construct with code change actions (i.e., UPDate, DELete and INSert). 
% Therefore, to generate patches with these fix patterns,
% If a node matches the context of a fix pattern(Shape)(i.e., same AST node types),the fix pattern will be applied to generate patch candidates by mutating the matched code entiy following the recipe in the fix pattern.(Action)''
\underline{\em Pattern Validation.}
Once a patch candidate is generated, it is applied to buggy program and will be validated against the test suite. If it can make the buggy program pass all test cases successfully, the patch candidate will be considered as a plausible patch and \apr stops trying other patch candidates for this bug. Otherwise, the pattern matching and patch generation steps are repeated until the entire suspicious code locations list is processed. Specifically, we consider only the first generated plausible patch for each bug to evaluate its correctness. For all plausible patches generated by \apr, we further manually check the equivalence between these patches and the oracle patch provided in Defects4J. If they are semantically similar to the developer-provided fix, we consider they as correct patches, otherwise remain as plausible.

% Therefore, in practice, no particular patch ordering is performed. During the iteration, when several fix pattern contexts match the node types, only their actions differ: TBar prioritizes UPDate over INSert which is prioritized over DELete.

% Additionally, our repair pipeline needs donor code for patch generation with fix patterns that is searched from the local file which contains the bug. The donor code search space is further restricted with the context information in the buggy code (such as, data type(s) of variable(s), expression types).

\noindent
{\bf Oracle.} We use Defects4J\footnote{Version 1.2.0 - \url{https://github.com/rjust/defects4j/releases/tag/v1.2.0}}~\cite{just2014defects4j} dataset which is widely used as a benchmark for Java-targeted APR research~\cite{martinez2016astor,le2016history,chen2017contract, martinez2017automatic}. The dataset contains 357 bugs with their corresponding developer fixes and test cases covering the bugs. Table~\ref{tab:defects4j} details statistics on the benchmark.

\begin{table}[!h]
	\scriptsize
	\centering
	\setlength\tabcolsep{2pt}
	\caption{Details of the benchmark.}
	\label{tab:defects4j}
	\begin{threeparttable}
			\begin{tabular}{L{4.5cm}R{1.2cm}R{1.5cm}R{1.5cm}}
			\toprule
			{\bf Project} & {\bf Bugs} & {\bf LOC} & {\bf Tests} \\
			\hline
			JFreechart (Chart, C)   & 26   & 96K & 2,205\\
			Apache commons-lang (Lang, L)    & 65 & 22K  & 2,245\\
			Apache commons-math (Math, M)    & 106 & 85K & 3,602\\
			Joda-Time (Time, T)    & 27   & 28K  & 4,130 \\
			Closure compiler (Closure, Cl) & 133   & 90K  & 7,927\\
			\hline
			Total & 357 & 321K & 20,109\\
			\bottomrule 
		\end{tabular}
		{\footnotesize $^\dagger$ In the table, column ``Bugs'' denotes the total number of bugs in Defects4J benchmark, column ``LOC'' denotes the number of thousands of lines of code, and column ``Tests'' denotes the total number of test cases for each project.}
		\end{threeparttable}
\end{table}

\noindent
{\bf Results.} Overall, we implemented the 31 fix patterns (i.e., Action patterns) mined by \toolname, focusing only on the top-50 clusters (in terms of size). 
% Among these 31 fix patterns: 21 patterns are from UPDate changes, 8 are from INSert changes, 1 is from DELete changes, and the last one is from MOVe changes.\ak{we need to say in order way.}

We compare the performance of \apr against 13 state-of-the-art APR tools which have also used Defects4J benchmark for evaluating their repair performance. Table~\ref{tab:comparison} illustrates the comparative results in terms of numbers of {\em plausible} (i.e., that passes all the test cases) and {\em correct} (i.e., that is eventually manually validated as semantically similar to the developer-provided fix) patches. Note that although HDRepair manuscript counts 23 bugs for which "correct" fixes are generated (and among which a correct fix is ranked number one for 13 bugs), the authors labeled fixes as "verified ok" for only 6 bugs (see artefact page \footnote{https://github.com/xuanbachle/bugfixes/blob/master/fixed.txt}). We consider these 6 bugs in our comparison.

Overall, we find that \apr successfully repaired 26 bugs from the Defects4J benchmark by generating correct patches. This performance is only surpassed to date by SimFix~\cite{jiang2018shaping} that was concurrently developed with \apr. 

\begin{table}[!h]
	\scriptsize
	\centering
	\setlength\tabcolsep{2pt}
	\caption{Number of bugs fixed by different APR tools.}
	\label{tab:comparison}
	\resizebox{1.00\linewidth}{!}
    {
    	\begin{threeparttable}
			\begin{tabular}{l|>{\columncolor[gray]{0.8}}c|c|c|c|c|c|c|c|c|c|c|c|c|c}
			\toprule
			{\bf Proj.} & {\bf \apr} & {\bf kPAR} & {\bf jGenProg} & {\bf jKali} & {\bf jMutRepair} & {\bf Nopol} & {\bf HDRepair} & {\bf ACS} & {\bf ssFix} & {\bf ELIXIR} & {\bf JAID} & {\bf SketchFix} & {\bf CapGen} & {\bf SimFix} \\
			\hline
			Chart   & {\bf 5/8} & 3/10 & 0/7  & 0/6 & 1/4  & 1/6   & 0/2 & 2/2   & 3/7   & 4/7   & 2/4 & 6/8 & 4/4   & 4/8\\
			Lang    & {\bf 2/3} & 1/8 & 0/0  & 0/  & 0/1  & 3/7    & 2/6 & 3/4   & 5/12  & 8/12  & 1/8 & 3/4 & 5/5   & 9/13\\
			Math    & {\bf 13/15} & 7/18 & 5/18 & 1/14& 2/11 & 1/21 & 4/7 & 12/16 & 10/26 & 12/19 & 1/8 & 7/8 & 12/16 & 14/26\\
			Time    & {\bf 1/1} & 1/2 & 0/2  & 0/2 & 0/1  & 0/1   & 0/1 & 1/1   & 0/4   & 2/3   & 0/0 & 0/1 & 0/0   & 1/1\\
%			Mockito & 0/0   & 0/0  & 0/0 & 0/0 & & 0/0 & 0/0   & 0/0   & 0/0   & 0/0 & 0/0 & 0/0   & 0/0\\
			Closure & {\bf 5/5} & 5/9  & 0/0  & 0/0 & 0/0 & 0/0   & 0/7 & 0/0  & 2/11  & 0/0   & 5/11& 3/5 & 0/0   & 6/8\\
			\hline
			\hline
			Total & {\bf 26/32} & 17/47& 5/27 & 1/22 & 3/17 & 5/35 & 6/23 & 18/23 & 20/60 & {\bf 26/41} & 9/31 & 19/26 & 21/25 & {\bf 34/56}\\
			\hline
			P(\%) & {\bf 81.3} & 36.2 & 18.5 & 4.5  & 17.7 & 14.3 & 26.1  & 78.3  & 33.3  & 63.4  & 29.0 & 73.1  & {\bf 84.0}  & 60.7 \\
			\bottomrule 
		\end{tabular}
		{\footnotesize $^\dagger$ In each column, we provide $x/y$ numbers: $x$ is the number of correctly fixed bugs; $y$ is the number of bugs for which a plausible patch is generated by the APR tool (i.e., a patch that makes the program pass all test cases). Precision (P) means the precision of correctly fixed bugs in bugs fixed by each APR tool. kPAR~\cite{liu2019you} is the Java implementation of PAR. The data about jGenProg, jKali and Nopol are extracted from the experimental results reported by Martinez et al.~\cite{martinez2017automatic}. The data of HDRepair~\cite{le2016history} is collected from its author's reply. And the results of other tools are obtained from their papers in the literature (jMutRepair~\cite{martinez2016astor}, ACS~\cite{xiong2017Precise}, ssFix~\cite{xin2017leveraging}, ELIXIR~\cite{saha2017elixir}, JAID~\cite{chen2017contract}, SketchFix(SF)~\cite{hua2018towards}, CapGen~\cite{wen2018context} and SimFix~\cite{jiang2018shaping}). The same for the data presented in Table~\ref{tab:results.apr}.}
		\end{threeparttable}
	}
\end{table}

Nevertheless, while these tools generate more correct patches than \apr, they also generate many more plausible patches which are however not correct. In order to comparatively assess the different tools, we resort to a Precision metric (P), which is the probability of correctness of the generated patches. P(\%) is defined as the ratio of the number of bugs for which a correct fix is generated first (i.e., before any other plausible patch) against the number of bugs for which a plausible (but incorrect) patch is generated first. For example, 81\% of \apr's plausible patches are actually correct, while it is the case for 63\% and 60\% of respectively ELIXIR and SimFix plausible patches are correct. To date only CapGen~\cite{wen2018context} achieves similar performance at yielding patches with slighter higher probability (at 84\%) to be correct. The high performance of CapGen confirms their intuition that context-awareness, which we provide with \ediff, is essential for improving patch correctness.

Table~\ref{tab:results.apr} enumerates 128 bugs that are currently fixed (both correct and plausible) in the literature.  89 of them can be correctly fixed by at least one APR tool. \apr generates correct patches for 26 bugs. Among the bugs in the used version of Defects4J benchmark, 267 bugs have not yet been fixed by any tools in the literature, which still is a big challenge for automated program repair research. 

\begin{figure}[!h]
\def\firstcircle{[name path=firstcircle] (0,0) circle (1cm)}
\def\thirdcircle{[name path=thirdcircle] (0:1cm) circle (1cm)}

\centering
\resizebox{0.4\textwidth}{!}
    {

% Now we can draw the sets:
\begin{tikzpicture}
    \draw \firstcircle node[below,name=A] at (-0.5,0.3) {$63$};
    \draw \thirdcircle node [below,name=C]at (1.5,0.3) {$6$};
    \path [ name intersections = {of = firstcircle and thirdcircle } ] (intersection-1) -- (intersection-2) node [pos=0.5] {$20$};
    \node at (-1.5,1.2) {\textbf{Existing\ APR\ Tools}};
    \node at (2,1.2) {\apr};
\end{tikzpicture}
}
\caption{Overlap of the correct patches by \apr and other APR tools.}\label{fig:comparison}
\end{figure}
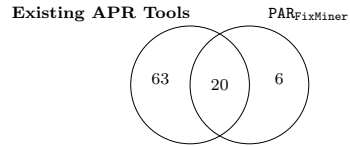

Finally, we find that, thanks to its automatically mined patterns, \apr is able to fix six (6) bugs which have not been fixed by any state-of-the-art APR tools (cf. Figure~\ref{fig:comparison}).

%\begin{figure}[!h]
%	\includegraphics[width=\linewidth]{figs/overlaps.pdf}
%	\caption{The overlaps between FixMiner and existing ARP tools. \textcolor{red}{ TODO: update to the final results.}}
%	\label{fig:comparison}
%\end{figure}

\begin{table}

%\scriptsize
%        \centering
        \setlength\tabcolsep{2pt}
\caption{Defects4J bugs fixed by different APR tools.}{\footnotesize  ``\cmark'' indicates that the bug is correctly fixed, ``\xmark'' indicates the produced patch is plausible but not correct. ``(\cmark)'' indicates that a correct patch is generated by JAID, but is not the first plausible patch to be generated)''. }

%\resizebox{1.0\linewidth}{!}{
%\resizebox{\textwidth}{!}{%
\begin{tabular}{cc}
\begin{minipage}{.5\linewidth}	

\resizebox{1\linewidth}{!}{
\begin{tabular}{|l|>{\columncolor[gray]{0.8}}c|c|c|c|c|c|c|c|c|c|c|c|c|c|c|c|c|c|c|c|c|c|c|c|c|c|c|c|c|c }
\hline
%\rotatebox[origin=l]{45}{Project}   &       \rotatebox[origin=l]{45}{SimFix}  &       \rotatebox[origin=l]{45}{CapGen}  &       \rotatebox[origin=l]{45}{SketchFix}       &       \rotatebox[origin=l]{45}{JAID}    &       \rotatebox[origin=l]{45}{ssFix}   &       \rotatebox[origin=l]{45}{ACS}     &       \rotatebox[origin=l]{45}{ELIXIR}  &       \rotatebox[origin=l]{45}{HDRepair}        &       \rotatebox[origin=l]{45}{xPAR}    &       \rotatebox[origin=l]{45}{jGenProg}        &       \rotatebox[origin=l]{45}{jKali}   &       \rotatebox[origin=l]{45}{jMutRepair}      &       \rotatebox[origin=l]{45}{Nopol}   &       \rotatebox[origin=l]{45}{MiningFixer} 
\rotatebox[origin=l]{0}{Proj.}	&\rotatebox[origin=l]{90}{FixMiner}&\rotatebox[origin=l]{90}{kPAR}	&\rotatebox[origin=l]{90}{SimFix}	&\rotatebox[origin=l]{90}{CapGen}	&\rotatebox[origin=l]{90}{SketchFix}	&\rotatebox[origin=l]{90}{JAID}	&\rotatebox[origin=l]{90}{ssFix}	&\rotatebox[origin=l]{90}{ACS}	&\rotatebox[origin=l]{90}{ELIXIR}	&\rotatebox[origin=l]{90}{HDRepair}	&\rotatebox[origin=l]{90}{jGenProg}	&\rotatebox[origin=l]{90}{jKali}	&\rotatebox[origin=l]{90}{jMutRepair}	&\rotatebox[origin=l]{90}{Nopol}\\
       
\hline
\hline
C-1	&\cmark	&\cmark	&\cmark	&\cmark	&\cmark	&(\cmark)&\cmark	&		&\cmark	&\xmark	&\xmark	&\xmark	&\cmark	&	\\
C-3	&		&\xmark	&\cmark	&		&		&		&		&		&\xmark	&		&\xmark	&		&		&\xmark\\
C-4	&\cmark	&\cmark	&		&		&		&		&		&		&		&		&		&		&		&	\\
C-5	&		&\xmark	&		&		&		&		&		&		&		&		&\xmark	&\xmark	&		&\cmark\\
C-7	&		&\cmark	&\cmark	&		&		&		&		&		&		&		&\xmark	&		&\xmark	&	\\
C-8	&		&       &		&\cmark	&\cmark	&		&		&		&\cmark	&\xmark	&		&		&		&	\\
C-9	&		&       &		&		&\cmark	&(\cmark)&		&		&\cmark	&		&		&		&		&	\\
C-11&\cmark	&		&		&\cmark	&\cmark	&		&		&		&\cmark	&		&		&		&		&	\\
C-12&\xmark	&\xmark	&\xmark	&		&		&		&		&		&		&		&		&		&		&	\\
C-13&\xmark	&\xmark	&		&		&\xmark	&		&		&		&\xmark	&		&\xmark	&\xmark	&		&\xmark\\
C-14&		&		&\xmark	&		&		&		&		&\cmark	&		&		&		&		&		&	\\
C-15&		&\xmark	&		&		&		&		&		&		&		&		&\xmark	&\xmark	&		&	\\
C-17&		&		&		&		&		&		&		&		&\xmark	&		&		&		&		&	\\
C-18&		&		&\xmark	&		&		&		&		&		&		&		&		&		&		&	\\
C-19&		&		&		&		&		&		&		&\cmark	&		&		&		&		&		&	\\
C-20&		&		&\cmark	&		&\cmark	&		&\cmark	&		&		&		&		&		&		&	\\
C-21&		&		&		&		&		&		&		&		&		&		&		&		&		&\xmark\\
C-22&		&		&\xmark	&		&		&		&		&		&		&		&		&		&		&	\\
C-24&\cmark	&		&		&\cmark	&\cmark	&\cmark	&\cmark	&		&		&		&		&		&		&	\\
C-25&\xmark	&\xmark	&		&		&		&		&		&		&		&		&\xmark	&\xmark	&\xmark	&\xmark\\
C-26&\cmark	&\xmark	&		&		&\xmark	&\cmark	&		&		&		&		&		&\xmark	&\xmark	&\xmark\\
\hline
Cl-2    &       &\cmark &		&		&		&       &		&		&		&		&		&		&		&	\\
Cl-5    &		&		&		&		&		&\xmark	&		&		&		&		&		&		&		&	\\
Cl-10   &\cmark	&		&		&		&		&		&		&		&		&\xmark	&		&		&		&	\\
Cl-14	&		&		&\cmark	&		&\cmark	&		&\cmark	&		&		&\xmark	&		&		&		&	\\
Cl-18	&		&		&		&		&		&\cmark	&		&		&		&		&		&		&		&	\\
Cl-21   &       &\xmark &		&		&		&       &		&		&		&		&		&		&		&	\\
Cl-22   &       &\xmark &		&		&		&       &		&		&		&		&		&		&		&	\\
Cl-31	&		&       &		&		&		&(\cmark)&		&		&		&		&		&		&		&	\\
Cl-33	&		&       &		&		&		&\cmark	&		&		&		&		&		&		&		&	\\
Cl-38	&\cmark	&\cmark	&		&		&		&		&		&		&		&		&		&		&		&	\\
Cl-40	&		&		&		&		&		&\cmark	&		&		&		&		&		&		&		&	\\
Cl-51	&		&		&		&		&		&		&		&		&		&\xmark	&		&		&		&	\\
Cl-57	&		&		&\cmark	&		&		&		&		&		&		&		&		&		&		&	\\
Cl-62	&\cmark	&\cmark &\cmark	&		&\cmark	&(\cmark)&		&		&		&\xmark	&		&		&		&	\\
Cl-63	&\cmark	&\cmark &\cmark	&		&		&(\cmark)&		&		&		&		&		&		&		&	\\
Cl-70	&		&		&		&		&\xmark	&\cmark	&		&		&		&\xmark	&		&		&		&	\\
Cl-73	&\cmark	&\cmark &\cmark	&		&\xmark	&\cmark	&		&		&		&\xmark	&		&		&		&	\\
Cl-79	&		&		&\xmark	&		&		&		&		&		&		&		&		&		&		&	\\
Cl-106	&		&		&\xmark	&		&		&		&		&		&		&		&		&		&		&	\\
Cl-109  &       &\xmark &		&		&		&       &		&		&		&		&		&		&		&	\\
Cl-115	&		&		&\cmark	&		&		&		&\cmark	&		&		&		&		&		&		&	\\
Cl-125	&		&		&		&		&		&\xmark	&		&		&		&		&		&		&		&	\\
Cl-126	&		&\xmark &		&		&\cmark	&(\cmark)&		&		&		&\xmark	&		&		&		&	\\
\hline
L-6	&		&		&		&\cmark	&\cmark	&		&\cmark	&		&\cmark	&\cmark	&		&		&		&	\\
L-7	&		&		&		&		&		&		&		&\cmark	&		&		&		&		&		&	\\
L-10	&		&		&\xmark	&		&		&		&		&		&		&\xmark	&		&		&		&	\\
L-16	&		&		&\cmark	&		&		&		&		&		&		&		&		&		&		&	\\
L-21	&		&		&		&		&		&		&\cmark	&		&		&		&		&		&		&	\\
L-24	&		&		&		&		&		&\xmark	&		&\cmark	&\cmark	&		&		&		&		&	\\
L-26	&		&		&		&\cmark	&		&		&		&		&\cmark	&		&		&		&		&	\\
L-27	&		&		&\cmark	&		&		&		&		&		&		&		&		&		&\xmark	&	\\
L-33	&		&		&\cmark	&		&		&\cmark	&\cmark	&		&\cmark	&		&		&		&		&	\\
L-35	&		&		&		&		&		&		&		&\cmark	&		&		&		&		&		&	\\
L-38	&		&		&		&		&		&(\cmark)&		&		&\cmark	&		&		&		&		&	\\
L-39	&		&		&\cmark	&		&		&\xmark	&		&\xmark	&\xmark	&		&		&		&		&\xmark\\
L-41	&		&		&\cmark	&		&		&		&		&		&		&		&		&		&		&	\\
L-43	&\xmark	&\xmark	&\cmark	&\cmark	&		&		&\cmark	&		&\cmark	&\xmark	&		&		&		&	\\
L-44	&		&\xmark	&\xmark	&		&		&		&		&		&\xmark	&		&		&		&		&\cmark\\
L-45	&		&\xmark	&\xmark	&		&		&(\cmark)&		&		&		&		&		&		&		&	\\
L-46	&		&		&		&		&		&		&		&		&		&		&		&		&		&\xmark\\
L-50	&		&		&\cmark	&		&		&		&		&		&		&		&		&		&		&	\\
L-51	&		&\xmark	&		&		&\xmark	&(\cmark)&		&		&\xmark	&\cmark	&		&		&		&\xmark\\
L-53	&		&\xmark	&		&		&		&		&		&		&		&		&		&		&		&\xmark\\
L-55	&		&		&		&		&\cmark	&(\cmark)&		&		&		&		&		&		&		&\cmark\\
L-57	&\cmark	&\xmark	&		&\cmark	&		&		&		&		&\cmark	&\xmark	&		&		&		&	\\

\hline
\end{tabular}
}

\end{minipage} &
\begin{minipage}{.5\linewidth}
\vspace{-1mm}
\resizebox{1\linewidth}{!}{
\begin{tabular}{|l|>{\columncolor[gray]{0.8}}c|c|c|c|c|c|c|c|c|c|c|c|c|c|c|c|c|c|c|c|c|c|c|c|c|c|c|c|c|c }
\hline
%\rotatebo\xmark[origin=l]{45}{Project}   &       \rotatebo\xmark[origin=l]{45}{SimFi\xmark}  &       \rotatebo\xmark[origin=l]{45}{CapGen}  &       \rotatebo\xmark[origin=l]{45}{SketchFi\xmark}       &       \rotatebo\xmark[origin=l]{45}{JAID}    &       \rotatebo\xmark[origin=l]{45}{ssFi\xmark}   &       \rotatebo\xmark[origin=l]{45}{ACS}     &       \rotatebo\xmark[origin=l]{45}{ELI\xmarkIR}  &       \rotatebo\xmark[origin=l]{45}{HDRepair}        &       \rotatebo\xmark[origin=l]{45}{\xmarkPAR}    &       \rotatebo\xmark[origin=l]{45}{jGenProg}        &       \rotatebo\xmark[origin=l]{45}{jKali}   &       \rotatebo\xmark[origin=l]{45}{jMutRepair}      &       \rotatebo\xmark[origin=l]{45}{Nopol}   &       \rotatebo\xmark[origin=l]{45}{MiningFi\xmarker} 
\rotatebox[origin=l]{0}{Proj.}	&\rotatebox[origin=l]{90}{FixMiner}&\rotatebox[origin=l]{90}{kPAR}	&\rotatebox[origin=l]{90}{SimFix}	&\rotatebox[origin=l]{90}{CapGen}	&\rotatebox[origin=l]{90}{SketchFix}	&\rotatebox[origin=l]{90}{JAID}	&\rotatebox[origin=l]{90}{ssFix}	&\rotatebox[origin=l]{90}{ACS}	&\rotatebox[origin=l]{90}{ELIXIR}	&\rotatebox[origin=l]{90}{HDRepair}	&\rotatebox[origin=l]{90}{jGenProg}	&\rotatebox[origin=l]{90}{jKali}	&\rotatebox[origin=l]{90}{jMutRepair}	&\rotatebox[origin=l]{90}{Nopol}\\
       
%\hline
\hline
\hline

L-58	&		&\xmark	&\cmark	&		&		&		&		&		&\xmark	&		&		&		&		&\cmark\\
L-59	&\cmark	&\cmark	&		&\cmark	&\cmark	&		&\cmark	&		&\cmark	&\xmark	&		&		&		&	\\
L-60	&		&		&\cmark	&		&		&		&		&		&		&		&		&		&		&	\\
L-61	&		&		&		&		&		&\xmark	&		&		&		&		&		&		&		&	\\
L-63	&		&		&\xmark	&		&		&		&		&		&		&		&		&		&		&	\\
\hline
M-1		&		&\xmark	&\xmark	&		&		&		&		&		&		&		&		&		&		&	\\
M-2		&		&		&		&		&		&		&		&		&\xmark	&		&\xmark	&\xmark	&\xmark	&	\\
M-3		&		&		&		&		&		&		&		&\cmark	&		&		&		&		&		&	\\
M-4		&		&		&		&		&		&		&		&\cmark	&		&		&		&		&		&	\\
M-5		&		&		&\cmark	&\cmark	&\cmark	&\cmark	&		&\cmark	&\cmark	&\cmark	&\cmark	&		&		&	\\
M-6		&		&		&\xmark	&		&		&		&		&		&		&		&		&		&		&	\\
M-8	    &		&\xmark	&\xmark	&		&		&		&		&		&		&		&\xmark	&\xmark	&		&	\\
M-10	&\cmark	&		&		&		&		&		&		&		&		&		&		&		&		&	\\
M-15	&		&\cmark	&		&		&		&		&		&		&		&		&		&		&		&	\\
M-20	&		&		&\xmark	&		&		&		&		&		&\xmark	&		&		&		&		&	\\
M-22	&\cmark	&		&		&		&		&		&		&		&		&\cmark	&		&		&		&	\\
M-25	&		&		&		&		&		&		&		&\cmark	&		&		&		&		&		&	\\
M-28	&		&\xmark &\xmark	&		&		&		&		&\xmark	&		&		&\xmark	&\xmark	&\xmark	&	\\
M-30	&\cmark	&		&		&\cmark	&		&		&\cmark	&		&\cmark	&		&		&		&		&	\\
M-32	&		&		&		&		&		&(\cmark)&		&		&\xmark	&		&		&\xmark	&		&\xmark\\
M-33	&\cmark	&\cmark	&\cmark	&\cmark	&\cmark	&		&\cmark	&		&\cmark	&		&		&		&		&\xmark\\
M-34	&\cmark	&		&		&		&		&		&		&		&\cmark	&\xmark	&		&		&		&	\\
M-35	&\cmark &		&\cmark	&		&		&		&		&\cmark	&		&		&		&		&		&	\\
M-40	&		&		&		&		&		&		&		&		&		&		&\xmark	&\xmark	&\xmark	&\xmark\\
M-41	&		&		&\cmark	&		&		&		&\cmark	&		&		&		&		&		&		&	\\
M-42	&		&		&		&		&		&		&		&		&		&		&		&		&		&\xmark\\
M-49	&		&\xmark &		&		&		&		&		&		&		&		&\xmark	&\xmark	&		&\xmark\\
M-50	&		&\xmark &\cmark	&		&\cmark	&(\cmark)&\cmark	&		&\cmark	&\cmark	&\cmark	&\cmark	&\xmark	&\cmark\\
M-51	&		&\xmark &    	&		&		&		&		&		&		&		&		&		&		&	\\
M-53	&		&		&\cmark	&\cmark	&		&(\cmark)&\cmark	&		&		&\cmark	&\cmark	&		&		&	\\
M-57	&\cmark	&		&\cmark	&\cmark	&		&		&\cmark	&		&\cmark	&		&		&		&\xmark	&\xmark\\
M-58	&\cmark	&\cmark &		&\cmark	&		&		&		&		&\cmark	&		&		&		&\xmark	&\xmark\\
M-59	&		&		&\cmark	&\cmark	&\cmark	&		&\cmark	&		&\cmark	&		&		&		&		&	\\
M-61	&		&		&		&		&		&		&		&\cmark	&		&		&		&		&		&	\\
M-62	&		&\xmark &    	&		&		&		&		&		&		&		&		&		&		&	\\
M-63	&		&\xmark &\cmark	&\cmark	&		&		&		&		&\xmark	&		&		&		&		&	\\
M-65	&		&		&		&\cmark	&		&		&		&		&		&		&		&		&		&	\\
M-69	&		&		&		&		&		&		&		&		&		&		&		&		&		&\xmark\\
M-70	&\cmark	&\cmark	&\cmark	&\cmark	&\cmark	&		&\cmark	&		&\cmark	&\xmark	&\cmark	&		&		&	\\
M-71	&		&		&\cmark	&		&		&		&		&		&		&		&\xmark	&		&		&\xmark\\
M-72	&		&		&\xmark	&		&		&		&		&		&		&		&		&		&		&	\\
M-73	&		&		&\xmark	&		&\xmark	&		&		&\xmark	&\xmark	&		&\cmark	&		&		&\xmark\\
M-75	&\cmark	&\cmark	&\cmark	&\cmark	&		&		&		&		&\cmark	&		&		&		&		&	\\
M-78	&		&		&		&		&		&		&		&		&		&		&\xmark	&\xmark	&		&\xmark\\
M-79	&\cmark	&		&\cmark	&\xmark	&		&		&\cmark	&		&		&		&		&		&		&	\\
M-80	&		&		&\xmark	&\xmark	&		&(\cmark)&\cmark	&		&\xmark	&		&\xmark	&\xmark	&		&\xmark\\
M-81	&\xmark	&\xmark	&\xmark	&\xmark	&		&		&		&\xmark	&		&		&\xmark	&\xmark	&\xmark	&\xmark\\
M-82	&\cmark	&\xmark	&\xmark	&\xmark	&\cmark	&(\cmark)&		&\cmark	&\cmark	&\xmark	&\xmark	&\xmark	&\cmark	&\xmark\\
M-84	&\xmark	&\xmark	&		&		&		&		&		&		&		&		&\xmark	&\xmark	&\xmark	&	\\
M-85	&\cmark	&\cmark	&\xmark	&\cmark	&\cmark	&(\cmark)&		&\cmark	&\cmark	&		&\xmark	&\xmark	&\cmark	&\xmark\\
M-87	&		&		&		&		&		&		&		&		&		&		&		&		&		&\xmark\\
M-88	&		&		&\xmark	&		&		&		&		&		&		&		&		&		&\xmark	&\xmark\\
M-89	&		&\cmark &		&		&		&		&		&\cmark	&		&		&		&		&		&	\\
M-90	&		&		&		&		&		&		&		&\cmark	&		&		&		&		&		&	\\
M-93	&		&		&		&		&		&		&		&\cmark	&		&		&		&		&		&	\\
M-95	&		&		&		&		&		&		&		&		&		&		&\xmark	&\xmark	&		&	\\
M-97	&		&		&		&		&		&		&		&\xmark	&		&		&		&		&		&\xmark\\
M-98	&		&		&\cmark	&		&		&		&		&		&		&		&		&		&		&	\\
M-99	&		&		&		&		&		&		&		&\cmark	&		&		&		&		&		&	\\
M-104	&		&		&		&		&		&		&		&		&\xmark	&		&		&		&		&\xmark\\
M-105	&		&		&		&		&		&\xmark	&		&		&		&		&		&		&		&\xmark\\
\hline
T-4		&		&		&		&		&\xmark	&		&		&		&\cmark	&		&\xmark	&\xmark	&		&	\\
T-7		&		&\cmark	&\cmark	&		&		&		&		&		&		&		&		&		&		&	\\
T-11	&		&\xmark	&		&		&		&		&		&		&\xmark	&		&\xmark	&\xmark	&\xmark	&\xmark\\
T-15	&		&		&		&		&		&		&		&\cmark	&\cmark	&		&		&		&		&	\\
T-19	&\cmark	&		&		&		&		&		&		&		&		&\xmark	&		&		&		&	\\
\hline
\end{tabular}
}

 \end{minipage} 

\end{tabular}
%}
\label{tab:results.apr}

\end{table}

\find{{\bf RQ3}$\blacktriangleright$ Fix patterns (i.e., Action Patterns) yielded by \toolname can be directly used in automated program repair pipelines and generates correct patches for buggy programs effectively. Additionally, the repair performance of \apr, which uses fix patterns yielded by \toolname, is comparable to the state-of-the-art APR tools.}

%\ak{comment the following block or move the step where we discuss patterns}
%In our automated repair pipeline, the Shape patterns and Token patterns not taken into consideration. 
%Shape patterns only contain the structure of AST nodes without corresponding code change actions, which makes it impossible to directly use them in automated program repair. If we randomly assign code change actions to each AST node, it will increase the search space of patches explosively, which will cause the explosive search space problem in automated program repair~\cite{liu2018lsrepair, wen2018context}.
%Token patterns contain the code raw tokens, which makes them specific. Thus, they only can be used to address exact the same issues but not the common similar ones. In this work, we observe that only two bugs (i.e., Closure-92 and Closure-93) can be fixed by a Token pattern.

% To sum up, considering the explosive search space problem and specific problem, only the second-level fix patterns are implemented into our automated program repair pipeline.
% \kui{Any more discussion?}

\section{Discussions and Threats to Validity}
\label{sec:discussion}
\paragraph{Runtime performance.} To run the experiments with \toolname, we leveraged a computing system with 24 Intel Xeon E5-2680 v3 cores with 2.GHz per core and 3TB RAM.
The construction of the \ediff{s} took about 17 minutes.
\ediff{s} are cached in memory to reduce disk access during the computation of identical trees. Nevertheless, we recorded that comparing 1\,108\;060 pairs of trees took about 18 minutes.

% \paragraph{Limitations.} The main limitation of \toolname currently stems from our support for Enhanced AST Diffs that are performing a single repair action (either {\tt UPD}, {\tt INS}, {\tt DEL}, {\tt MOV} in all nodes rather than the mix of them). This decision was taken to focus on a specific and pure set of changes so as to ensure a reliable manual assessment. In future work, \toolname will investigate more complex changes combining several repair action types.

\label{sec:threats}
\paragraph{Threats to external validity.}
The selection of our bug-fix datasets carries some threats to external validity that we have limited by considering known projects, and heuristics used in previous studies. We also make our best effort to link commits with bug reports as tagged by developers. Some false positives may be included if one considers a strict and formal definition of what constitutes a bug.

%\ak{more threats}
\paragraph{Threats to construct validity} arise when checking the compatibility of \toolname's patterns against the patterns used by literature APR systems. Indeed, for the comparison, we do not conduct exact mapping where the elements should be the same, given that literature patterns can be more abstract than the ones yielded by \toolname. For example, {\em Modify Method Name} (i.e., FP10.1) is a sub-fix pattern of {\em Mutate Method Invocation Expression} (i.e., FP10), which is about replacing the method name of a method invocation expression with another appropriate method name~\cite{liu2019tbar}. This fix pattern can be matched to any statement that contains a method name under method invocation expression. However, in this paper, the similar fix patterns yielded by \toolname have more context information. Therefore, we consider context information to check the compatibility of \toolname's patterns against the patterns used by literature APR systems. For example, the fix pattern shown in Figure~\ref{fig:fpEg} is to modify the buggy method name of a method invocation expression with another appropriate method name which is inside a {\tt Return-Statement}. As the context information refers to a {\tt Return-Statement} the fix pattern shown in Figure~\ref{fig:fpEg} considered as compatible with 
 {\em Mutate Return Statement} (i.e., FP12.). Nevertheless, the mapping is conservative in the sense that we consider that a \toolname pattern matches a pattern from the literature as long as it can fit with the literature pattern.

%Eventually, this fine-grained fix pattern further constraints to the buggy {\tt Return-Statement}s resulting a reduction on the possibility of exploding the search space problem of matching fix patterns.\ak{i dont understand this sentence}
%
%In practice, the extra information has revealed to be useful in the automated program pipeline, since it can contribute to yield higher precision in generating correct patches than most state-of-the-art APR tools (cf. Section~\ref{sec:RQ3}).

\section{Related Work}
\label{sec:relatedwork}

\noindent
\paragraph{Automated Program Repair.}
Patch generation is one of the key tasks in software maintenance since it is time-consuming and tedious. If this task is automated, the cost and time of developers for maintenance will be dramatically reduced. To address the issue, many automated techniques have been proposed for program repair~\cite{monperrus2018automatic}. GenProg~\cite{claire2012GenProg}, which leverages genetic programming, is a pioneering work on program repair. It relies on mutation operators that insert, replace, or delete code elements. Although these mutations can create a limited number of variants, GenProg could fix several bugs (in their evaluation, test cases were passed for 55 out of 105 real program bugs) automatically, although most of them have been found to be incorrect patches later. PACHIKA~\cite{dallmeier2009generating} leverages object behavior models. SYDIT~\cite{meng2011systematic} and LASE~\cite{meng2013lase} automatically extracts an edit script from a program change. While several techniques have focused on fixability, Kim et al.~\cite{kim2013automatic} pointed out that patch acceptability should be considered as well in program repair. Automatically generated patches often have nonsensical structures and logic even though those patches can fix program bugs with respect to program behavior (i.e., w.r.t. test cases). To address this issue, they proposed PAR, which leverages manually-crafted fix patterns. Similarly Long and Rinard proposed Prophet~\cite{long2016automatic} and Genesis~\cite{long2017automatic} which generates patches by leveraging fix patterns extracted from the history of changes in repositories. Recently, several approaches~\cite{bhatia2016automated,gupta2017deepfix} leveraging deep learning have been proposed for learning to fix bugs. Even recent APR approaches that target bug reports rely on fix templates to generate patches. iFixR~\cite{koyuncu2019ifixr} is such an example which builds on top of the templates built TBar~\cite{liu2019tbar} templates. Overall, we note that the  community is going in the direction of implementing repair strategies based on fix patterns or templates. Our work is thus essential in this direction as it provides a scalable, accurate and actionable tool to mine such relevant patterns for automated program repair.

\noindent
\paragraph{Code differencing.}
Code differencing is an important research and practice concern in software engineering. Although commonly used by human developers in manual tasks, differencing at the text line level granularity~\cite{myers1986ano} is generally unsuitable for automated analysis of changes and the associated semantics. AST differencing work has benefited in the last decade for the extensive investigations that the research community has performed for general tree differencing~\cite{bille2005survey,chawathe1996change,chilowicz2009syntax, al2005diffx}. ChangeDistiller~\cite{fluri2007change} and GumTree~\cite{falleri2014Fine} constitute the current state-of-the-art for AST differencing in Java. In this work, we have selected GumTree as the base tool for the computation of edit scripts as its results have been validated by humans, and it has been shown to be more accurate and fine-grained edit scripts. Nevertheless, we have further enhanced the edit script yielding an algorithm that keeps track of contextual information. Our approach echoes a recently published work by Huang et al.~\cite{huang2018cldiff}: their CLDIFF tool similarly enriches the AST produced by GumTree to enable the generation of concise code differences. The tool however was not available at the time of our experiments. Thus, to satisfy the input requirements of our fix pattern mining approach, we implement \ediff, to enrich GumTree-yielded edit scripts by retaining more contextual information. 
% We plan to  that when their tool be

\paragraph{Change patterns.} The literature includes a large body of work on mining change patterns.

\noindent
\paragraph{Mining-based approaches.}
In recent years, several approaches have built upon the idea of mining patterns or leveraging templates. Fluri et al., based on edit scripts computed by their ChangeDistiller AST difference, have used hierarchical clustering to discover unknown change types in three Java applications~\cite{fluri2008discovering}. They have limited themselves however to considering only changes implementing the 41 basic change types that they had previously identified~\cite{fluri2006classifying}. Kreutzer et al. have developed C3 to automatically detect groups of similar code changes in code repositories with the help of clustering algorithms~\cite{Kreutzer2016}. Martinez and Monperrus~\cite{martinez2015mining} assessed the relationship between the types of bug fixes and automatic program repair. They perform extensive large scale empirical investigations on the nature of human bug fixes based on fine-grained abstract syntax tree differences by ChangeDistiller. Their experiments show that the mined models are more effective for driving the search compared to random search.  Their models however remain at a high level and may not carry any actionable patterns to be used by other template-based APR. Our work however also targets systematizing and automating the “mining of actionable fix patterns” to feed pattern-based program repair tools.

An example application is related to work by Livshits and Zimmermann~\cite{livshits2005dynamine} who discovered application-specific repair templates by using association rule mining on two Java projects. More recently, Hanam et al.~\cite{hanam2016discovering} have developed the BugAID technique for discovering most prevalent repair templates in JavaScript. They use AST differencing and unsupervised learning algorithms.  Our objective is similar to theirs, focusing on Java programs with different abstraction levels of the patterns. FixMiner builds on a three-fold clustering strategy where we iteratively discover recurrent changes preserving surrounding code context.

\noindent
\paragraph{Studies on code change redundancies.}
A number of empirical studies have confirmed that code changes are repeatedly performed in software code bases~\cite{kim2009discovering, kim2006memories, molderez2017mining,yue2017characterization}. Same changes are prevalent because multiple occurrences of the same bug require the same change. Similarly, when an API evolves, or when migrating to a new library/framework, all calling code must be adapted by same collateral changes~\cite{padioleau2008documenting}. Finally, code refactoring or routine code cleaning can lead to similar changes. In a manual investigation, Pan et al.~\cite{pan2009toward} have identified 27 extractable repair templates for Java software. Among other findings, they observed that if-condition changes are the most frequently applied to fix bugs. Their study, however, does not discuss whether most bugs are related to If-condition or not. This is important as it clarifies the context to perform if-related changes. Recently, Nguyen et al.~\cite{nguyen2010recurring} have empirically found that 17-45\% of bug fixes are recurring.  Our focus in this paper is to provide tool-support automated approach to inferring change patterns in a dataset to drive repair patterns to guide APR mutation. Moreover, our patterns are less generic than the ones in previous works (e.g., as in \cite{pan2009toward, nguyen2010recurring}).

 Concurrently to our work, Jiang et al. have proposed SimFix~\cite{jiang2018shaping}, and Wen et al. CapGen~\cite{wen2018context}  which implements a similar idea of leveraging code redundancies using contextual information for shaping the program repair space. In \toolname however, the pattern mining phase is independent from the patch generation phase, and the resulting patterns are tractable and reusable as input to other APR systems.
 
\paragraph{Generic and semantic patch inference.}
Ideally, \toolname is a tool that aims at performing towards finding a generic patch that can be leveraged by automated program repair to correctly update a collection of buggy code fragments. This problem has been recently studied by approaches such as {\tt spdiff}~\cite{andersen2010generic,andersen2012semantic} which work on the inference of generic and semantic patches. This approach, however, is known to be poorly scalable and has constraints of producing ready-to-use semantic patches that can be used by the Coccinelle matching and transformation engine~\cite{brunel2009foundation}. There have however a number of prior work that tries to detect and summarize program changes. A seminal work by Chawathe et al. describes a method to detect changes to structured information
based on an ordered tree and its updated version~\cite{chawathe1996change}. The goal was to derive a compact description of the changes with the notion of minimum
cost edit script which has been used in the recent ChangeDistiller and GumTree tools. The representations of edit operations, however, are either
often too overfit to a particular code change or abstract very loosely the change so that it cannot be easily instantiated.
Neamtiu et al.~\cite{neamtiu2005understanding} proposed an approach for identifying changes, additions and deletions of C program elements based on structural matching of syntax trees. Two trees that are structurally identical
but have differences in their nodes are considered to represent matching program
fragments. Kim et al.~\cite{kim2007automatic} have later proposed a method to infer ``change-rules'' that capture
many changes. They generally express changes related to program headers (method
headers, class names, package names, etc.). 
Weissgerber et al.~\cite{weissgerber2006identifying} have also proposed a technique to identify likely refactorings in the changes
that have been performed in Java programs. Overall, these generic patch inference approaches address the challenges of how the patterns that will be leveraged in practice. Our work goes in that direction by yielding different kinds of patterns for different purposes: shape-based patterns reduce the context of code to match; action patterns are the ones that correspond to fix patterns used in the repair community; token patterns are used for inferring collateral evolutions.

\section{Conclusion}
\label{sec:conclusion}
We have presented \toolname, a systematic and automated approach to mine relevant and actionable fix patterns for automated program repair. The approach builds on an iterative and three-fold clustering strategy, where in each round forming clusters of identical trees representing recurrent patterns.    

We assess the consistency of the mined patterns with the patterns in the literature. 
We further demonstrate with the implementation of an automated repair pipeline that the patterns mined by our approach are relevant for generating correct patches for 26 bugs in the Defects4J benchmark. These correct patches correspond to 81\% of all plausible patches generated by the tool.

\noindent {\bf Availability} All the data and tool support is available at :
\begin{center}
	{\tt \url{https://github.com/SerVal-DTF/fixminer-core}}
\end{center}.

% \paragraph*{Acknowledgments}

% This work is supported by the Fonds National de la Recherche
% (FNR), Luxembourg, through RECOMMEND 15/IS/10449467 and
% FIXPATTERN C15/IS/9964569.

\begin{acknowledgements}
This work is supported by the Fonds National de la Recherche (FNR), Luxembourg, through RECOMMEND 15/IS/10449467 and FIXPATTERN C15/IS/9964569.
\end{acknowledgements}
\balance

\bibliographystyle{spmpsci}
\bibliography{bib/references.bib}

\begin{thebibliography}{10}
\providecommand{\url}[1]{{#1}}
\providecommand{\urlprefix}{URL }
\expandafter\ifx\csname urlstyle\endcsname\relax
  \providecommand{\doi}[1]{DOI~\discretionary{}{}{}#1}\else
  \providecommand{\doi}{DOI~\discretionary{}{}{}\begingroup
  \urlstyle{rm}\Url}\fi

\bibitem{abreu2007accuracy}
Abreu, R., Zoeteweij, P., Van~Gemund, A.J.: On the accuracy of spectrum-based
  fault localization.
\newblock In: Testing: Academic and Industrial Conference Practice and Research
  Techniques-MUTATION (TAICPART-MUTATION 2007), pp. 89--98. IEEE (2007)

\bibitem{al2005diffx}
Al-Ekram, R., Adma, A., Baysal, O.: diffx: an algorithm to detect changes in
  multi-version xml documents.
\newblock In: Proceedings of the 2005 conference of the Centre for Advanced
  Studies on Collaborative research, pp. 1--11. IBM Press (2005)

\bibitem{andersen2010generic}
Andersen, J., Lawall, J.L.: Generic patch inference.
\newblock Automated software engineering \textbf{17}(2), 119--148 (2010)

\bibitem{andersen2012semantic}
Andersen, J., Nguyen, A.C., Lo, D., Lawall, J.L., Khoo, S.C.: Semantic patch
  inference.
\newblock In: Automated Software Engineering (ASE), 2012 Proceedings of the
  27th IEEE/ACM International Conference on, pp. 382--385. IEEE (2012)

\bibitem{bhatia2016automated}
Bhatia, S., Singh, R.: Automated correction for syntax errors in programming
  assignments using recurrent neural networks.
\newblock arXiv preprint arXiv:1603.06129  (2016)

\bibitem{bille2005survey}
Bille, P.: A survey on tree edit distance and related problems.
\newblock Theoretical computer science \textbf{337}(1-3), 217--239 (2005)

\bibitem{brunel2009foundation}
Brunel, J., Doligez, D., Hansen, R.R., Lawall, J.L., Muller, G.: A
  {{Foundation}} for {{Flow}}-based {{Program Matching}}: {{Using Temporal
  Logic}} and {{Model Checking}}.
\newblock In: Proceedings of the 36th {{Annual ACM SIGPLAN}}-{{SIGACT
  Symposium}} on {{Principles}} of {{Programming Languages}}, {{POPL}} '09, pp.
  114--126. {ACM}, {New York, NY, USA} (2009).
\newblock \doi{10.1145/1480881.1480897}

\bibitem{campos2012GZoltar}
Campos, J., Riboira, A., Perez, A., Abreu, R.: Gzoltar: an eclipse plug-in for
  testing and debugging.
\newblock In: Proceedings of the 27th IEEE/ACM International Conference on
  Automated Software Engineering, pp. 378--381. ACM (2012)

\bibitem{chawathe1996change}
Chawathe, S.S., Rajaraman, A., {Garcia-Molina}, H., Widom, J.: Change
  {{Detection}} in {{Hierarchically Structured Information}}.
\newblock In: Proceedings of the 1996 {{ACM SIGMOD International Conference}}
  on {{Management}} of {{Data}}, {{SIGMOD}} '96, pp. 493--504. {ACM}, {New
  York, NY, USA} (1996).
\newblock \doi{10.1145/233269.233366}

\bibitem{chen2017contract}
Chen, L., Pei, Y., Furia, C.A.: Contract-based program repair without the
  contracts.
\newblock In: Proceedings of the 32nd IEEE/ACM International Conference on
  Automated Software Engineering, pp. 637--647. IEEE, Urbana, IL, USA (2017)

\bibitem{chilowicz2009syntax}
Chilowicz, M., Duris, E., Roussel, G.: Syntax tree fingerprinting for source
  code similarity detection.
\newblock In: Program Comprehension, 2009. ICPC'09. IEEE 17th International
  Conference on, pp. 243--247. IEEE (2009)

\bibitem{coker2013program}
Coker, Z., Hafiz, M.: Program transformations to fix c integers.
\newblock In: Proceedings of the International Conference on Software
  Engineering, pp. 792--801. IEEE, San Francisco, CA, USA (2013)

\bibitem{dallmeier2009generating}
Dallmeier, V., Zeller, A., Meyer, B.: Generating fixes from object behavior
  anomalies.
\newblock In: Proceedings of the 2009 IEEE/ACM International Conference on
  Automated Software Engineering, pp. 550--554. IEEE Computer Society (2009)

\bibitem{duley2012vdiff}
Duley, A., Spandikow, C., Kim, M.: Vdiff: a program differencing algorithm for
  verilog hardware description language.
\newblock Automated Software Engineering \textbf{19}(4), 459--490 (2012)

\bibitem{durieux2017dynamic}
Durieux, T., Cornu, B., Seinturier, L., Monperrus, M.: Dynamic patch generation
  for null pointer exceptions using metaprogramming.
\newblock In: Proceedings of the 24th International Conference on Software
  Analysis, Evolution and Reengineering, pp. 349--358. IEEE (2017)

\bibitem{gumtree}
Falleri, J.R.: {GumTree}.
\newblock \url{https://github.com/GumTreeDiff/gumtree} (Last Access: Mar.
  2018.)

\bibitem{falleri2014Fine}
Falleri, J.R., Morandat, F., Blanc, X., Martinez, M., Monperrus, M.:
  Fine-grained and accurate source code differencing.
\newblock In: Proceedings of {ACM/IEEE} International Conference on Automated
  Software Engineering, pp. 313--324. ACM, Vasteras, Sweden (2014)

\bibitem{fischer2003populating}
Fischer, M., Pinzger, M., Gall, H.: Populating a release history database from
  version control and bug tracking systems.
\newblock In: Proceeding of the 19th ICSM, pp. 23--32. IEEE (2003)

\bibitem{fluri2006classifying}
Fluri, B., Gall, H.C.: Classifying change types for qualifying change
  couplings.
\newblock In: Program Comprehension, 2006. ICPC 2006. 14th IEEE International
  Conference on, pp. 35--45. IEEE (2006)

\bibitem{fluri2008discovering}
Fluri, B., Giger, E., Gall, H.C.: Discovering patterns of change types.
\newblock In: Proceedings of the 23rd IEEE/ACM International Conference on
  Automated Software Engineering, pp. 463--466. IEEE, L'Aquila, Italy (2008)

\bibitem{fluri2007change}
Fluri, B., Wuersch, M., PInzger, M., Gall, H.: Change distilling: Tree
  differencing for fine-grained source code change extraction.
\newblock IEEE Transactions on software engineering \textbf{33}(11) (2007)

\bibitem{gupta2017deepfix}
Gupta, R., Pal, S., Kanade, A., Shevade, S.: Deepfix: Fixing common c language
  errors by deep learning.
\newblock In: AAAI, pp. 1345--1351 (2017)

\bibitem{hanam2016discovering}
Hanam, Q., Brito, F.S.d.M., Mesbah, A.: Discovering bug patterns in javascript.
\newblock In: Proceedings of the 2016 24th ACM SIGSOFT International Symposium
  on Foundations of Software Engineering, pp. 144--156. ACM (2016)

\bibitem{hashimoto2008diff}
Hashimoto, M., Mori, A.: Diff/ts: A tool for fine-grained structural change
  analysis.
\newblock In: 2008 15th Working Conference on Reverse Engineering, pp.
  279--288. IEEE (2008)

\bibitem{kim2013Impact}
Herzig, K., Zeller, A.: The impact of tangled code changes.
\newblock In: Proceedings of the 10th Working Conference on Mining Software
  Repositories, {MSR} '13, pp. 121--130. IEEE, San Francisco, CA, USA (2013)

\bibitem{hovemeyer2004finding}
Hovemeyer, D., Pugh, W.: Finding bugs is easy.
\newblock ACM Sigplan Notices \textbf{39}(12), 92--106 (2004)

\bibitem{hua2018towards}
Hua, J., Zhang, M., Wang, K., Khurshid, S.: Towards practical program repair
  with on-demand candidate generation.
\newblock In: Proceedings of the 40th International Conference on Software
  Engineering, pp. 12--23. ACM (2018)

\bibitem{huang2018cldiff}
Huang, K., Chen, B., Peng, X., Zhou, D., Wang, Y., Liu, Y., Zhao, W.: Cldiff:
  generating concise linked code differences.
\newblock In: Proceedings of the 33rd ACM/IEEE International Conference on
  Automated Software Engineering, pp. 679--690. ACM (2018)

\bibitem{jaro1989advances}
Jaro, M.A.: Advances in record-linkage methodology as applied to matching the
  1985 census of tampa, florida.
\newblock Journal of the American Statistical Association \textbf{84}(406),
  414--420 (1989)

\bibitem{jiang2018shaping}
Jiang, J., Xiong, Y., Zhang, H., Gao, Q., Chen, X.: Shaping program repair
  space with existing patches and similar code.
\newblock In: Proceedings of the 27th ACM SIGSOFT International Symposium on
  Software Testing and Analysis, pp. 298--309. ACM (2018)

\bibitem{just2014defects4j}
Just, R., Jalali, D., Ernst, M.D.: Defects4j: A database of existing faults to
  enable controlled testing studies for java programs.
\newblock In: Proceedings of the 2014 International Symposium on Software
  Testing and Analysis, pp. 437--440. ACM, San Jose, CA, {USA} (2014)

\bibitem{ke2015repairing}
Ke, Y., Stolee, K.T., Le~Goues, C., Brun, Y.: Repairing programs with semantic
  code search.
\newblock In: Proceedings of the 30th IEEE/ACM International Conference on
  Automated Software Engineering (ASE), pp. 295--306. IEEE, Lincoln, NE, USA
  (2015)

\bibitem{kim2013automatic}
Kim, D., Nam, J., Song, J., Kim, S.: Automatic patch generation learned from
  human-written patches.
\newblock In: Proceedings of the 2013 International Conference on Software
  Engineering, pp. 802--811. IEEE Press (2013)

\bibitem{kim2009discovering}
Kim, M., Notkin, D.: Discovering and representing systematic code changes.
\newblock In: Proceedings of the 31st International Conference on Software
  Engineering, pp. 309--319. IEEE Computer Society (2009)

\bibitem{kim2007automatic}
Kim, M., Notkin, D., Grossman, D.: Automatic inference of structural changes
  for matching across program versions.
\newblock In: ICSE, vol.~7, pp. 333--343. Citeseer (2007)

\bibitem{kim2006memories}
Kim, S., Pan, K., Whitehead~Jr, E.: Memories of bug fixes.
\newblock In: Proceedings of the 14th ACM SIGSOFT international symposium on
  Foundations of software engineering, pp. 35--45. ACM (2006)

\bibitem{koyuncu2017impact}
Koyuncu, A., Bissyand{\'e}, T., Kim, D., Klein, J., Monperrus, M., Le~Traon,
  Y.: Impact of {Tool} {Support} in {Patch} {Construction}.
\newblock In: Proceedings of the 26th {ACM} {SIGSOFT} {International}
  {Symposium} on {Software} {Testing} and {Analysis}, pp. 237--248. ACM, New
  York, NY, USA (2017)

\bibitem{koyuncu2019d}
Koyuncu, A., Bissyand{\'e}, T.F., Kim, D., Liu, K., Klein, J., Monperrus, M.,
  Traon, Y.L.: D\&c: A divide-and-conquer approach to ir-based bug
  localization.
\newblock arXiv preprint arXiv:1902.02703  (2019)

\bibitem{koyuncu2019ifixr}
Koyuncu, A., Liu, K., Bissyand{\'e}, T.F., Kim, D., Monperrus, M., Klein, J.,
  Le~Traon, Y.: ifixr: bug report driven program repair.
\newblock In: Proceedings of the 2019 27th ACM Joint Meeting on European
  Software Engineering Conference and Symposium on the Foundations of Software
  Engineering, pp. 314--325. ACM (2019)

\bibitem{Kreutzer2016}
Kreutzer, P., Dotzler, G., Ring, M., Eskofier, B.M., Philippsen, M.: Automatic
  clustering of code changes.
\newblock In: Proceedings of the 13th International Conference on Mining
  Software Repositories, MSR '16, pp. 61--72. ACM, New York, NY, USA (2016).
\newblock \doi{10.1145/2901739.2901749}.
\newblock
  \urlprefix\url{http://doi.acm.org.proxy.bnl.lu/10.1145/2901739.2901749}

\bibitem{le2017s3}
Le, X.B.D., Chu, D.H., Lo, D., Le~Goues, C., Visser, W.: S3: syntax-and
  semantic-guided repair synthesis via programming by examples.
\newblock In: Proceedings of the 11th Joint Meeting on Foundations of Software
  Engineering, pp. 593--604. ACM, Paderborn, Germany (2017)

\bibitem{le2016enhancing}
Le, X.B.D., Le, Q.L., Lo, D., Le~Goues, C.: Enhancing automated program repair
  with deductive verification.
\newblock In: Proceedings of the International Conference on Software
  Maintenance and Evolution (ICSME), pp. 428--432. IEEE, Raleigh, NC, USA
  (2016)

\bibitem{le2016history}
Le, X.D., Lo, D., {Le Goues}, C.: History driven program repair.
\newblock In: Proceedings of the 23rd International Conference on Software
  Analysis, Evolution, and Reengineering, vol.~1, pp. 213--224. IEEE (2016)

\bibitem{le2012genprog}
Le~Goues, C., Nguyen, T., Forrest, S., Weimer, W.: {GenProg}: A generic method
  for automatic software repair.
\newblock TSE \textbf{38}(1), 54--72 (2012)

\bibitem{claire2012GenProg}
{Le Goues}, C., Nguyen, T., Forrest, S., Weimer, W.: Genprog: {A} generic
  method for automatic software repair.
\newblock {IEEE} Trans. Software Eng. \textbf{38}(1), 54--72 (2012)

\bibitem{lee2018bench4bl}
Lee, J., Kim, D., Bissyand{\'e}, T.F., Jung, W., Le~Traon, Y.: Bench4bl:
  reproducibility study on the performance of ir-based bug localization.
\newblock In: Proceedings of the 27th ACM SIGSOFT International Symposium on
  Software Testing and Analysis, pp. 61--72. ACM (2018)

\bibitem{lin2016empirical}
Lin, W., Chen, Z., Ma, W., Chen, L., Xu, L., Xu, B.: An empirical study on the
  characteristics of python fine-grained source code change types.
\newblock In: Software Maintenance and Evolution (ICSME), 2016 IEEE
  International Conference on, pp. 188--199. IEEE (2016)

\bibitem{liu2018mining2}
Liu, K., Kim, D., Bissyand{\'e}, T.F., Yoo, S., Le~Traon, Y.: Mining fix
  patterns for findbugs violations.
\newblock IEEE Transactions on Software Engineering  (2018)

\bibitem{liu2018closer}
Liu, K., Kim, D., Koyuncu, A., Li, L., Bissyand{\'e}, T.F., Le~Traon, Y.: A
  closer look at real-world patches.
\newblock In: 2018 IEEE International Conference on Software Maintenance and
  Evolution, pp. 275--286. IEEE (2018)

\bibitem{liu2019you}
Liu, K., Koyuncu, A., Bissyand{\'e}, T.F., Kim, D., Klein, J., Le~Traon, Y.:
  You cannot fix what you cannot find! an investigation of fault localization
  bias in benchmarking automated program repair systems.
\newblock In: 2019 12th IEEE Conference on Software Testing, Validation and
  Verification (ICST), pp. 102--113. IEEE (2019)

\bibitem{liu2019avatar}
Liu, K., Koyuncu, A., Kim, D., Bissyand{\'e}, T.F.: Avatar: Fixing semantic
  bugs with fix patterns of static analysis violations.
\newblock In: Proceedings of the IEEE 26th International Conference on Software
  Analysis, Evolution and Reengineering, pp. 456--467. IEEE (2019)

\bibitem{liu2019tbar}
Liu, K., Koyuncu, A., Kim, D., Bissyand{\'e}, T.F.: {TBar:} revisiting
  template-based automated program repair.
\newblock In: Proceedings of the 28th International Symposium on Software
  Testing and Analysis (2019)

\bibitem{liu2018lsrepair}
Liu, K., Koyuncu, A., Kim, K., Kim, D., Bissyand\'e, T.F.: {LSRepair}: Live
  search of fix ingredients for automated program repair.
\newblock In: Proceedings of the 25th Asia-Pacific Software Engineering
  Conference, pp. 658--662 (2018)

\bibitem{liu2018mining}
Liu, X., Zhong, H.: Mining stackoverflow for program repair.
\newblock In: Proceedings of the 25th International Conference on Software
  Analysis, Evolution and Reengineering, pp. 118--129. IEEE (2018)

\bibitem{livshits2005dynamine}
Livshits, B., Zimmermann, T.: {{DynaMine}}: {{Finding Common Error Patterns}}
  by {{Mining Software Revision Histories}}.
\newblock In: Proceedings of the 10th {{European Software Engineering
  Conference Held Jointly}} with 13th {{ACM SIGSOFT International Symposium}}
  on {{Foundations}} of {{Software Engineering}}, {{ESEC}}/{{FSE}}-13, pp.
  296--305. {ACM}, {New York, NY, USA} (2005).
\newblock \doi{10.1145/1081706.1081754}

\bibitem{long2017automatic}
Long, F., Amidon, P., Rinard, M.: Automatic inference of code transforms for
  patch generation.
\newblock In: Proceedings of the 11th Joint Meeting on Foundations of Software
  Engineering, pp. 727--739. ACM, Paderborn, Germany (2017)

\bibitem{long2015staged}
Long, F., Rinard, M.: Staged program repair with condition synthesis.
\newblock In: Proceedings of the 2015 10th Joint Meeting on Foundations of
  Software Engineering, pp. 166--178. ACM, Bergamo, Italy (2015)

\bibitem{long2016automatic}
Long, F., Rinard, M.: Automatic patch generation by learning correct code.
\newblock In: Proceedings of the 43rd Annual ACM SIGPLAN-SIGACT Symposium on
  Principles of Programming Languages, pp. 298--312. ACM, St. Petersburg, FL,
  USA (2016)

\bibitem{martinez2013automatically}
Martinez, M., Duchien, L., Monperrus, M.: Automatically extracting instances of
  code change patterns with ast analysis.
\newblock In: Software Maintenance (ICSM), 2013 29th IEEE International
  Conference on, pp. 388--391. IEEE (2013)

\bibitem{martinez2017automatic}
Martinez, M., Durieux, T., Sommerard, R., Xuan, J., Monperrus, M.: Automatic
  repair of real bugs in java: A large-scale experiment on the defects4j
  dataset.
\newblock Empirical Software Engineering \textbf{22}(4), 1936--1964 (2017)

\bibitem{martinez2015mining}
Martinez, M., Monperrus, M.: Mining software repair models for reasoning on the
  search space of automated program fixing.
\newblock Empirical Software Engineering \textbf{20}(1), 176--205 (2015)

\bibitem{martinez2016astor}
Martinez, M., Monperrus, M.: Astor: A program repair library for java.
\newblock In: Proceedings of the 25th International Symposium on Software
  Testing and Analysis, pp. 441--444. ACM, Saarbr{\"{u}}cken, Germany (2016)

\bibitem{martinez2018ultra}
Martinez, M., Monperrus, M.: Ultra-large repair search space with automatically
  mined templates: The cardumen mode of astor.
\newblock In: Proceedings of the 10th SSBSE, pp. 65--86. Springer (2018)

\bibitem{mechtaev2015directfix}
Mechtaev, S., Yi, J., Roychoudhury, A.: Directfix: Looking for simple program
  repairs.
\newblock In: Proceedings of the 37th International Conference on Software
  Engineering-Volume 1, pp. 448--458. IEEE, Florence, Italy (2015)

\bibitem{meng2011systematic}
Meng, N., Kim, M., McKinley, K.S.: Systematic editing: generating program
  transformations from an example.
\newblock ACM SIGPLAN Notices \textbf{46}(6), 329--342 (2011)

\bibitem{meng2013lase}
Meng, N., Kim, M., McKinley, K.S.: Lase: locating and applying systematic edits
  by learning from examples.
\newblock In: Proceedings of the 2013 International Conference on Software
  Engineering, pp. 502--511. IEEE Press (2013)

\bibitem{molderez2017mining}
Molderez, T., Stevens, R., De~Roover, C.: Mining change histories for unknown
  systematic edits.
\newblock In: Proceedings of the 14th International Conference on Mining
  Software Repositories, pp. 248--256. IEEE Press (2017)

\bibitem{monperrus2018automatic}
Monperrus, M.: Automatic software repair: a bibliography.
\newblock ACM Computing Surveys (CSUR) \textbf{51}(1), 17 (2018)

\bibitem{myers1986ano}
Myers, E.W.: Ano (nd) difference algorithm and its variations.
\newblock Algorithmica \textbf{1}(1-4), 251--266 (1986)

\bibitem{neamtiu2005understanding}
Neamtiu, I., Foster, J.S., Hicks, M.: Understanding source code evolution using
  abstract syntax tree matching.
\newblock ACM SIGSOFT Software Engineering Notes \textbf{30}(4), 1--5 (2005)

\bibitem{nguyen2013filtering}
Nguyen, H.A., Nguyen, A.T., Nguyen, T.N.: Filtering noise in mixed-purpose
  fixing commits to improve defect prediction and localization.
\newblock In: 2013 IEEE 24th International Symposium on Software Reliability
  Engineering (ISSRE), pp. 138--147. IEEE (2013)

\bibitem{nguyen2013semfix}
Nguyen, H.D.T., Qi, D., Roychoudhury, A., Chandra, S.: {SemFix}: program repair
  via semantic analysis.
\newblock In: Proceedings of the 35th ICSE, pp. 772--781. IEEE (2013)

\bibitem{nguyen2010recurring}
Nguyen, T.T., Nguyen, H.A., Pham, N.H., Al-Kofahi, J., Nguyen, T.N.: Recurring
  bug fixes in object-oriented programs.
\newblock In: Software Engineering, 2010 ACM/IEEE 32nd International Conference
  on, vol.~1, pp. 315--324. IEEE (2010)

\bibitem{osman2014mining}
Osman, H., Lungu, M., Nierstrasz, O.: Mining frequent bug-fix code changes.
\newblock In: Software Maintenance, Reengineering and Reverse Engineering
  (CSMR-WCRE), 2014 Software Evolution Week-IEEE Conference on, pp. 343--347.
  IEEE (2014)

\bibitem{oumarou2015identifying}
Oumarou, H., Anquetil, N., Etien, A., Ducasse, S., Taiwe, K.D.: Identifying the
  exact fixing actions of static rule violation.
\newblock In: Software Analysis, Evolution and Reengineering (SANER), 2015 IEEE
  22nd International Conference on, pp. 371--379. IEEE (2015)

\bibitem{padioleau2008documenting}
Padioleau, Y., Lawall, J., Hansen, R.R., Muller, G.: Documenting and
  {{Automating Collateral Evolutions}} in {{Linux Device Drivers}}.
\newblock In: Proceedings of the 3rd {{ACM SIGOPS}}/{{EuroSys European
  Conference}} on {{Computer Systems}} 2008, Eurosys '08, pp. 247--260. {ACM},
  {New York, NY, USA} (2008).
\newblock \doi{10.1145/1352592.1352618}

\bibitem{pan2009toward}
Pan, K., Kim, S., Whitehead, E.J.: Toward an understanding of bug fix patterns.
\newblock Empirical Software Engineering \textbf{14}(3), 286--315 (2009)

\bibitem{park2012empirical}
Park, J., Kim, M., Ray, B., Bae, D.H.: An empirical study of supplementary bug
  fixes.
\newblock In: Proceedings of the 9th IEEE Working Conference on Mining Software
  Repositories, pp. 40--49. IEEE Press (2012)

\bibitem{pawlik2011rted}
Pawlik, M., Augsten, N.: Rted: a robust algorithm for the tree edit distance.
\newblock Proceedings of the VLDB Endowment \textbf{5}(4), 334--345 (2011)

\bibitem{rolim2018learning}
Rolim, R., Soares, G., Gheyi, R., D'Antoni, L.: Learning quick fixes from code
  repositories.
\newblock arXiv preprint arXiv:1803.03806  (2018)

\bibitem{saha2017elixir}
Saha, R.K., Lyu, Y., Yoshida, H., Prasad, M.R.: Elixir: Effective
  object-oriented program repair.
\newblock In: Automated Software Engineering (ASE), 2017 32nd IEEE/ACM
  International Conference on, pp. 648--659. IEEE (2017)

\bibitem{skiena1997stony}
Skiena, S.S.: The stony brook algorithm repository.
\newblock URL http://www. cs. sunysb. edu/algorith/implement/nauty/implement.
  shtml  (1997)

\bibitem{defects4J-dissection}
Sobreira, V., Durieux, T., Madeiral, F., Monperrus, M., Maia, M.A.: {Dissection
  of a Bug Dataset: Anatomy of 395 Patches from Defects4J}.
\newblock In: Proceedings of SANER (2018)

\bibitem{tan2015relifix}
Tan, S.H., Roychoudhury, A.: relifix: Automated repair of software regressions.
\newblock In: Proceedings of the 37th International Conference on Software
  Engineering-Volume 1, pp. 471--482. IEEE Press (2015)

\bibitem{tao2015partitioning}
Tao, Y., Kim, S.: Partitioning composite code changes to facilitate code
  review.
\newblock In: 2015 IEEE/ACM 12th Working Conference on Mining Software
  Repositories, pp. 180--190. IEEE (2015)

\bibitem{thomas2013impact}
Thomas, S.W., Nagappan, M., Blostein, D., Hassan, A.E.: The impact of
  classifier configuration and classifier combination on bug localization.
\newblock TSE \textbf{39}(10), 1427--1443 (2013)

\bibitem{tian2012identifying}
Tian, Y., Lawall, J., Lo, D.: Identifying linux bug fixing patches.
\newblock In: Proceedings of the 34th International Conference on Software
  Engineering, pp. 386--396. IEEE Press (2012)

\bibitem{westley2009automatically}
Weimer, W., Nguyen, T., {Le Goues}, C., Forrest, S.: Automatically finding
  patches using genetic programming.
\newblock In: Proceedings of the 31st International Conference on Software
  Engineering, May 16-24,, pp. 364--374. {IEEE}, Vancouver, Canada (2009)

\bibitem{weissgerber2006identifying}
Weissgerber, P., Diehl, S.: Identifying refactorings from source-code changes.
\newblock In: Automated Software Engineering, 2006. ASE'06. 21st IEEE/ACM
  International Conference on, pp. 231--240. IEEE (2006)

\bibitem{wen2018context}
Wen, M., Chen, J., Wu, R., Hao, D., Cheung, S.C.: Context-aware patch
  generation for better automated program repair.
\newblock In: Proceedings of the 40th International Conference on Software
  Engineering, pp. 1--11. ACM (2018)

\bibitem{wen2016locus}
Wen, M., Wu, R., Cheung, S.C.: Locus: Locating bugs from software changes.
\newblock In: 2016 31st IEEE/ACM International Conference on Automated Software
  Engineering (ASE), pp. 262--273. IEEE (2016)

\bibitem{winkler1990string}
Winkler, W.E.: String comparator metrics and enhanced decision rules in the
  fellegi-sunter model of record linkage.  (1990)

\bibitem{xin2017leveraging}
Xin, Q., Reiss, S.P.: Leveraging syntax-related code for automated program
  repair.
\newblock In: Proceedings of the 32nd IEEE/ACM International Conference on
  Automated Software Engineering, pp. 660--670. IEEE (2017)

\bibitem{xiong2017Precise}
Xiong, Y., Wang, J., Yan, R., Zhang, J., Han, S., Huang, G., Zhang, L.: Precise
  condition synthesis for program repair.
\newblock In: Proceedings of the 39th International Conference on Software
  Engineering, pp. 416--426. IEEE, Buenos Aires, Argentina (2017)

\bibitem{xuan2017Nopol}
Xuan, J., Martinez, M., DeMarco, F., Clement, M., Marcote, S.L., Durieux, T.,
  Le~Berre, D., Monperrus, M.: Nopol: Automatic repair of conditional statement
  bugs in java programs.
\newblock IEEE Transactions on Software Engineering \textbf{43}(1), 34--55
  (2017)

\bibitem{ying2004predicting}
Ying, A.T., Murphy, G.C., Ng, R., Chu-Carroll, M.C.: Predicting source code
  changes by mining change history.
\newblock IEEE transactions on Software Engineering \textbf{30}(9), 574--586
  (2004)

\bibitem{yue2017characterization}
Yue, R., Meng, N., Wang, Q.: A characterization study of repeated bug fixes.
\newblock In: Software Maintenance and Evolution (ICSME), 2017 IEEE
  International Conference on, pp. 422--432. IEEE (2017)

\end{thebibliography}

\begin{wrapfigure}{l}{0.2\textwidth}
    \includegraphics[width=1in,height=1.25in,clip,keepaspectratio]{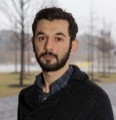}%
\end{wrapfigure}\par

\textbf{Anil Koyuncu} is a PhD student at the University of Luxembourg. He received a Master's degree from Politecnico di Milano, Italy. His research interest includes automatic patch repair, fault localization.
\newline
\par
\mbox{} \\

\begin{wrapfigure}{l}{0.2\textwidth}
    \includegraphics[width=1in,height=1.25in,clip,keepaspectratio]{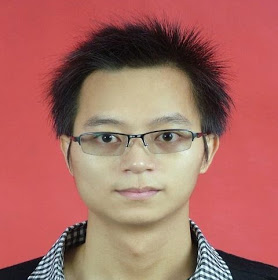}%
\end{wrapfigure}\par
\textbf{Kui Liu} received the master degree in computer engineering from Southwest University, China, in 2013. He is working toward the PhD degree in software engineering at the University of Luxembourg from 2016. His current research focuses on automated program repair. 
\newline
\par
\mbox{} \\

\begin{wrapfigure}{l}{0.2\textwidth}
   \includegraphics[width=1in,height=1.25in,clip,keepaspectratio]{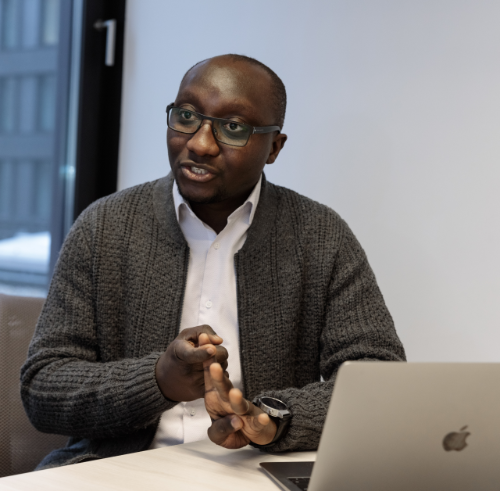}%
\end{wrapfigure}\par
\textbf{Tegawend{\'e} F. Bissyand{\'e}} 
is research scientist with the Interdisciplinary Center for Security, Reliability and Trust at the University of Luxembourg. He holds a PhD in computer from the Universit{\'e} de Bordeaux in 2013, and an engineering degree (MSc) from ENSEIRB. His research interests are in debugging, including bug localization and program repair, as well as code search, including code clone detection and code classification. He has published research results in all major venues in Software engineering (ICSE, ESEC/FSE, ASE, ISSTA, EMSE, TSE). His research is supported by FNR (Luxembourg National Research Fund). Dr. Bissyand{\'e} is the PI of the CORE RECOMMEND project on program repair, under which the current work has been performed.
\par

\mbox{} \\
% \newpage
% \newpage
\begin{wrapfigure}{l}{0.2\textwidth}
    \includegraphics[width=1in,height=1.25in,clip,keepaspectratio]{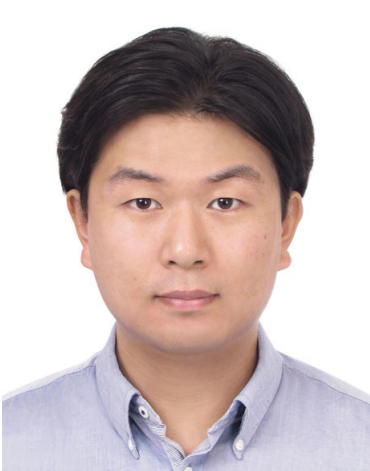}%
\end{wrapfigure}\par
\textbf{Dongsun Kim} is a Software Engineer at Furiosa.ai. He was formerly a research associate at the University of Luxembourg and a post-doctoral fellow at the Hong Kong University of Science and Technology. His research interest includes testing AI systems, automatic patch generation, fault localization, static analysis, and search-based software engineering. In particular, automated debugging is his current focus. His recent work has been recognized by several awards such as a featured article of the IEEE Transactions on Software Engineering (TSE) and ACM SIGSOFT Distinguished Paper of the International Conference on Software Engineering (ICSE). He is leading the FIXPATTERN project funded by FNR (Luxembourg National Research Fund) CORE programme.
\par
\mbox{} \\

\begin{wrapfigure}{l}{0.2\textwidth}
    \includegraphics[width=1in,height=1.25in,clip,keepaspectratio]{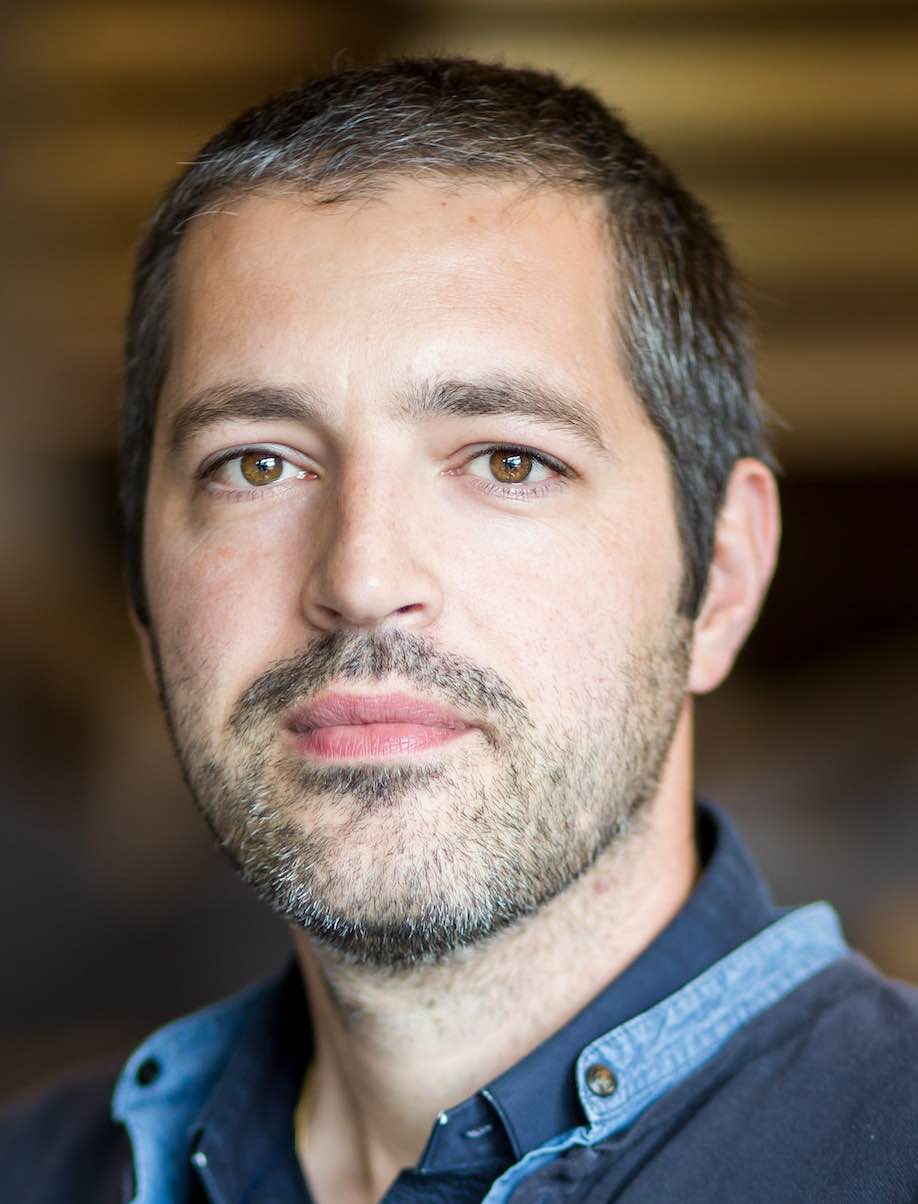}%
\end{wrapfigure}\par
\par
\textbf{Jacques Klein} is senior research scientist at the University of Luxembourg, and at the Interdisciplinary
Centre for Security, Reliability and Trust (SnT). He received his Ph.D. degree in Computer Science from the
University of Rennes, France in 2006. His main areas of expertise are threefold: (1) Mobile Security (malware
detection, prevention and dissection, static analysis for security, vulnerability detection, etc.); (2) Software
Reliability (software testing, semi-automated and fully-automated program repair, etc.); (3) Data Analytics
(multi-objective reasoning and optimization, model-driven data analytics, time series pattern recognition, text
mining, etc.). In addition to academic achievements, Dr. Klein has also standing experience and expertise on
successfully running industrial projects with several industrial partners in various domains by applying data
analytics, software engineering, information retrieval, etc., to their research problems.
\par
\mbox{} \\

\begin{wrapfigure}{l}{0.2\textwidth}
    \includegraphics[width=1in,height=1.25in,clip,keepaspectratio]{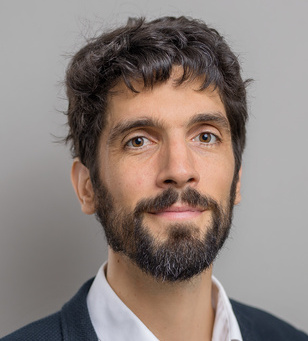}%
\end{wrapfigure}
\par
\textbf{Martin Monperrus} is Professor of Software Technology at KTH Royal Institute of Technology. He was previously associate professor at the University of Lille and adjunct researcher at Inria. He received a Ph.D. from the University of Rennes, and a Master's degree from the Compiègne University of Technology. His research lies in the field of software engineering with a current focus on automatic program repair, program hardening and chaos engineering. 
\par

\mbox{} \\

\begin{wrapfigure}{l}{0.2\textwidth}
    \includegraphics[width=1in,height=1.25in,clip,keepaspectratio]{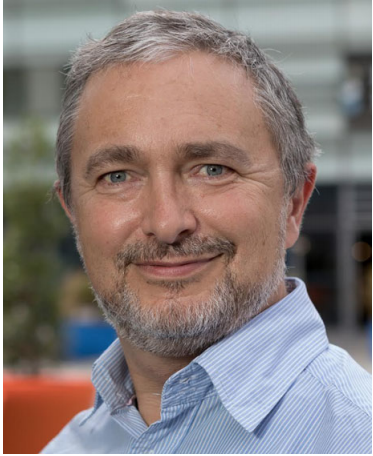}%

\end{wrapfigure}

\par
\textbf{Yves Le Traon}
 is professor at University of Luxembourg, in the domain of software engineering, testing,
security and model-driven engineering. He received his engineering degree and his PhD in Computer Science at the “Institut National Polytechnique” in Grenoble, France, in 1997. From 1998 to 2004, he was an
associate professor at the University of Rennes, in Brittany, France. From 2004 to 2006, he was an expert
in Model-Driven Architecture and Validation at “France Te le com R\&D”. In 2006, he became professor at 
Telecom Bretagne (Ecole Nationale des Tlcommunications de Bretagne). He is currently the head of the
CSC Research Unit (e.g. Department of Computer Science) at University of Luxembourg. He is a member of
the Interdisciplinary Centre for Security, Reliability and Trust (SnT), where he leads the research group SERVAL (SEcurity Reasoning and VALidation). His research interests include software testing, model-driven
engineering, model based testing, evolutionary algorithms, software security, security policies and Android
security. The current key-topics he explores are related to Internet of things (IoT), Big Data (stress testing,
multi-objective optimization and data protection), and mobile security and reliability. He is author of more
than 140 publications in international peer-reviewed conferences and journals.
\par
\end{document}